\documentclass{article}

\usepackage{microtype}
\usepackage{graphicx}
\usepackage{subcaption}
\usepackage{booktabs} 

\usepackage{hyperref}
\usepackage{xurl}   
\usepackage{url}
\usepackage{enumitem}
\usepackage{wrapfig}
\usepackage{makecell} 
\usepackage{array}
\usepackage{siunitx}
\usepackage{tabularx}
\usepackage{xcolor}
\usepackage{threeparttable}
\usepackage{adjustbox}

\usepackage[dvipsnames,table]{xcolor}
 
\usepackage{subcaption} 

\usepackage{tcolorbox}
\usepackage{algorithmic}
\usepackage{amsmath}
\usepackage{amssymb}

\usepackage{fontawesome5}
\newcommand{\ftcheck}{\faCheck}
\newcommand{\ftcross}{\faTimes}

\usepackage{booktabs,longtable}

\usepackage[preprint]{icml2026}

\usepackage{amsmath}
\usepackage{amssymb}
\usepackage{mathtools}
\usepackage{amsthm}

\usepackage[capitalize,noabbrev]{cleveref}

\theoremstyle{plain}

\theoremstyle{definition}

\theoremstyle{remark}

\usepackage[textsize=tiny]{todonotes}

\icmltitlerunning{From Trace to Line: LLM Agent for Real-World OSS Vulnerability Localization}

\begin{document}

\twocolumn[
  \icmltitle{From Trace to Line: LLM Agent for Real-World OSS Vulnerability Localization}



  \icmlsetsymbol{equal}{*}

  
  \begin{icmlauthorlist}
      \icmlauthor{Haoran Xi}{equal,tandon}
      \icmlauthor{Minghao Shao}{equal,tandon,ad}
      \icmlauthor{Brendan Dolan-Gavitt}{xbow}
      \icmlauthor{Muhammad Shafique}{ad}
      \icmlauthor{Ramesh Karri}{tandon}
  \end{icmlauthorlist}
    
  \icmlaffiliation{tandon}{NYU Tandon School of Engineering}
  \icmlaffiliation{ad}{NYU Abu Dhabi}
  \icmlaffiliation{xbow}{XBOW}

  \icmlcorrespondingauthor{Haoran Xi}{hx759@nyu.edu}
  \icmlcorrespondingauthor{Minghao Shao}{shao.minghao@nyu.edu}




  \icmlkeywords{Machine Learning, ICML}

  \vskip 0.3in
]



\printAffiliationsAndNotice{\icmlEqualContribution}

\begin{abstract}

Large language models show promise for vulnerability discovery, yet prevailing methods inspect code in isolation, struggle with long contexts, and focus on coarse function- or file-level detections that offer limited guidance to engineers who need precise line-level localization for targeted patches. We introduce T2L, an executable framework for project-level, line-level vulnerability localization that progressively narrows scope from repository modules to exact vulnerable lines via AST-based chunking and evidence-guided refinement. We provide a baseline agent with an Agentic Trace Analyzer (ATA) that fuses runtime evidence such as crash points and stack traces to translate failure symptoms into actionable diagnoses. To enable rigorous evaluation, we introduce T2L-ARVO, an expert-verified 50-case benchmark spanning five crash families in real-world projects. On T2L-ARVO, our baseline achieves up to 58.0\% detection and 54.8\% line-level localization rate. Together, T2L framework advance LLM-based vulnerability detection toward deployable, precision diagnostics in open-source software workflows. Our framework and benchmark are publicly available as open source at \url{https://github.com/haoranxi/T2LAgent}.



\end{abstract}

\section{Introduction}

Software vulnerabilities are widespread and costly, yet automated methods often provide coarse, file- or function-level predictions that are hard to turn into targeted fixes. Large language models (LLMs) have shown promise for code comprehension and vulnerability discovery, but today's LLM-based approaches still face practical barriers for real remediation. In 2023, more than 29,000 CVEs were recorded \cite{cve_details_2024}. The economic toll is mounting: software supply chain attacks are projected to cost the global economy 80.6 billion annually by 2026 \cite{darkreading_supply_chain}. Critically, 14\% of breaches in 2024 began with vulnerability exploitation, nearly triple the prior year. Despite advances in automated detection, localization remains underexplored \cite{zhang2024empirical}, leaving developers with coarse predictions while 32\% of critical vulnerabilities remain unpatched for over 180 days.

\textbf{Motivation.} The rise of LLMs has accelerated AI-driven automation across software development and cybersecurity \cite{zhang_et_al_2025}, from code assistants \cite{github_copilot_2021} to automated security analysis \cite{nunez2024autosafecoder}. LLMs show strong aptitude for understanding complex codebases and flagging potential issues \cite{divakaran2024llms}, naturally extending to vulnerability detection through fine-tuning and agentic frameworks \cite{awesome_llm4cybersecurity_2024}.

Yet today's LLM-based methods face practical barriers. Most operate at the function level, asking whether a fragment is vulnerable rather than pinpointing where the flaw lies \cite{zhang2024empirical, 10.1145/3715758}. These tasks are defined mainly on static code, ignoring runtime behavior and project-level context. Evaluations rely on synthetic datasets that miss the complexity of production systems \cite{springer_vulnerability_analysis}, while real-world debugging requires project-scale reasoning across multiple files. Engineers must navigate large repositories with cross-module dependencies and need line-level localization to craft targeted patches. Moreover, vulnerability analysis is often posed as static classification, ignoring diagnostic evidence such as crashes, sanitizer reports, and stack traces. As a result, the research-practice gap is most visible in localization: engineers need line-level guidance to craft patches, yet current systems rarely provide precise, actionable locations in complex repositories.

To address these gaps, we introduce T2L (Trace-to-Line), which reframes vulnerability localization as a two-tier problem: (a) coarse-grained detection: flagging suspicious code chunks, and (b) fine-grained localization: pinpointing exact vulnerable lines. This separation enables systematic evaluation of LLM capabilities from repository-scale narrowing to expert-level line pinpointing.

\textbf{Contribution.} This work makes three contributions: \textbf{(1)} \textit{New task formulation.} We define runtime-trace-guided, project-scale vulnerability localization as two structured subtasks: chunk-level detection and line-level localization, enabling evidence-guided refinement from coarse predictions to exact lines. \textbf{(2)} \textit{T2L-ARVO benchmark.} We present the first benchmark for agentic fine-grained localization, featuring 50 expert-verified cases across five vulnerability types with balanced category distribution. \texttt{T2L-ARVO} enables realistic, project-scale evaluation of LLM-based systems. \textbf{(3)} \textit{Evaluation framework and agent baseline.} We introduce an evaluation framework that standardizes environment setup, evidence collection, and line-level metrics, together with an agent baseline integrating an Agentic Trace Analyzer, iterative hypothesis generation, and two-stage refinement. On \texttt{T2L-ARVO}, our baseline achieves up to \textbf{58\%} chunk-level detection and \textbf{54.8\%} exact line localization.

\vspace{-2mm}
\section{Background}
\vspace{-2mm}


\textbf{Vulnerability Localization.} Classical localization combines static and dynamic analyses with retrieval and graph methods. Static slicing selects statements affecting a slicing criterion for debugging~\cite{10.5555/800078.802557}. Dynamic slicing derives input-specific slices from execution-time dependence graphs~\cite{agrawal1990dynamic}. Information-retrieval approaches rank files/methods by textual and historical signals~\cite{zhou2012should,6693093,10.1145/2597008.2597148}. Code Property Graphs unify program-analysis concepts for scalable pattern searches~\cite{yamaguchi2014modeling}.

\textbf{AI for Cybersecurity.} AI is widely applied to cybersecurity with tool execution in runnable environments~\cite{yang2023intercode}. Offensive security agents~\cite{shao2024empirical} reason about attack chains and interact with benchmarks~\cite{shao2024nyu, zhang2024cybench} to simulate CTF players~\cite{udeshi2025d,shao2025craken,abramovichenigma}. Pen-testing systems automate reconnaissance and exploitation~\cite{deng2024pentestgpt,shen2025pentestagent}. Agent-based frameworks localize vulnerabilities and synthesize fixes~\cite{yu2025patchagent,xue2025pagent}, while LLM-augmented red/blue teaming assists threat hunting~\cite{abuadbba2025promise,liu2025benchmarking}.

\textbf{LLM Agentic Systems.} Beyond cybersecurity, LLMs automate workflows across diverse domains~\cite{shao2024survey}, including scientific research~\cite{bran2023chemcrow,basit2025pennylang,jin2024prollm}, software engineering~\cite{yang2024swe}, healthcare~\cite{isern2016systematic}, and hardware design~\cite{wang2024llms}. Systems integrate memory for retaining states across sessions~\cite{zhang2025survey}. LLM-as-judge evaluates actions and provides feedback~\cite{li2025generation}. Agents employ self-reflection to refine reasoning~\cite{renze2024self}, and multi-agent architectures enable collaboration on complex subtasks~\cite{li2024survey}.

\begin{figure*}[!t]
    \centering
    \includegraphics[width=0.9\linewidth]
    {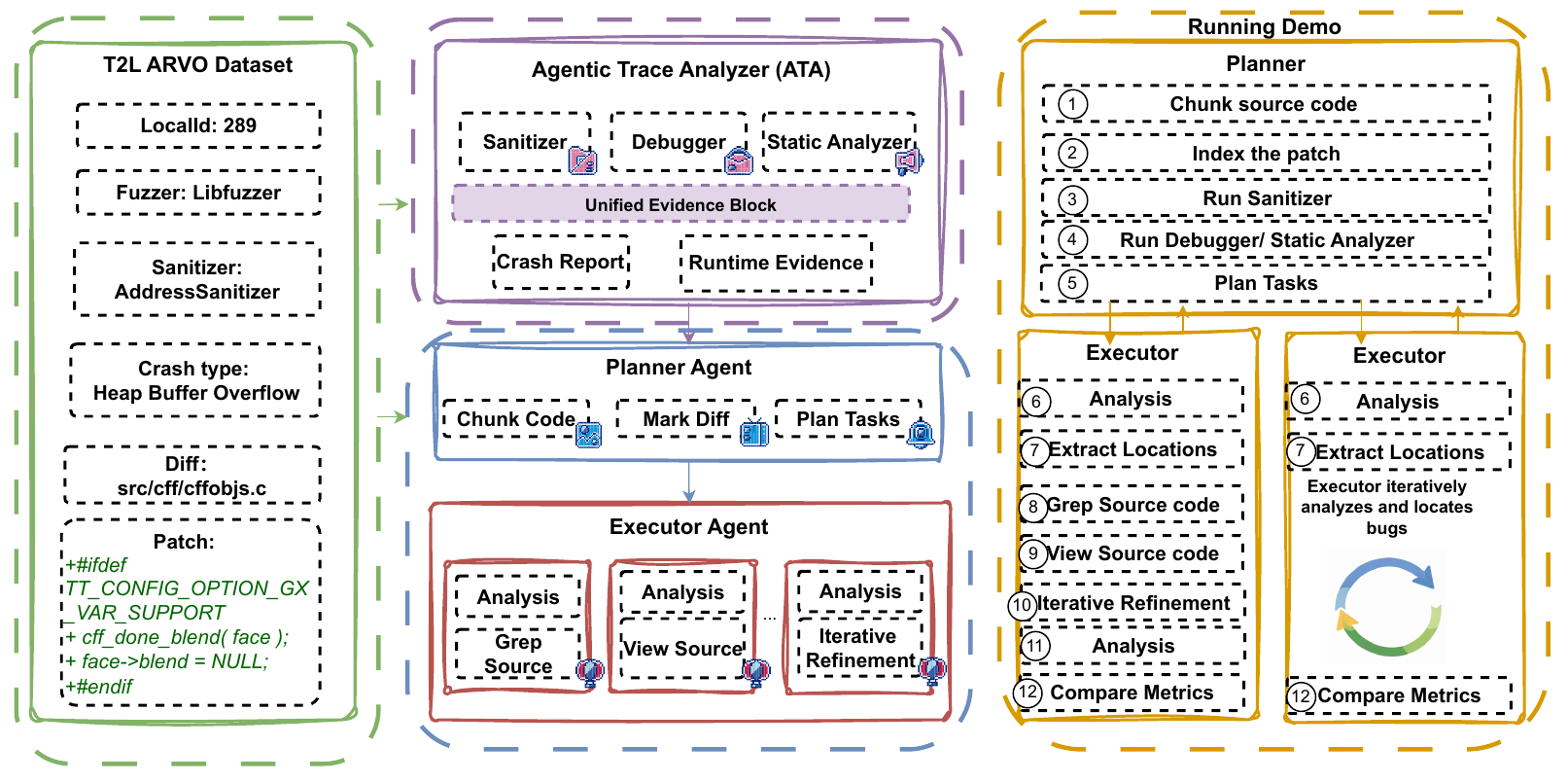}
    \caption{T2L-Agent Framework overview. Left: input from T2L ARVO dataset. Center: Agentic Trace Analyzer (ATA) with Planner and Executor agents. Right: running demo showing the iterative planning-execution workflow for vulnerability localization.}
    \label{fig:overview_flow}
\end{figure*}

\section{Related Work} 
LLMs show strong potential for vulnerability localization. LLMAO~\cite{yang2024large} fine-tunes LLMs on manually curated buggy programs, while BAP~\cite{stein2025s} learns localization without direct labels, outperforming baselines over eight benchmarks. However, data issues in Big-Vul and Devign~\cite{zhou2019devign} raise concerns about evaluation validity~\cite{croft2023data}.

Many works frame localization at file or function level, offering limited guidance for developers. GenLoc~\cite{asad2025leveraging} identifies vulnerable files from bug reports iteratively. AgentFL~\cite{qin2024agentfl} applies multi-agent frameworks for function-level localization. CoSIL~\cite{jiang2025cosil} narrows search space using module call graphs. AutoFL~\cite{kang2024quantitative} prompts LLMs via function-call navigation, showing multi-step reasoning mitigates context limits.

Recent studies shift toward line-level localization. LineVul~\cite{9796256} uses Transformer-based classifiers, LOVA~\cite{li2024attention} introduces self-attention scoring, MatsVD~\cite{10.1145/3671016.3674807} adds dependency-aware attention, and xLoc~\cite{10.1145/3660804} learns multilingual knowledge. LLM4FL~\cite{rafi2024multi} leverages graph-based retrieval, and MemFL~\cite{yeo2025improving} incorporates external memory for repository-scale systems. However, most still rely on limited runtime evidence or benchmarks lacking realistic project settings.

T2L formalizes trace-guided, project-scale, line-level localization. Unlike LLMAO, BAP, and LineVul that rely on static signals and operate on single files, our task targets training-free pipelines on multi-file codebases, incorporating runtime evidence including crash logs, stack traces, and sanitizer reports to pinpoint exact vulnerable lines. This formulation reflects real-world triage workflows where developers rely on both static inspection and dynamic behavior.

\begin{table}[htbp]
  \centering
  \scriptsize
  \renewcommand{\arraystretch}{0.75}
  \setlength{\tabcolsep}{1.6pt} 
  \renewcommand{\arraystretch}{1.05}
  \caption{Related Works}
  \begin{tabular}{@{}lccccccccccccc@{}}
    \toprule
    & \rotatebox{90}{LLMAO~\citeyear{yang2024large}}
    & \rotatebox{90}{BAP~\citeyear{stein2025s}}
    & \rotatebox{90}{GenLoc~\citeyear{asad2025leveraging}}
    & \rotatebox{90}{AgentFL~\citeyear{qin2024agentfl}}
    & \rotatebox{90}{CoSIL~\citeyear{jiang2025cosil}}
    & \rotatebox{90}{AutoFL~\citeyear{kang2024quantitative}}
    & \rotatebox{90}{LineVul~\citeyear{9796256}}
    & \rotatebox{90}{LOVA~\citeyear{li2024attention}}
    & \rotatebox{90}{MatsVD~\citeyear{10.1145/3671016.3674807}}
    & \rotatebox{90}{xLoc~\citeyear{10.1145/3660804}}
    & \rotatebox{90}{LLM4FL~\citeyear{rafi2024multi}}
    & \rotatebox{90}{MemFL~\citeyear{yeo2025improving}}
    & \rotatebox{90}{\textbf{T2L (ours)}} \\
    \midrule

    \textbf{Line Level Localization}
    & \ftcheck & \ftcheck & \ftcross & \ftcross & \ftcross & \ftcross & \ftcheck & \ftcheck & \ftcross & \ftcheck & \ftcross & \ftcross & \ftcheck \\

    \textbf{Multiple Agents}
    & \ftcross & \ftcross & \ftcross & \ftcheck & \ftcross & \ftcheck & \ftcross & \ftcross & \ftcross & \ftcross & \ftcheck & \ftcross & \ftcheck \\

    \textbf{Runtime Evidence}
    & \ftcross & \ftcross & \ftcheck & \ftcheck & \ftcross & \ftcheck & \ftcross & \ftcross & \ftcross & \ftcross & \ftcheck & \ftcheck & \ftcheck \\

    \textbf{Iterative loop}
    & \ftcross & \ftcross & \ftcheck & \ftcheck & \ftcheck & \ftcheck & \ftcross & \ftcross & \ftcross & \ftcross & \ftcheck & \ftcheck & \ftcheck \\

    \bottomrule
  \end{tabular}
  \label{tab:related_work_comparison}
  \vspace{-3mm}
\end{table}

\vspace{-1mm}
\section{Methods}
\vspace{-2mm}
\subsection{T2L Baseline Solver}
\label{sec:t2lagent}
To support the proposed trace-to-line evaluation task, we provide \texttt{T2L-Agent} as a baseline solver that implements an end-to-end workflow from runtime traces to line-level localization under a reproducible evaluation framework. The solver uses a hierarchical planner–executor design \cite{udeshi2025d} that splits localization into evidence gathering, hypothesis generation, and iterative refinement. Unlike single-pass analyzers, it uses a human-like process: gather runtime signals, link them to code, and narrow the search space. Figure~\ref{fig:overview_flow} provides an overview of the end-to-end T2L workflow, including structure of T2L ARVO dataset, trace-driven evidence collection via ATA, Planner and Executor Agent, and the Evaluation running demo.
Details of the tool list that T2L originally supports are provided in ~\ref{app:toolkit}.


\textbf{Evidence Tracing} T2L Planner coordinates repository analysis and runtime evidence capture in a structured pipeline. First, code-structure analysis partitions the repositories using tree-sitter into function-aligned, semantically meaningful chunks that preserve syntactic relationships while fitting LLM context windows. Patch locations are indexed for evaluation. Next, the solver collects runtime evidence as the primary signal for trace-to-line localization task.

We collect multiple types of runtime and static evidence, including crash logs, sanitizer reports(memory-violation patterns and allocation context), stack traces, debugger snapshots, and static analyzer results. All evidence is merged into a unified textual block and fed to the LLM. This design ensures that the LLM forms its reasoning based on global evidence context, rather than aggregating per-evidence predictions. The integrated workflow is a key design in our evaluation framework and baseline solver as detailed in Agentic Trace Analyzer.

\textbf{Agentic Trace Analyzer (ATA)}
\label{sec:ata}
The Agentic Trace Analyzer (ATA) is the core module of \texttt{T2L-Agent} baseline solver. Its purpose is to connect runtime behavior with relevant source code regions by running targets in reproducible Docker, and instrumenting executions with analysis toolkit such as Sanitizers, Debuggers, and Static analyzers to collect stack traces, memory errors, and other runtime cues. The resulting evidence is constructed into a unified ``evidence block" which is then passed to \texttt{T2L-Agent} baseline solver’s subsequent hypothesis generation and refinement steps. This ATA is designed to match the characteristics of the trace-guided, repo-scale, line-level localization task, where crash points alone are often insufficient to identify the root-cause lines. 


\textbf{Hypothesis Generation}
A ``hypothesis" is a candidate vulnerable location proposed by the LLM given the unified evidence block for a repository. The LLM proposes a list of \texttt{file:line} candidate pairs ranked by confidence. Each hypothesis is independently checked against the ground-truth patch diff, and candidates do not affect one another during evaluation.




\textbf{Feedback Control} Each iteration outputs a brief task summary with success indicators and confidence. The Planner uses these signals to decide the next step: continue refining or stop, preventing premature termination and avoiding over-analysis under a fixed cost budget.

\begin{figure*}
  \centering
  \begin{subfigure}[t]{0.32\linewidth}
    \centering
    \includegraphics[width=\linewidth]{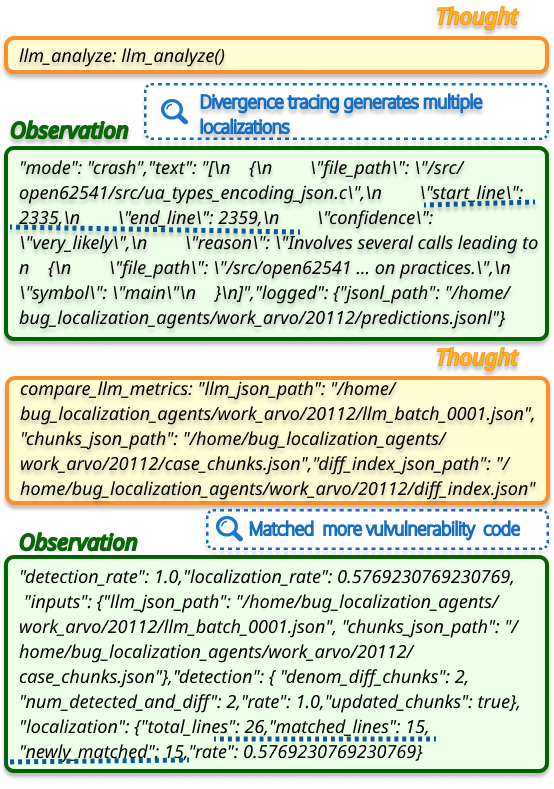}
    \caption{Divergence Tracing}
  \end{subfigure}\hfill
  \begin{subfigure}[t]{0.32\linewidth}
    \centering
    \includegraphics[width=\linewidth]
    {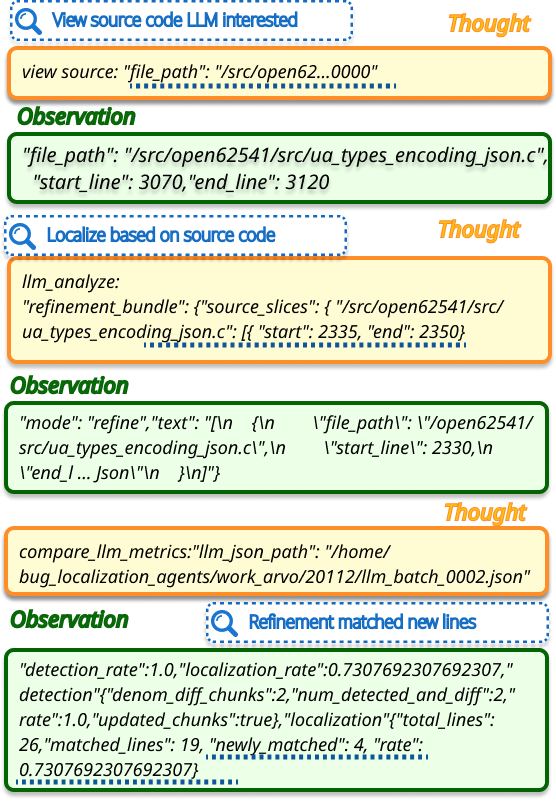}
    \caption{Detection Refinement}
  \end{subfigure}\hfill
  \begin{subfigure}[t]{0.32\linewidth}
    \centering
    \includegraphics[width=\linewidth]{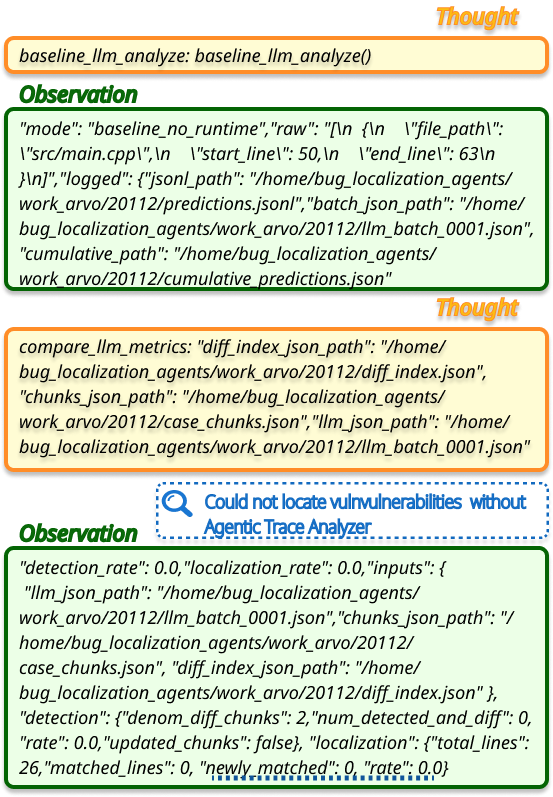}
    \caption{Agentic Trace Analyzer}
  \end{subfigure}\hfill
  \caption{Partial \texttt{T2L-Agent} logs to show the how the three proposed technique on \texttt{T2L-Agent} work and help the task: Divergence Tracing, Detection Refinement and Agentic Trace Analyzer.}
  \label{fig:arvo-3features}
  \vspace{-3mm}
\end{figure*}

\subsection{T2L Features}

\textbf{Divergence Tracing.} Recognizing that complex vulnerabilities often involves multiple files and functions, \texttt{T2L-Agent} uses divergence tracing to improve coverage. 
Rather than committing to a single chain of thought, it explores multiple parallel reasoning branches from the same evidence block and aggregates them into a ranked list of candidates 
This helps recover correct localizations that may not appear in the top hypothesis of a single run and is especially helpful for bugs that span multiple modules. From Figure~\ref{fig:arvo-3features} (a), we can observe that the divergence tracing generates more localization candidates and matched more vulnerable lines in this round.

\textbf{Detection Refinement.} To reflect the coarse-to-fine nature of the localization task, T2L baseline solver uses a iterative two-stage refinement loop. It begins with \texttt{coarse-grained filtering}: the system derives broad candidate regions from crash logs, using LLM code comprehension to link symptoms to likely causes and produce a ranked list of suspicious location candidates with confidence and rationale. Each LLM-generated candidate gets a confidence label from \texttt{very unlikely} to \texttt{very likely}. We use these labels to sort and prioritize candidates for subsequent inspection.

Instead of exhaustively scanning long code contexts, T2L selects targeted code slices guided by the crash log, stack trace information, and common vulnerability patterns such as missing bounds checks, uninitialized variables, or improper memory management. Concretely, we extract code fragments that the LLM explicitly flagged as relevant in the first pass, and enter the \texttt{fine-grained refinement} stage: these extracted fragments are appended back to the unified evidence as additional context providing richer context surrounding the initially flagged locations for a second LLM pass, which updates the candidate set. Newly proposed locations are merged with existing candidates. This refinement loop continues until no new candidates emerge or the cost budget is reached.

The refinement process operates iteratively to help the agent correct early mistakes and discover vulnerabilities that are not obvious from traces, particularly for complex vulnerabilities involving memory corruption bugs where the crash point can be far from the actual vulnerability. As Figure~\ref{fig:arvo-3features}(b) shows, incorporating the LLM-identified source snippets enables refinement to discover new vulnerable lines that were missed in the initial pass using only the runtime evidence.

\vspace{-2mm}
\subsection{T2L-ARVO Benchmark}
\label{sec:benchmarks}


The original \texttt{ARVO} dataset contains over 4{,}993 reproducible vulnerabilities across 250{+} C/C{++} projects, but its human-oriented design and build-centric structure do not directly support evaluating \emph{agentic} vulnerability-localization systems. To bridge this gap, we developed \texttt{T2L-ARVO}, a 50-case benchmark derived from \texttt{ARVO} for assessing LLM agents in evidence-guided vulnerability localization, rather than human bug reproduction.\texttt{T2L-ARVO} maintains a moderate, calibrated difficulty while keeping the original crash-type diversity. Its construction employs a dual validation pipeline combining manual expert review with LLM-assisted checks to ensure each case is reproducible and appropriately challenging for automated agents. This yields a realistic, balanced benchmark for rigorous end-to-end evaluation of LLM-based vulnerability localization. In \texttt{T2L-ARVO}, each of the cases has multiple ground-truth bug lines. The average accuracy overall is denoted as $\textit{Accuracy} = \textit{correct localized vulnerable lines} / \textit{total vulnerable lines}$.


\textbf{T2L-ARVO Composition} We analyzed \(4{,}993\) ARVO instances and grouped them by underlying failure mechanism: \emph{Buffer Overflows} \(49.9\%\) \((n=2{,}490)\), \emph{Uninitialized Access \& Unknown States} \(35.4\%\) \((n=1{,}768)\), \emph{Memory Lifecycle Errors} \(11.5\%\) \((n=573)\), \emph{Type Safety \& Parameter Validation} \(2.9\%\) \((n=147)\), and \emph{System \& Runtime Errors} \(0.3\%\) \((n=15)\). Each family subsumes concrete subtypes (e.g., \texttt{heap-buffer-overflow}, \texttt{use-of-uninitialized-value}, \texttt{heap-use-after-free}, \texttt{bad-cast}). \texttt{T2L-ARVO} deliberately mirrors this distribution to avoid bias toward any single failure mode.

The final benchmark comprises 50 vulnerabilities samples, evenly sampled across five crash families (10 each) for broad yet controlled difficulty. Each family includes representative subtypes, covering single-file defects and cross-module interactions to prevent overflow bias and exercise diverse failure modes observed in real repositories.

\begin{table}[t]
  \centering
  \caption{Crash types in \texttt{T2L-ARVO} Bench.}
  \label{tab:crash-types}
  \scriptsize
  \renewcommand{\arraystretch}{0.95}
  \setlength{\tabcolsep}{5pt}
  \begin{tabular}{ll}
    \toprule
    \textbf{Crash Family} & \textbf{Brief Description} \\
    \midrule
    Buffer Overflow & Violations of memory bounds (heap/stack) \\
    Uninitialized Access & Reads from undefined or indeterminate state \\
    Memory Lifecycle & Use-after-free / double-free / lifetime bugs \\
    Type Safety & Bad casts, invalid args, contract violations \\
    System Runtime & Environment and runtime interaction faults \\
    \bottomrule
  \end{tabular}
  \vspace{-5mm}
\end{table}

\textbf{Verification Process} 
Our verification follows two main considerations. First, ARVO crash families contain multiple subtypes. To reduce redundancy and improve representativeness, we select only a few typical subtypes from each family (see~\ref{app:crashtype}), preserving diversity without overloading the benchmark with similar vulnerabilities. Second, experts evaluate each sample’s complexity to keep \texttt{T2L-ARVO} at a moderate, calibrated difficulty. We score candidates using diff-based metrics (e.g., files changed, architectural spread, directory depth) and semantic factors (e.g., cross-module coupling, interface changes), ensuring that \texttt{T2L-ARVO} remains balanced in both coverage and difficulty.

\vspace{-2mm}
\section{Experiment Setup}

\textbf{Metrics.} We report two complementary scores for project-level OSS vulnerability studies. \emph{Detection} asks whether the agent flags a vulnerability within the correct module/chunk. \emph{Localization} requires exact line matches to ground-truth patches. Together, they separate "finding the neighborhood" from "pinpointing the line", mirroring real debugging.

\textbf{Data Preparation.} We evaluate on the full \texttt{T2L-ARVO} set: 50 verified challenges derived from \texttt{ARVO}, spanning diverse domains (e.g., imaging, networking) and balanced complexity so results reflect production patterns. Because ARVO lacks detection-ready chunking, \texttt{T2L-ARVO} adds AST-based segmentation: projects are partitioned into semantically meaningful units for scoring coarse \emph{detection}, while exact line matches assess fine \emph{localization}.

\textbf{Model Selection.} We assess state-of-the-art language models, both open-source and commercial, to probe generality and robustness across architectures and scales, including Qwen3 Next, Qwen3 235B, DeepSeek 3.1, LLaMA 4, Claude4 Sonnet, GPT-5, GPT-4.1, GPT-4o-mini, Gemini 2.5 Pro, and Gemini 2.5 Flash with a maximum budget \$1.0. We use API keys from official providers and Together.ai for open-source models.

\textbf{Implementation.} We build \texttt{T2L-Agent} from scratch without LangChain, DSPy, or LlamaIndex to keep the core lightweight and retain fine-grained control. The \emph{Agentic Trace Analyzer} compiles targets with ASAN and collects crashes, stack traces, and allocation metadata. Following ARVO's layout, we maintain per-project environments and provide both vulnerable and patched revisions in containers. Our harness runs dockerized \texttt{T2L-Agent}'s via the Docker SDK for Python, orchestrating build–run–reproduce cycles and ensuring deterministic reproduction.

\textbf{Comparison with existing localization agent architectures.} We compare against representative agent architectures from AgentFL and AutoFL. Rather than reusing their original systems, we implement architecture preserving transfer baselines with minimal adaption mappings to our crash-driven trace-to-line protocol, including replacing test signals with trace-derived evidence and generating line predictions for unified scoring.

\vspace{-2mm}
\section{Evaluation}

\subsection{Baseline Benchmarking}

\begin{table*}[!t]
\centering
\footnotesize
\caption{Base \texttt{T2L-Agent} Localization and Detection Rate Performance Across Different Models.}
\label{tab:Benchmark_Performance_Comparison}
\setlength{\tabcolsep}{4pt}
\renewcommand{\arraystretch}{0.75}
\begin{tabular}{l r r c c c c c c c c c c}
\toprule
& \multicolumn{2}{c}{\textbf{\% Avg.}} & \multicolumn{2}{c}{\textbf{Buffer}} & \multicolumn{2}{c}{\textbf{Initialize}} & \multicolumn{2}{c}{\textbf{Memory}} & \multicolumn{2}{c}{\textbf{Parameter}} & \multicolumn{2}{c}{\textbf{Runtime}} \\
\cmidrule(lr){2-3} \cmidrule(lr){4-5} \cmidrule(lr){6-7} \cmidrule(lr){8-9} \cmidrule(lr){10-11} \cmidrule(lr){12-13}
& \textbf{Det} & \textbf{Loc} & \textbf{Det} & \textbf{Loc} & \textbf{Det} & \textbf{Loc} & \textbf{Det} & \textbf{Loc} & \textbf{Det} & \textbf{Loc} & \textbf{Det} & \textbf{Loc} \\
\midrule
GPT-5    & 44.3 & \textbf{41.7} & 57.5 & \textbf{53.8} & 35.6 & 35.5 & \textbf{60.8} & \textbf{55.9} & 36.5 & 39.4 & 11.2 & 10.0 \\
GPT-4.1            & \textbf{48.0} & 38.5 & \textbf{60.8} & 36.5 & 50.6 &\textbf{46.4} & \textbf{60.8} & 46.1 & 26.5 & 29.9 & 21.2 & \textbf{20.5} \\
GPT-4o-mini        & 44.3 & 22.6 & \textbf{60.8} & 20.2 & 48.1 & 12.5 & 55.8 & 24.7 & 28.7 & 26.7 & 11.2 & 10.0 \\
Claude 4 Sonnet & 45.9 & 30.5 & 57.5 & 50.6 & \textbf{60.8} & 36.1 & 1.3  & 31.6 & \textbf{37.8} & \textbf{46.1}& \textbf{24.8} & 0.5   \\
Gemini2.5 Pro &17.4 &10.5& 25.0 &	5.6 &	25.0 & 20.0 &	11.3 &	10.8 &	5.4 &	11.9 &	10.0 &	10.5 \\
Qwen3 235B         & 25.9 &  9.2 & 25.8 & 6.5 & 23.1 & 1.7 & 40.8 & 16.7 & 28.7 & 17.3 & 7.9 & 0.0 \\
Qwen3 Next 80B             & 37.4 &  5.9 & 54.2 & 3.7 & 33.1 & 0.4 & 55.8 & 14.5 & 29.1 & 6.8 & 1.2 & 0.0 \\
\bottomrule
\end{tabular}
\vspace{-4mm}
\end{table*}


We evaluate \texttt{T2L-Agent} on \texttt{T2L-ARVO} and report Detection Rate and Localization Rate. Table~\ref{tab:Benchmark_Performance_Comparison} covers seven models: GPT-5, GPT-4.1, GPT-4o-mini, Claude 4 Sonnet, Gemini2.5 Pro, Qwen3 235B, and Qwen 3 Next 80B, running under identical per-case budgets, environments, and AST-based chunking. Overall, detection is higher than localization. Under this setting, GPT-5 leads localization at 41.7\%, followed by GPT-4.1 at 38.5\% and Claude 4 Sonnet at 30.5\%; open-source models trail below 10\%. For detection, GPT-4.1 is highest at 48.0\%, with GPT-5 and Claude 4 Sonnet close behind; Gemini2.5 Pro shows limited effectiveness at 17.4\%.

Family-wise patterns are consistent. \emph{Buffer} and \emph{Memory} are easier due to concrete runtime cues: GPT-5 reaches 53.8\% and 55.9\% localization, and most models cluster near the mid-50s for detection. \emph{Initialize} sits mid-range and benefits from multi-step reasoning. \emph{Parameter} is often solvable from interface/call-site context. \emph{Runtime} remains uniformly hardest: detection hovers around 11–21\% even for top configurations, reflecting sparse, unstable traces.

Taken together, equal budgets surface clear profiles. GPT-5 converts evidence into the strongest line-level localization, while GPT-4.1 extracts slightly more coarse-grained signal at the chunk level. GPT-4o-mini shows competitive detection but weak localization, suggesting higher recall that does not always translate to precise line hits. Open-source models lag on both metrics, indicating gaps in code understanding and tool use. Overall, improvements track the availability of concrete runtime evidence, and structured, tool-grounded reasoning appears more impactful than generation settings for end-to-end vulnerability localization.

\vspace{-2mm}
\subsection{Discussion 1: Feature-wise Evaluation}

\textbf{Agentic Trace Analyzer.}
This table ~\ref{tab:Combined_Performance_Comparison} demonstrates the critical effectiveness of our proposed Agentic Trace Analyzer (ATA) through ablation experiments. Without ATA, GPT-5 and Claude 4 Sonnet show 0.0\% detection and localization across all families. This performance breakdown validates that our ATA component successfully bridges the gap between crash symptoms and vulnerability locations, addresses the fundamental challenge of vulnerability localization in complex codebases. 

Like human debugging, \texttt{T2L-Agent} baseline solver seeds candidates via static–dynamic correlation and iteratively refines them with source–feedback loops. ATA brings fewer single shot failures, improved compute efficiency, and behavior based decisions that enable precise, line-level localization even in large, tightly coupled codebases. With ATA disabled, the LLM localizes on its own, as shown in Figure~\ref{fig:arvo-3features}(c).

\textbf{Detection Refinement.} Compared with Tab ~\ref{tab:Benchmark_Performance_Comparison}, Tab.~\ref{tab:Combined_Performance_Comparison} shows broad, across-the-board gains after enabling refinement. Strong proprietary models improve steadily, while open-source models jump the most—Qwen3~235B’s localization rises by roughly sevenfold. Improvements vary by crash family: \emph{Initialize} bugs benefit most (they demand multi-step reasoning), whereas \emph{Buffer} and \emph{Memory} see smaller lifts because concrete runtime evidence already anchors the search. \emph{Runtime} cases remain hard—when traces are sparse, refinement offers limited benefit. Net effect: higher recall and more precise line-level hits with minimal tuning. Several additional models show promising performance. Deepseek V3.1 achieves the highest overall results with 53.9\% detection and 53.4\% localization rate. LLaMa 4 demonstrates balanced capabilities on both metrics, and Gemini 2.5 Flash shows variable performance across crash families.

\begin{table*}[t]
\centering
\footnotesize
\caption{Localization and Detection Rate Performance of \texttt{T2L-Agent} with Refinement and Divergence Tracing Across Different Models.}
\label{tab:Combined_Performance_Comparison}
\setlength{\tabcolsep}{4pt}
\renewcommand{\arraystretch}{0.75}
\begin{tabular}{l r r r r r r r r r r r r r r}
\toprule
& \multicolumn{4}{c}{\textbf{\% Avg.}} & \multicolumn{2}{c}{\textbf{Buffer}} & \multicolumn{2}{c}{\textbf{Initialize}} & \multicolumn{2}{c}{\textbf{Memory}} & \multicolumn{2}{c}{\textbf{Parameter}} & \multicolumn{2}{c}{\textbf{Runtime}} \\
\cmidrule(lr){2-5} \cmidrule(lr){6-7} \cmidrule(lr){8-9} \cmidrule(lr){10-11} \cmidrule(lr){12-13} \cmidrule(lr){14-15}
& \textbf{Det} & \textbf{Loc} & \textbf{$\Delta$Det} & \textbf{$\Delta$Loc} & \textbf{Det} & \textbf{Loc} & \textbf{Det} & \textbf{Loc} & \textbf{Det} & \textbf{Loc} & \textbf{Det} & \textbf{Loc} & \textbf{Det} & \textbf{Loc} \\
\midrule
\multicolumn{15}{c}{\textbf{w/ Detection Refinement}} \\
\midrule
GPT-5            & 52.4 & 44.5 & +8.1\textcolor{ForestGreen}{$\uparrow$} & +2.8\textcolor{ForestGreen}{$\uparrow$} & 57.5 & 55.0 & 55.6 & 43.1 & 60.8 & 41.3 & 43.5 & 48.2 & 21.2 & 20.5 \\
GPT-4.1          & 48.3 & 40.8 & +0.3\textcolor{ForestGreen}{$\uparrow$} & +2.3\textcolor{ForestGreen}{$\uparrow$} & 60.8 & 51.9 & 53.1 & 44.9 & 57.5 & 46.1 & 39.8 & 42.9 & 6.7 & 0.0 \\
GPT-4o-mini      & 34.6 & 29.1 & -9.7\textcolor{red}{$\downarrow$} & +6.5\textcolor{ForestGreen}{$\uparrow$} & 45.8 & 43.6 & 30.6 & 20.2 & 45.8 & 33.6 & 22.4 & 21.2 & 0.0 & 0.0 \\

Claude 4 Sonnet         & 44.8 & 41.4 & -1.1\textcolor{red}{$\downarrow$} & +10.9\textcolor{ForestGreen}{$\uparrow$} & 57.5 & 54.3 & 40.6 & 43.2 & 61.7 & 52.5 & 26.1 & 29.4 & 14.6 & 10.5 \\

Gemini2.5 Pro    & 14.1 & 11.4 & -3.3\textcolor{red}{$\downarrow$} & -0.9\textcolor{red}{$\downarrow$} & 10.0 & 8.6  & 20.0 & 16.2 & 40.0 & 32.1 & 0.4  & 0.3  & 0.0  & 0.0 \\

Qwen3 Next 80B   & 42.9 & 39.5 & +5.5\textcolor{ForestGreen}{$\uparrow$} & +33.6\textcolor{ForestGreen}{$\uparrow$} & 60.8 & 55.1 & 50.6 & 44.4 & 60.8 & 48.6 & 25.7 & 29.2 & 0.0 & 0.0 \\
Qwen3 235B       & 34.1 & 26.7 & +8.2\textcolor{ForestGreen}{$\uparrow$} & +17.5\textcolor{ForestGreen}{$\uparrow$} & 34.2 & 32.0 & 30.6 & 34.5 & 57.5 & 38.8 & 23.3 & 13.3 & 5.0 & 8.2 \\
Gemini 2.5 Flash  & 22.5 & 18.4 & -- & -- & 34.2 & 0.6  & 40.8 & 25.7 & 7.9  & 27.6 & 0.4  & 33.4 & 24.6 & 0.6 \\
Llama4           & 28.3 & 28.1 & -- & -- & 30.8 & 25.6 & 35.8 & 26.1 & 0.0  & 27.9 & 24.1 & 32.0 & 30.2 & 0.0 \\
Deepseek V3.1     & 53.9 & 53.4 & -- & -- & 60.8 & 55.6 & \textbf{62.5} & 47.5 & 16.3 & 58.1 & 55.0 & \textbf{60.2} & 47.6 & 19.4 \\
\midrule
\multicolumn{15}{c}{\textbf{w/ Divergence Tracing}} \\
\midrule

GPT-5            & \textbf{58.0} & 52.0 & +13.7\textcolor{ForestGreen}{$\uparrow$} & +10.3\textcolor{ForestGreen}{$\uparrow$} & 60.8 & 56.8 & 60.6 & 53.4 & \textbf{62.5} & 47.6 & 53.2 & 46.7 & 26.2 & \textbf{28.7} \\

GPT-4.1          & 52.0 & 49.9 & +4.0\textcolor{ForestGreen}{$\uparrow$} & +11.4\textcolor{ForestGreen}{$\uparrow$} & 60.8 & 53.5 & 60.6 & \textbf{57.3} & 57.5 & 53.0 & 43.2 & 37.1 & 21.2 & 20.1 \\
GPT-4o-mini      & 47.2 & 43.3 & +2.9\textcolor{ForestGreen}{$\uparrow$} & +20.7\textcolor{ForestGreen}{$\uparrow$} & \textbf{64.2} & 55.6 & 55.6 & 48.8 & 52.5 & 47.0 & 33.9 & 32.4 & 1.2 & 0.5 \\

Claude 4 Sonnet         & 48.7 & 49.8 & +2.8\textcolor{ForestGreen}{$\uparrow$} & +19.3\textcolor{ForestGreen}{$\uparrow$} & 60.8 & 55.6 & \textbf{62.5} & 39.8 & 11.3 & 57.5 & 53.7 & 57.6 & 46.3 & 10.6 \\

Qwen 3 Next 80B   & 51.2 & \textbf{54.8} & +13.8\textcolor{ForestGreen}{$\uparrow$} & +48.9\textcolor{ForestGreen}{$\uparrow$} & \textbf{64.2} & \textbf{58.1} & \textbf{62.5} & 43.2 & 11.3 & \textbf{63.2} & \textbf{58.6} & 57.7 & \textbf{48.8} & 21.2 \\

Qwen 3 235B       & 42.7 & 42.1 & +16.8\textcolor{ForestGreen}{$\uparrow$} & +32.9\textcolor{ForestGreen}{$\uparrow$} & 50.8 & 50.6 & 47.5 & 40.2 & 1.0  & 45.4 & 46.9 & 53.5 & 33.7 & 11.2 \\

\midrule
\multicolumn{15}{c}{\textbf{w/o Agentic Trace Analyzer}} \\
\midrule
GPT-5            & 0.0 & 0.0 & -44.3\textcolor{red}{$\downarrow$} & -41.7\textcolor{red}{$\downarrow$} & 0.0 & 0.0 & 0.0 & 0.0 & 0.0 & 0.0 & 0.0 & 0.0 & 0.0 & 0.0 \\
Claude 4 Sonnet         & 0.0 & 0.0 & -45.9\textcolor{red}{$\downarrow$} & -30.5\textcolor{red}{$\downarrow$} & 0.0 & 0.0 & 0.0 & 0.0 & 0.0 & 0.0 & 0.0 & 0.0 & 0.0 & 0.0 \\
\bottomrule
\end{tabular}
\vspace{-3mm}
\end{table*}

\textbf{Divergence Tracing.} Tab.~\ref{tab:Combined_Performance_Comparison} shows divergence tracing yields the strongest gains. All models improve on both metrics: GPT-5 is up 13.7\% in detection and 10.3 in localization; GPT-4.1 gains 4.0\% and 11.4\% respectively. Qwen3 Next see the largest jumps with adding 13.8\% in detection and 48.9\% in localization, while Qwen3 235B adds 16.8\% and 32.9\%. These consistent lifts across architectures highlight divergence tracing as a core algorithmic upgrade for vulnerability localization.

\vspace{-2mm}
\subsection{Discussion 2: Parameter Tuning}
\vspace{-1mm}

\textbf{Thinking Budget.} As shown in Tab. ~\ref{tab:parameter_tuning}, more thinking didn’t help. On GPT-5, the Medium budget outperforms High with 50.9\% detection vs 41.3\%, and 41.6\% localization vs 36.1\%. The Low setting trails Medium by a few points yet often matches or even exceeds High on key metrics, while sharply reducing compute and latency. This pattern suggests diminishing returns—and decision drag—at very high budgets: the model over-explores, delays commitment, and accumulates tool-use errors. In practice, Medium strikes the best accuracy–cost balance for vulnerability localization; Low is a strong option when throughput and responsiveness matter most.

\begin{table*}[htbp]
\centering
\footnotesize
\caption{Localization and Detection Rate Performance of \texttt{T2L-Agent} for Temperature Tuning and Chain of Thought Across Different Models.}
\label{tab:parameter_tuning}
\setlength{\tabcolsep}{4pt}
\renewcommand{\arraystretch}{0.75}
\begin{tabular}{l c r r r r r r r r r r r r}
\toprule
& & \multicolumn{2}{c}{\textbf{\% Avg.}} & \multicolumn{2}{c}{\textbf{Buffer}} & \multicolumn{2}{c}{\textbf{Initialize}} & \multicolumn{2}{c}{\textbf{Memory}} & \multicolumn{2}{c}{\textbf{Parameter}} & \multicolumn{2}{c}{\textbf{Runtime}} \\
\cmidrule(lr){3-4} \cmidrule(lr){5-6} \cmidrule(lr){7-8} \cmidrule(lr){9-10} \cmidrule(lr){11-12} \cmidrule(lr){13-14}
& \textbf{Config} & \textbf{Det} & \textbf{Loc} & \textbf{Det} & \textbf{Loc} & \textbf{Det} & \textbf{Loc} & \textbf{Det} & \textbf{Loc} & \textbf{Det} & \textbf{Loc} & \textbf{Det} & \textbf{Loc} \\
\midrule
\multicolumn{14}{c}{\textbf{Temperature}} \\
\midrule
GPT-4.1 & 0.2 & \textbf{51.0} & 43.6 & 57.5 & 44.4 & 55.6 & \textbf{51.1} & 55.8 & 45.7 & 43.2 & 35.4 & 21.2 & \textbf{20.2} \\
GPT-4.1 & 0.6 & 50.8 & 43.5 & \textbf{60.8} & 52.0 & 53.1 & 47.8 & 55.8 & 44.8 & 39.4 & \textbf{42.7} & 17.9 & 10.1 \\
Claude 4 Sonnet & 0.2 & 46.5 & 43.0 & 54.2 & \textbf{56.1} & 55.6 & 43.0 & \textbf{61.7} & 49.0 & 36.5 & 39.7 & 11.2 & 10.9 \\
Claude 4 Sonnet & 0.6 & 47.3 & \textbf{44.9} & 54.2 & \textbf{56.1} & 50.6 & 48.0 & 60.8 & \textbf{53.6} & 36.5 & 39.6 & 11.2 & 10.9 \\
\midrule
\multicolumn{14}{c}{\textbf{Reasoning Effect}} \\
\midrule
GPT-5 & High & 41.3 & 36.1 & 54.2 & 47.8 & 40.6 & 32.4 & 55.8 & 39.3 & 32.8 & 36.7 & 10.0 & 10.0 \\
GPT-5 & Medium   & 50.9 & 41.6 & \textbf{60.8} & 55.6 & \textbf{60.8} & 42.8 & 11.3 & 46.1 & \textbf{45.8} & 40.4 & \textbf{47.4} & 10.5 \\
\bottomrule
\end{tabular}
\vspace{-3mm}
\end{table*}

\textbf{Temperature.} Temperature changes barely matter from Tab. ~\ref{tab:parameter_tuning}. On GPT-4.1, detection is 51.0\% at 0.2 and 50.8\% at 0.6, with localization 43.6\% vs.\ 43.5\%. Claude 4 Sonnet shows the same pattern: 46.5\% vs.\ 47.3\% detection and 43.0\% vs.\ 44.9\% localization. Performance is stable in 0.2-0.6 range. The exception is \emph{Initialize} bugs, which are more temperature sensitive than \emph{Buffer} and \emph{Memory} cases that lean on concrete runtime evidence. Overall, precise localization benefits more from structured, tool-grounded reasoning than from extra sampling, making parameter choices simple.

\begin{figure*}
  \centering
  \begin{subfigure}[t]{0.32\linewidth}
    \centering
    \includegraphics[width=\linewidth]{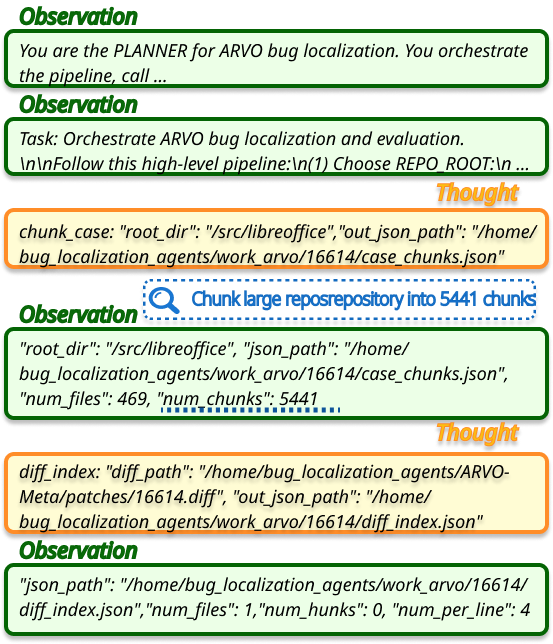}
  \end{subfigure}\hfill
  \begin{subfigure}[t]{0.32\linewidth}
    \centering
    \includegraphics[width=\linewidth]{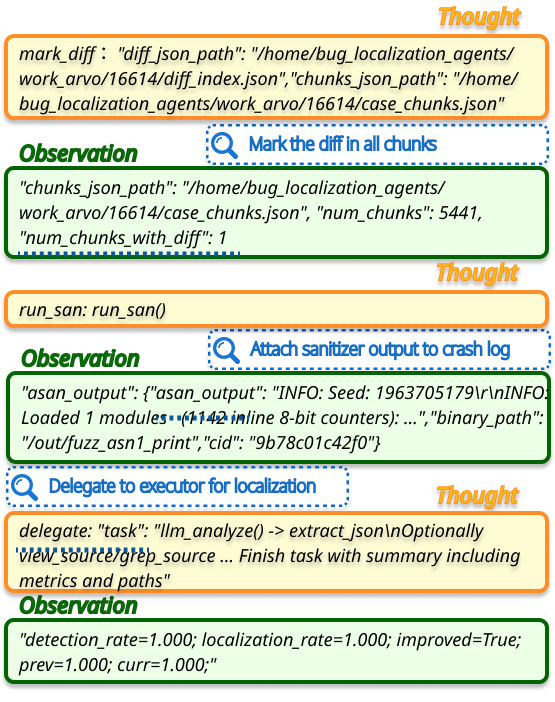}
  \end{subfigure}\hfill
  \begin{subfigure}[t]{0.32\linewidth}
    \centering
    \includegraphics[width=\linewidth]{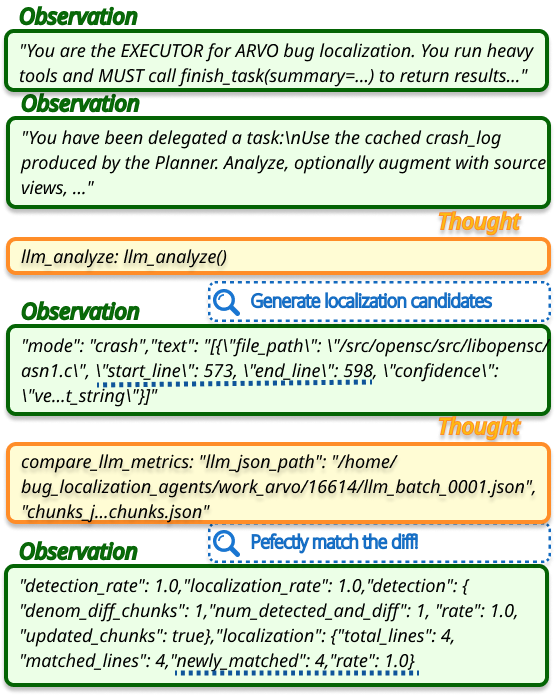}
  \end{subfigure}
  \caption{\texttt{T2L-Agent} pipeline visualization.}
  \label{fig:arvo-3pages}
\end{figure*}

\subsection{Discussion 3: Compared Frameworks}

We select AgentFL and AutoFL as representative agent-based localization frameworks combining LLM-guided navigation with code comprehension, making them natural baselines for our repo-scale trace-to-line task.

For AgentFL, we follow its three-stage workflow (fault comprehension, codebase navigation, fault confirmation) with minimal changes: replacing test-based inputs with crash logs, guiding navigation over relevant files, and applying fault confirmation via view source tools. The baseline outputs the top-1 method plus a line span. For AutoFL, we follow its two-stage design. In Stage 1, the agent queries covered classes (source files) and methods (functions), retrieving code for selected candidates. In Stage 2, navigation tools are locked and the agent outputs suspicious methods with line spans. Both baselines use identical evaluation metrics.

Table~\ref{tab:t2l_agentfl_autofl} shows T2L consistently outperforms both transferred frameworks across models in detection and localization rate, indicating that current agent pipelines struggle to turn coarse detection into reliable line-level localization.

AgentFL achieves acceptable detection but substantially lower localization than T2L. Its best results come from GPT-5 (29.8\%) and GPT-4.1 (26.2\%), compared with 41.7\% and 38.5\% for T2L. However, trends vary: for Qwen models, AgentFL reaches slightly higher localization than T2L, suggesting existing pipelines sometimes work but behave inconsistently across models. AutoFL's performance drops sharply---localization remains low (27.0\%) and near zero for smaller models like Qwen3 Next 80B (2.6\%).

These results suggest that while coarse chunks can sometimes be detected, accurately narrowing to vulnerable lines remains difficult without tailored trace-to-line methods. Effective systems must use runtime evidence to connect symptoms to root-cause lines. Overall, transferred baselines reach substantially lower performance than T2L, indicating current agent architectures are insufficient for crash-driven, line-level localization in complex repositories.

\begin{table}[t]
\centering
\scriptsize
\setlength{\tabcolsep}{6pt}
\renewcommand{\arraystretch}{0.75}
\begin{tabular}{lcccccc}
\toprule
 & \multicolumn{2}{c}{T2L} & \multicolumn{2}{c}{AgentFL} & \multicolumn{2}{c}{AutoFL} \\
\cmidrule(lr){2-3}\cmidrule(lr){4-5}\cmidrule(lr){6-7}
Model & Det & Loc & Det & Loc & Det & Loc \\
\midrule
GPT-5            & \textbf{44.3} & \textbf{41.7} & 33.1 & 29.8 & 33.9 & 27.0 \\
GPT-4.1          & \textbf{48.0} & \textbf{38.5} & 27.3 & 26.2 & 18.3 &  9.0 \\
GPT-4o-mini      & \textbf{44.3} & \textbf{22.6} & 21.6 & 13.8 &  6.0 &  4.0 \\
Claude 4 Sonnet  & \textbf{45.9} & \textbf{30.5} & 38.8 & 25.4 & 29.8 & 13.3 \\
Gemini2.5 Pro    & \textbf{17.4} & \textbf{10.5} & 17.1 & 10.3 & 12.3 &  3.5 \\
Qwen3 235B       & \textbf{25.9} &  9.2 & 14.6 & \textbf{12.5} & 20.3 &  8.7 \\
Qwen3 Next 80B   & \textbf{37.4} &  5.9 & 11.0 &  \textbf{7.5} & 11.5 &  2.6 \\
\bottomrule
\end{tabular}
\caption{Average detection (Det) and line-level localization (Loc) rates on T2L-ARVO (\%).}
\label{tab:t2l_agentfl_autofl}
\vspace{-8mm}
\end{table}

\subsection{Discussion 4: Case Study}
\vspace{-2mm}
Figure~\ref{fig:arvo-3pages} illustrates a full \texttt{T2L-Agent} workflow on a real case from \texttt{T2L-ARVO}, showcasing the iterative planner–executor architecture in action. The process starts with the Planner orchestrating code chunking (\texttt{chunk\_case}, 5441 chunks) and diff indexing (\texttt{diff\_index}), then running sanitized execution (\texttt{run\_san}) to collect crash logs before delegating reasoning to the Executor.

The Executor analyzes traces via \texttt{llm\_analyze}, extracts ranked candidates (\texttt{extract\_json}), and iteratively evaluates them against ground truth using \texttt{compare\_llm\_metrics}, combining static patterns and dynamic signals. This multi-round refinement achieves perfect localization (detection: 1.0, localization: 1.0), demonstrating \texttt{T2L-Agent}'s ability to convert crash symptoms into precise diagnostics. Each panel visualizes thought (function call) and observation (result) with hand-drawn borders and no hallucinated text. Together, they highlight data flow, role separation, and metric-driven validation across planner and executor components.



\vspace{-2mm}
\section{Limitation and Future Work}
\vspace{-2mm}

Our work has three key limitations. First, the \texttt{T2L-ARVO} benchmark includes only 50 manually verified cases. While it offers broad vulnerability coverage and balanced categories, the limited sample size constrains evaluation due to the human verification efforts. Also, ARVO's dataset structure is designed for human developers, which needs more fine-grained metadata that could benefit LLM-based localization. Second, although \texttt{T2L-Agent} improves localization accuracy from 0\% to 54.8\% through three innovations, cost efficiency remains a concern. The agent operates effectively under a \$1.0 budget via task-aware planning and early stopping, but large-scale deployment across thousands of vulnerabilities would demand significant optimization. Third, higher model thinking budgets fail to boost localization performance, indicating that increased compute alone is insufficient. This points to a need for smarter ways to exploit model reasoning. Future work should explore more efficient architectures. Such as model cascading to coordinate cheaper and stronger models, and specialized multi-agent systems where roles are tailored to tools like our Agentic Trace Analyzer. These strategies may retain quality while scaling to production workloads.

\vspace{-2mm}
\section{Conclusion}
\vspace{-2mm}
T2L addresses a key gap between LLM-based vulnerability localization and real-world practice. We contribute three advances that move from coarse identification to precise diagnostics. First, we propose a new formulation for LLM-based vulnerability detection: chunk-wise detection and line-level localization, enabling structured and fine-grained evaluation. Second, \texttt{T2L-ARVO} introduces a benchmark for agentic line-level localization, with 50 expert-verified cases across diverse vulnerabilities. Third, \texttt{T2L-Agent} improves performance via our Agentic Trace Analyzer, which fuses runtime and static signals, as well as Divergence Tracing and Detection Refinement in a feedback-driven workflow. \texttt{T2L-Agent} achieves 44–58\% detection and 38–54.8\% localization, marking a step toward deployable systems for real-world code security.

\newpage
\section*{Ethical Considerations and Reproducibility Statement}
\subsection*{Ethics Statement}

This work builds upon the ARVO dataset for vulnerability analysis. Our T2L-ARVO benchmark is constructed using its full-version data, and we have properly cited the original ARVO project to acknowledge its contribution and comply with copyright and attribution standards. No additional data collection, user studies, or ethically sensitive procedures involved in the construction or evaluation of T2L-Agent. All experiments were conducted on Linux servers with only open source dependencies, and our system does not involve any privacy-sensitive data, bias-sensitive decision-making, or potentially harmful applications. We also note that large language models were used solely for light editing and polishing of manuscript, with no involvement in system design, code generation, or experimental results. Their use was limited to improving readability and presentation clarity.

\subsection*{Reproducibility Statement}

To ensure reproducibility, we will publicly release the full T2L-Agent framework upon publish of this paper, including all module codebase, evaluation scripts, and benchmark data used in this paper. Our implementation does not rely on any proprietary components. Detailed descriptions of our methodology are provided in the main text (Sections~\ref{sec:t2lagent}, \ref{sec:benchmarks}) and supported by step-by-step examples in the appendix. All experimental configurations, including model versions, prompting strategies, and budget constraints, will be documented and made available upon publication to enable full replication of our results.

\bibliography{example_paper}
\bibliographystyle{icml2026}

\newpage
\appendix
\onecolumn
\section{Appendix}

\subsection{ARVO Crash Type}
\label{app:crashtype}
To understand where our vulnerability localization agent should focus, we analyze the crash type distribution in ARVO dataset. 
ARVO's crash type distribution is dominated by classic memory corruption: Buffer Overflow accounts for 49.9\% of crashes, led by heap-buffer-overflow (36.1\% overall) and followed by stack-buffer-overflow (6.2\%), index-out-of-bounds (3.3\%), and global-buffer-overflow (3.2\%), with underflows/containers in the long tail (around 1\% each). Uninitialized Access \& Unknown States is the second largest family at 35.4\%, primarily use-of-uninitialized-value (20.3\%), then UNKNOWN READ/WRITE (around 11.8\% combined). Memory Lifecycle Errors contribute 11.5\%, dominated by heap-use-after-free (7.8\% overall) plus double-free, use-after-poison, and invalid frees. Type Safety \& Parameter Validation is smaller (2.9\%)—notably bad-cast (1.3\%) and negative-size-param (0.8\%). System \& Runtime Errors are rare (0.3\%). Overall, around 85\% of ARVO crashes fall into Buffer Overflow or Uninitialized/Unknown categories. 

\begin{longtable}{l l r r r}
\caption{Crash Families and Subtypes Analysis}\\
\toprule
\tiny
Family & Subtype & Count & \% within family & \% of total \\
\midrule
\endfirsthead
\caption[]{Crash Families and Subtypes Analysis (continued)}\\
\toprule
Family & Subtype & Count & \% within family & \% of total \\
\midrule
\endhead
\midrule
\endfoot
\bottomrule
\endlastfoot

\multicolumn{5}{l}{\textbf{Buffer Overflow Vulnerabilities}} \\
& \textit{Total} & \textit{2490} & \textit{---} & \textit{49.9\%} \\
\addlinespace[0.5ex]
& Heap-buffer-overflow & 1802 & 72.4\% & 36.1\% \\
& Stack-buffer-overflow & 308 & 12.4\% & 6.2\% \\
& Index-out-of-bounds & 165 & 6.6\% & 3.3\% \\
& Global-buffer-overflow & 160 & 6.4\% & 3.2\% \\
& Container-overflow & 33 & 1.3\% & 0.7\% \\
& Stack-buffer-underflow & 13 & 0.5\% & 0.3\% \\
& Dynamic-stack-buffer-overflow & 9 & 0.4\% & 0.2\% \\
\addlinespace

\multicolumn{5}{l}{\textbf{Uninitialized Access \& Unknown States}} \\
& \textit{Total} & \textit{1768} & \textit{---} & \textit{35.4\%} \\
\addlinespace[0.5ex]
& Use-of-uninitialized-value & 1015 & 57.4\% & 20.3\% \\
& UNKNOWN READ & 462 & 26.1\% & 9.3\% \\
& Segv on unknown address & 134 & 7.6\% & 2.7\% \\
& UNKNOWN WRITE & 123 & 7.0\% & 2.5\% \\
& Null-dereference READ & 25 & 1.4\% & 0.5\% \\
& UNKNOWN & 8 & 0.5\% & 0.2\% \\
& Unknown-crash & 1 & 0.1\% & 0.0\% \\
\addlinespace

\multicolumn{5}{l}{\textbf{Memory Lifecycle Errors}} \\
& \textit{Total} & \textit{573} & \textit{---} & \textit{11.5\%} \\
\addlinespace[0.5ex]
& Heap-use-after-free & 389 & 67.9\% & 7.8\% \\
& Heap-double-free & 63 & 11.0\% & 1.3\% \\
& Use-after-poison & 48 & 8.4\% & 1.0\% \\
& Invalid-free & 29 & 5.1\% & 0.6\% \\
& Stack-use-after-return & 26 & 4.5\% & 0.5\% \\
& Stack-use-after-scope & 13 & 2.3\% & 0.3\% \\
& Bad-free & 5 & 0.9\% & 0.1\% \\
\addlinespace

\multicolumn{5}{l}{\textbf{Type Safety \& Parameter Validation}} \\
& \textit{Total} & \textit{147} & \textit{---} & \textit{2.9\%} \\
\addlinespace[0.5ex]
& Bad-cast & 65 & 44.2\% & 1.3\% \\
& Negative-size-param & 42 & 28.6\% & 0.8\% \\
& Memcpy-param-overlap & 20 & 13.6\% & 0.4\% \\
& Object-size & 9 & 6.1\% & 0.2\% \\
& Incorrect-function-pointer-type & 6 & 4.1\% & 0.1\% \\
& Non-positive-vla-bound-value & 3 & 2.0\% & 0.1\% \\
& Strcpy-param-overlap & 1 & 0.7\% & 0.0\% \\
& Strncpy-param-overlap & 1 & 0.7\% & 0.0\% \\
\addlinespace

\multicolumn{5}{l}{\textbf{System \& Runtime Errors}} \\
& \textit{Total} & \textit{15} & \textit{---} & \textit{0.3\%} \\
\addlinespace[0.5ex]
& Check failed & 6 & 40.0\% & 0.1\% \\
& Unknown signal & 6 & 40.0\% & 0.1\% \\
& Bad parameters to --sanitizer-annotate-contiguous-container & 2 & 13.3\% & 0.0\% \\
& Nested bug in the same thread, aborting. & 1 & 6.7\% & 0.0\% \\

\end{longtable}

\subsection{ARVO Dataset Profiling}
\label{app:dataset_profiling}
We profiled the full ARVO corpus with all 4,993 vulnerabilities across 288 projects to guide \texttt{T2L-ARVO}'s design and document its coverage. The analysis maps distributional patterns, project traits, and crash-type frequencies, and clarifies how our 50-case subset aligns with the broader ARVO ecosystem. These profiles confirm that \texttt{T2L-ARVO} is representative across crash families, project complexity, and severity, providing a transparent baseline for extensions and alternative benchmarks. We include compact visualizations of these profiles to convey key ARVO factors—such as crash families, project complexity, and severity at a glance.

\begin{figure}[H]
  \centering
  \begin{subfigure}[t]{0.49\linewidth}
    \centering
    \includegraphics[width=\linewidth]{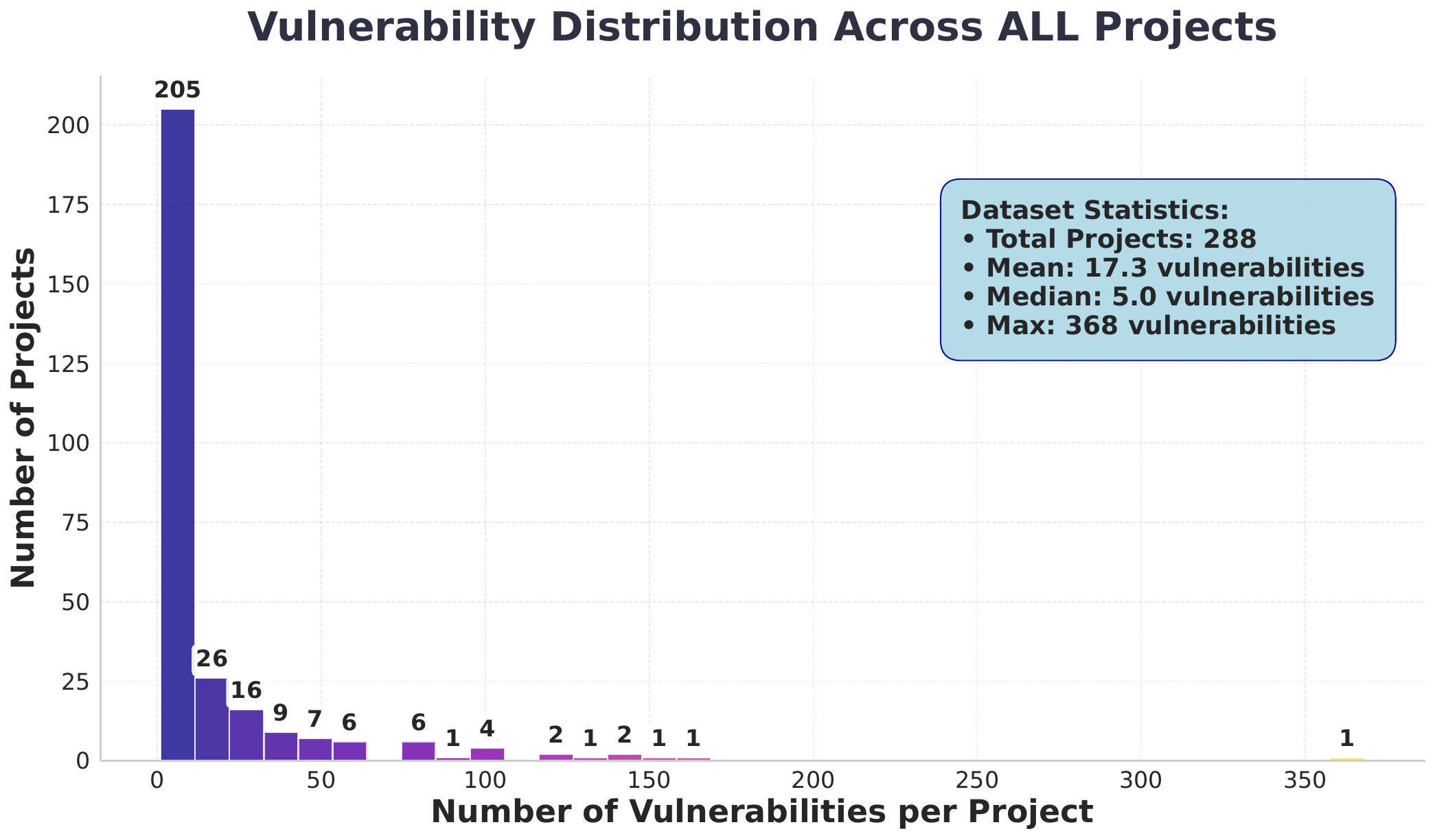}
  \end{subfigure}\hfill
  \begin{subfigure}[t]{0.49\linewidth}
    \centering
    \includegraphics[width=\linewidth]{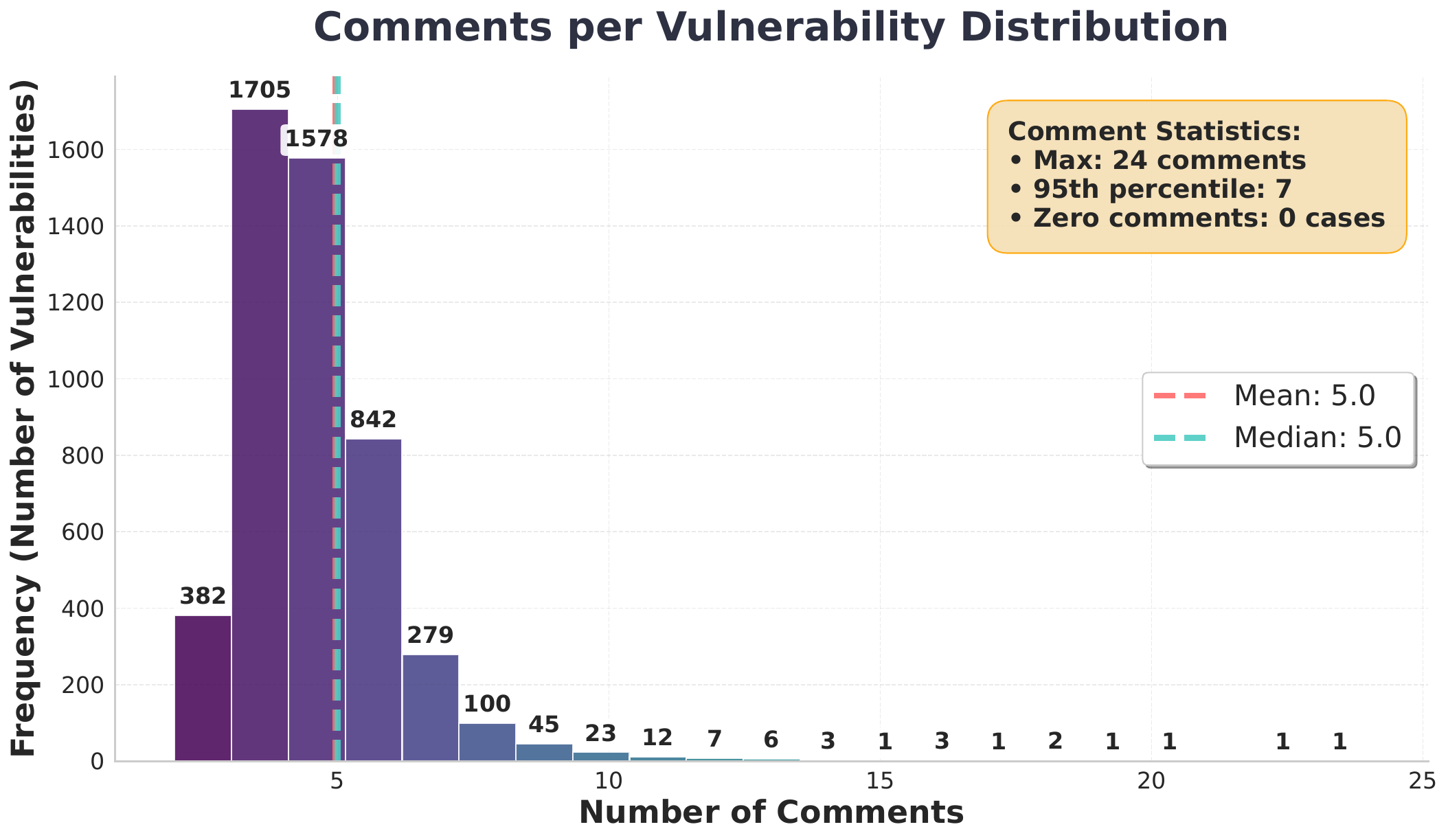}
  \end{subfigure}
  \caption{Distribution of vulnerability counts across the 288 ARVO projects. Most projects have under 25 vulnerabilities, with a long-tail of highly vulnerable ones.}
  \label{fig:prof1}
\end{figure}

\begin{figure}[H]
  \centering
  \begin{subfigure}[t]{0.42\linewidth}
    \centering
    \includegraphics[width=\linewidth]{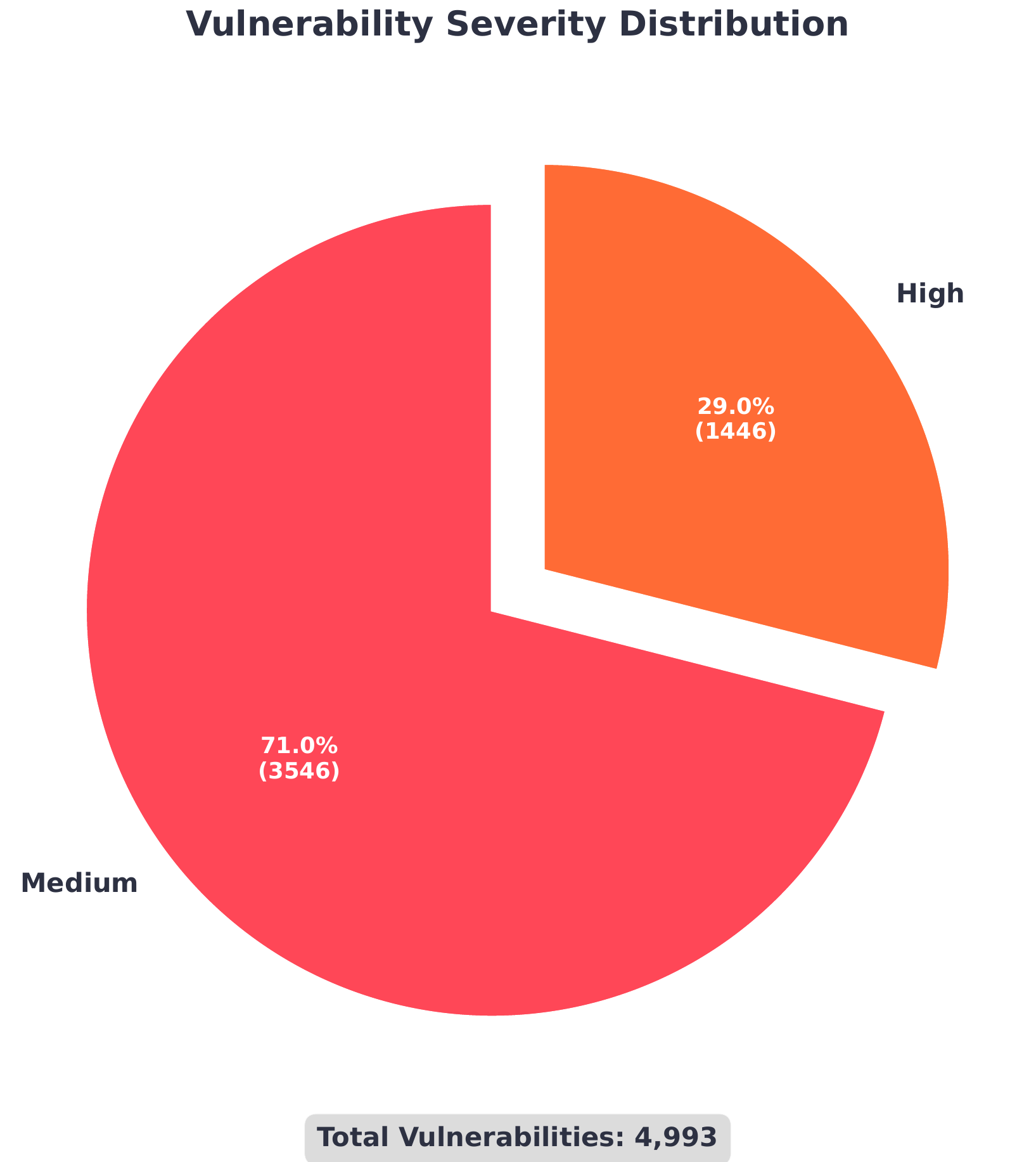}
  \end{subfigure}\hfill
  \begin{subfigure}[t]{0.42\linewidth}
    \centering
    \includegraphics[width=\linewidth]{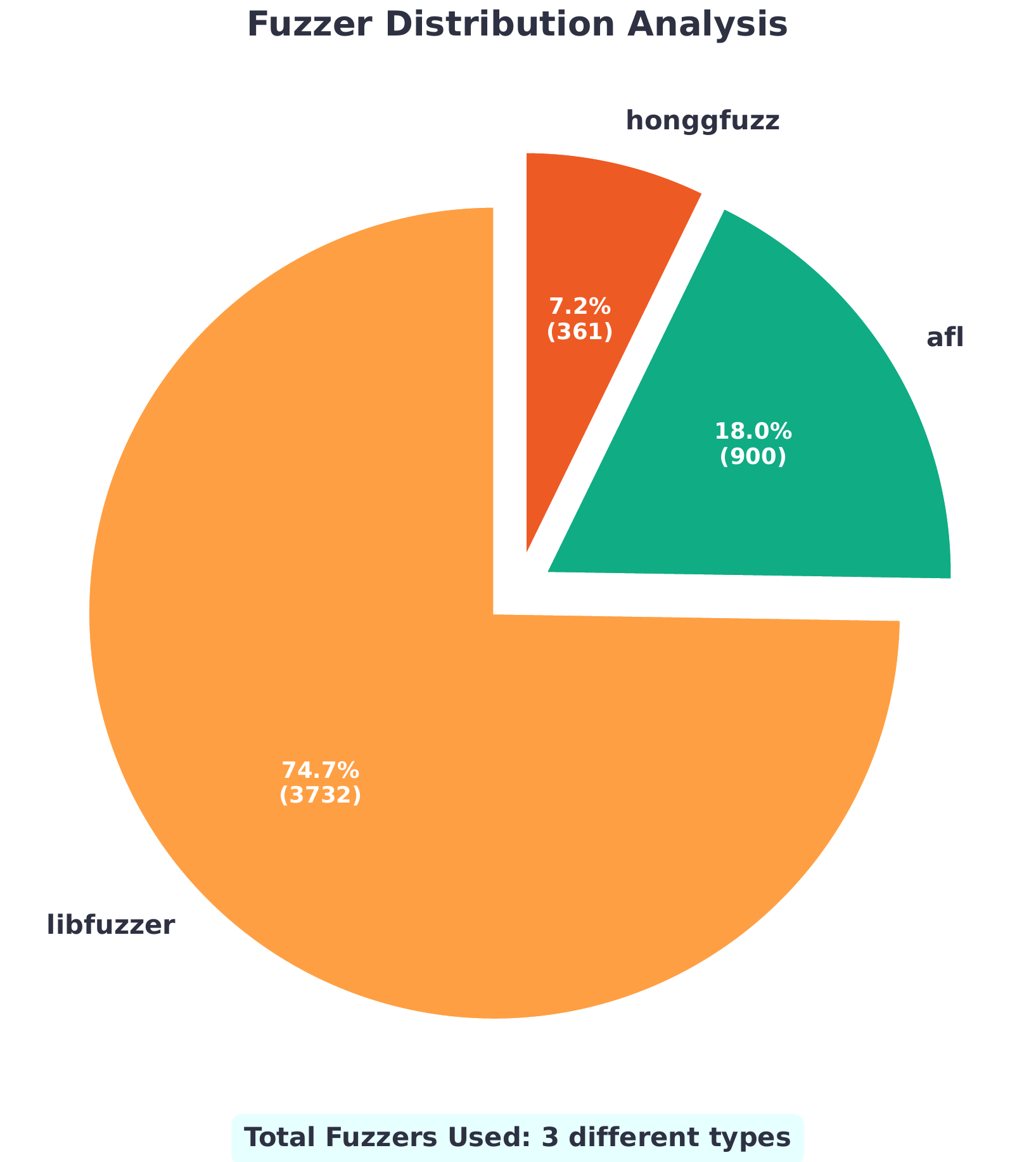}
  \end{subfigure}
  \caption{Breakdown of vulnerability severities in ARVO. Over 70\% are medium severity, while high severity cases account for the remaining 29\%.}
  \label{fig:prof2}
\end{figure}

\begin{figure}[H]
    \centering
    \includegraphics[width=0.93\linewidth]{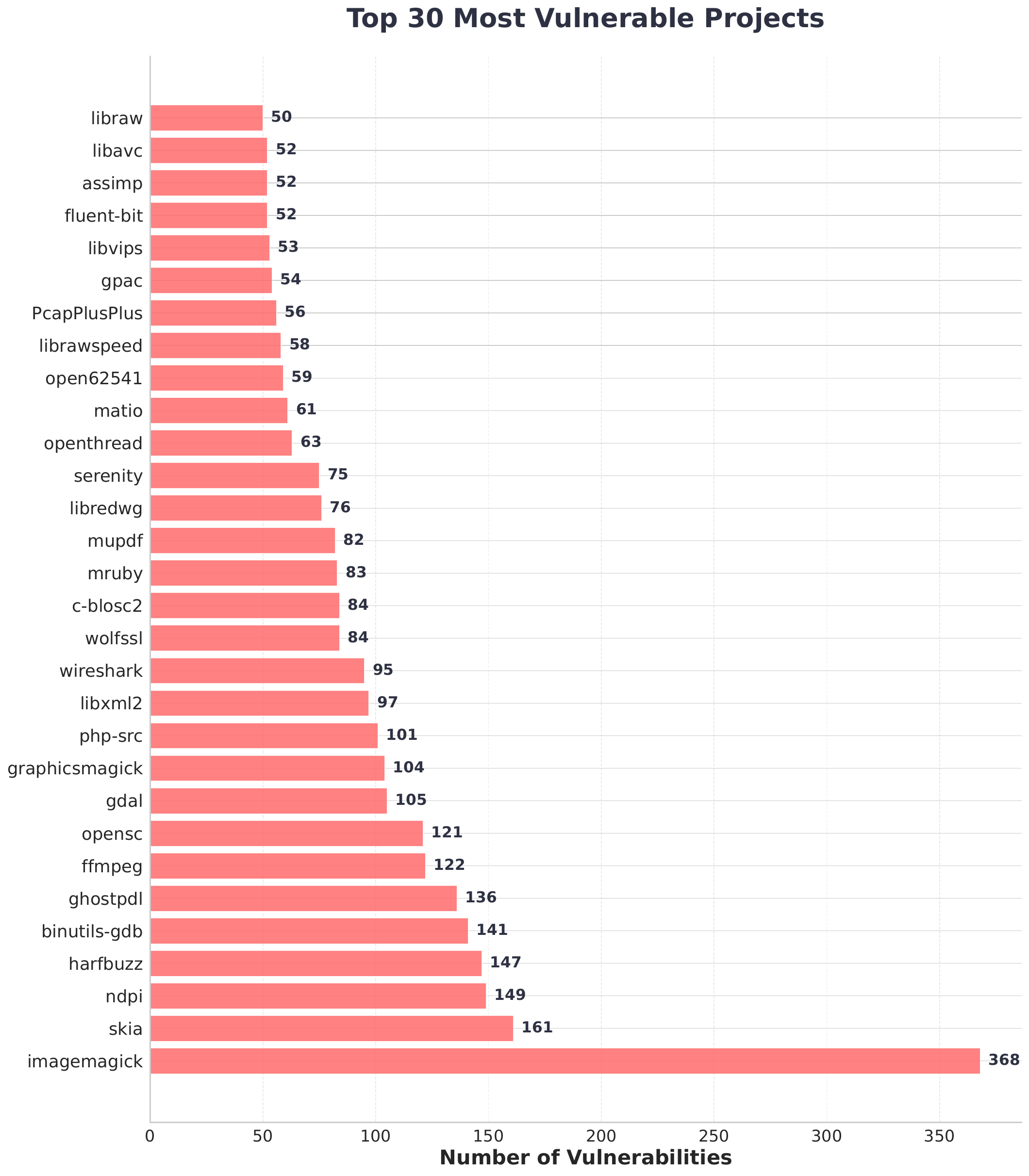}
    \caption{Analysis of fuzzing tools used in ARVO. libFuzzer dominates at 74.7\%, followed by AFL and honggfuzz.}
    \label{fig:prof3}
    \vspace{-3mm}
\end{figure}

\begin{figure}[H]
    \centering
    \includegraphics[width=\linewidth]{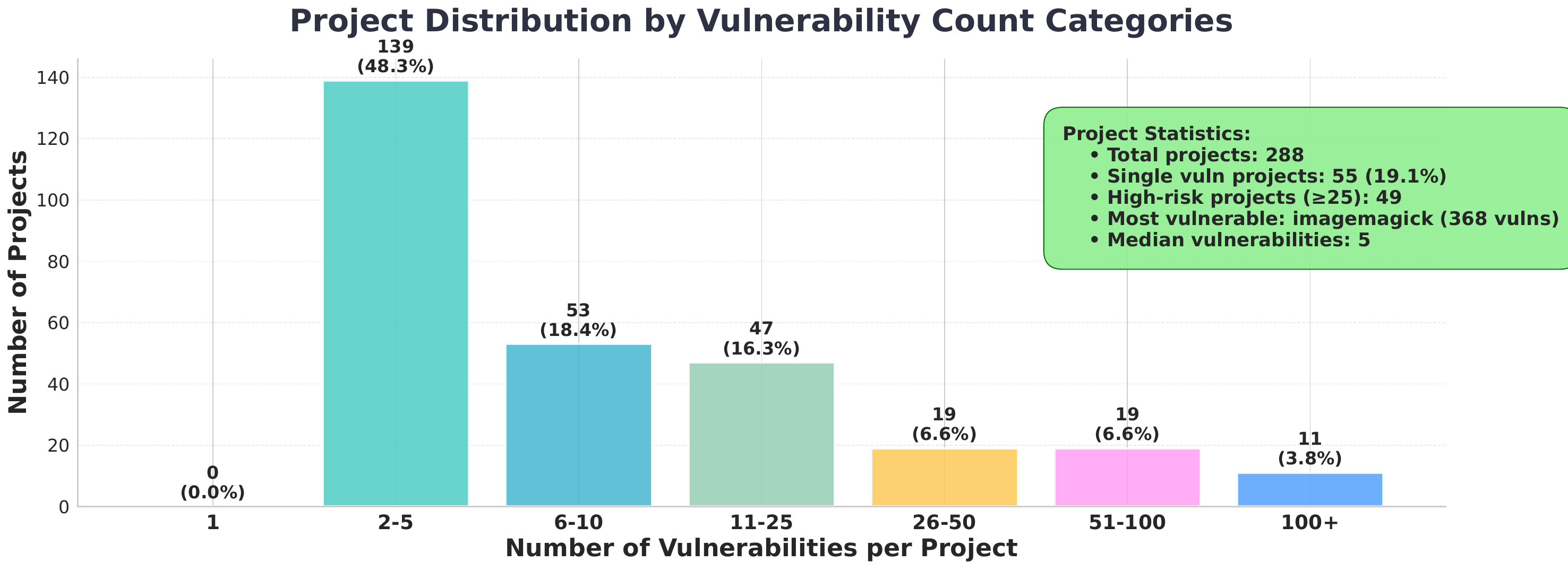}
    \vspace{-6mm}
    \caption{Visualization of the 30 most vulnerable projects in ARVO. ImageMagick leads with 368 vulnerabilities, with others showing diverse security footprints.}
    \label{fig:prof7}
\end{figure}

\begin{figure}[H]
    \centering
    \includegraphics[width=\linewidth]{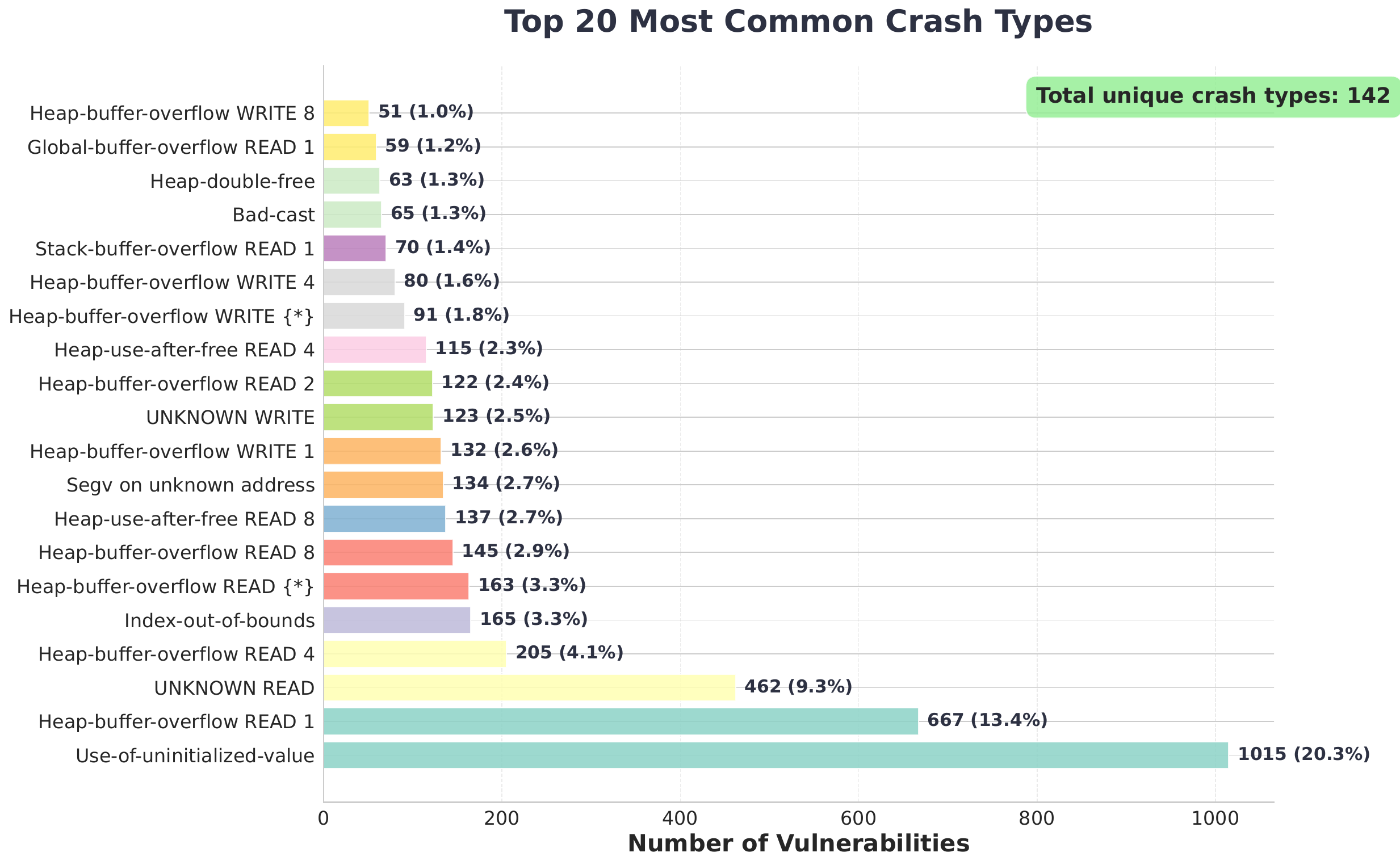}
    \vspace{-6mm}
    \caption{Project grouping by vulnerability count. Nearly half the projects have only 2–5 vulnerabilities, with very few exceeding 100.}
    \label{fig:prof6}
\end{figure}

\subsection{Model Failure Analysis}

\begin{figure}[H]
    \centering
    \includegraphics[width=0.9\linewidth]{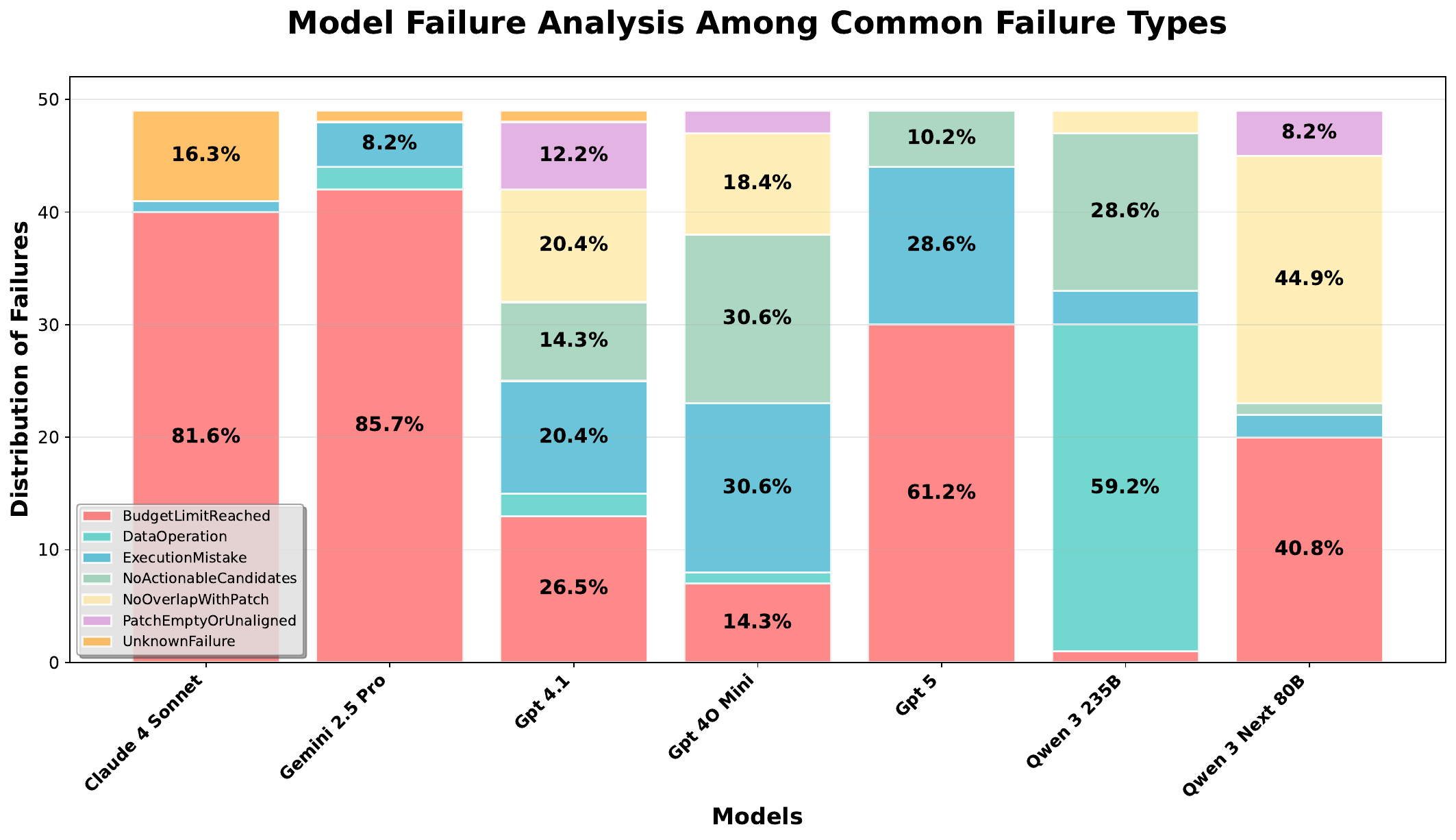}
    \caption{Model failure type distribution across five models on \texttt{T2L-ARVO}. GPT-5 commonly fails due to budget limits, while Qwen3 models often fail to generate actionable candidates.}
    \label{fig:failures}
\end{figure}

We conducted a targeted failure analysis on several baseline models to map out common failure modes as shown in Tab. ~\ref{fig:failures}. Claude 4 Sonnet and Gemini 2.5 Pro hit the budget ceiling in 81.6\% and 85.7\% of runs respectively, indicating efficient resource utilization. GPT-5 reaches 61.2\% with execution errors (28.6\%), while Qwen3 235B struggles with basic data operations (59.2\%). Open-source baselines stall early: Qwen 3 Next fails to surface actionable candidates in 44.9\% of trials. Execution errors remain common across older models (20–30\%), showing that tool use often breaks even when a plan exists. Net-net, while newer models show improved resource management, legacy models skew toward either incomplete exploration or difficulty navigating real-world code.

\subsection{T2L-ARVO Challenge List}
We provide the comprehensive list of \texttt{T2L-ARVO} benchmark we verified and collected in this work along with the key meta information for each challenge.
\begin{longtable}{p{1.2cm}p{1.5cm}p{1.5cm}p{2.2cm}p{4.5cm}p{1.0cm}}
\caption{Bug Analysis Results by Category} \label{tab:bug_analysis} \\
\toprule
\footnotesize
\centering
\textbf{Id} & \textbf{Fuzzer} & \textbf{Sanitizer} & \textbf{Project} & \textbf{Crash Type} & \textbf{Severity} \\
\midrule
\endfirsthead

\multicolumn{6}{c}%
{{\tablename\ \thetable{} -- continued from previous page}} \\
\toprule
\textbf{Id} & \textbf{Fuzzer} & \textbf{Sanitizer} & \textbf{Project} & \textbf{Crash Type} & \textbf{Severity} \\
\midrule
\endhead

\midrule \multicolumn{6}{r}{{Continued on next page}} \\
\endfoot

\bottomrule
\endlastfoot

\multicolumn{6}{l}{\textbf{System \& Runtime Errors}} \\
\midrule
16737 & libfuzzer & ubsan & graphicsmagick & Unknown signal & - \\
7966 & libfuzzer & ubsan & graphicsmagick & Unknown signal & - \\
7654 & libfuzzer & ubsan & graphicsmagick & Unknown signal & - \\
7639 & libfuzzer & ubsan & graphicsmagick & Unknown signal & - \\
59193 & libfuzzer & msan & faad2 & Check failed & - \\
49915 & libfuzzer & msan & ndpi & Check failed & - \\
48780 & libfuzzer & msan & libvpx & Check failed & - \\
7361 & afl & asan & ots & Bad parameters to sanitizer & - \\
32939 & libfuzzer & asan & rdkit & Bad parameters to sanitizer & - \\
\addlinespace

\multicolumn{6}{l}{\textbf{Buffer Overflow Vulnerabilities}} \\
\midrule
16614 & libfuzzer & asan & opensc & Heap-buffer-overflow & Med \\
13956 & libfuzzer & asan & yara & Heap-buffer-overflow & Med \\
16615 & libfuzzer & asan & opensc & Heap-buffer-overflow & Med \\
20856 & libfuzzer & asan & ndpi & Heap-buffer-overflow & Med \\
17330 & libfuzzer & asan & openthread & Stack-buffer-overflow & High \\
42454 & libfuzzer & asan & ghostpdl & Stack-buffer-overflow & High \\
17297 & libfuzzer & asan & openthread & Stack-buffer-overflow & Med \\
18562 & libfuzzer & asan & lwan & Global-buffer-overflow & - \\
30507 & honggfuzz & asan & serenity & Global-buffer-overflow & - \\
18231 & afl & asan & binutils-gdb & Global-buffer-overflow & - \\
\addlinespace

\multicolumn{6}{l}{\textbf{Uninitialized Access \& Unknown States}} \\
\midrule
49493 & libfuzzer & asan & mruby & Segv on unknown address & - \\
24290 & honggfuzz & asan & libvips & Segv on unknown address & - \\
57037 & libfuzzer & asan & mruby & Segv on unknown address & - \\
23778 & libfuzzer & msan & binutils-gdb & Use-of-uninitialized-value & Med \\
20112 & libfuzzer & msan & open62541 & Use-of-uninitialized-value & Med \\
47855 & libfuzzer & msan & harfbuzz & Use-of-uninitialized-value & Med \\
16857 & libfuzzer & msan & matio & Use-of-uninitialized-value & Med \\
43989 & libfuzzer & asan & ghostpdl & Null-dereference & - \\
2623 & libfuzzer & asan & h2o & Null-dereference & - \\
45320 & libfuzzer & asan & ghostpdl & Null-dereference & - \\
\addlinespace

\multicolumn{6}{l}{\textbf{Memory Lifecycle Errors}} \\
\midrule
42503 & libfuzzer & asan & php-src & Heap-use-after-free & High \\
38878 & libfuzzer & asan & harfbuzz & Heap-use-after-free & High \\
14245 & afl & asan & karchive & Heap-use-after-free & High \\
19723 & libfuzzer & asan & leptonica & Heap-use-after-free & Med \\
33750 & honggfuzz & asan & fluent-bit & Heap-double-free & High \\
34116 & honggfuzz & asan & fluent-bit & Heap-double-free & High \\
20785 & libfuzzer & asan & llvm-project & Use-after-poison & High \\
3505 & afl & asan & librawspeed & Use-after-poison & High \\
51687 & afl & asan & mongoose & Use-after-poison & High \\
31705 & afl & asan & c-blosc2 & Invalid-free & - \\
\addlinespace

\multicolumn{6}{l}{\textbf{Type Safety \& Parameter Validation}} \\
\midrule
2798 & libfuzzer & ubsan & gdal & Bad-cast & High \\
29267 & libfuzzer & ubsan & serenity & Bad-cast & High \\
33150 & libfuzzer & ubsan & libredwg & Object-size & Med \\
20217 & libfuzzer & ubsan & arrow & Object-size & Med \\
12679 & afl & asan & openthread & Memcpy-param-overlap & Med \\
23547 & honggfuzz & asan & php-src & Memcpy-param-overlap & Med \\
25357 & libfuzzer & asan & libsndfile & Negative-size-param & - \\
60605 & libfuzzer & asan & ndpi & Negative-size-param & Med \\
2692 & libfuzzer & ubsan & boringssl & Incorrect-function-pointer-type & Med \\
50623 & libfuzzer & ubsan & serenity & Non-positive-vla-bound-value & Med \\
\end{longtable}

\vspace{-2mm}
\subsection{T2L Toolkit List}
\vspace{-2mm}
\label{app:toolkit}
We list the tools used in \texttt{T2L-Agen} and their roles in the analysis workflow. The framework is modular and can be easily extended with new tools based on task requirements.
\footnotesize
\renewcommand{\arraystretch}{1.15}
\begin{longtable}{p{4.7cm}p{11.7cm}}
\caption{T2L Toolkit list and the usage description.} \label{tab:t2l-toolkit} \\
\hline
\textbf{Tool (NAME)} & \textbf{Description} \\
\hline
\endfirsthead

\multicolumn{2}{c}{\textbf{Table \thetable{} continued from previous page}} \\
\hline
\textbf{Tool (NAME)} & \textbf{Description} \\
\hline
\endhead

\hline
\multicolumn{2}{r}{Continued on next page} \\
\endfoot

\hline
\endlastfoot

\texttt{view\_source} & Preview a source file with line numbers. Optionally specify start\_line/end\_line. \\
\texttt{grep\_source} & Search code by regex under a root directory. Returns `file:line:match` lines. \\
\texttt{insert\_print} & Insert a single line of debug print before the given line number in a source file. \\
\texttt{build\_project} & Build the project inside container. Default workdir=/src. \\
\texttt{container\_exec} & Run an arbitrary shell command inside the container (advanced use). \\
\texttt{copy\_out} & Copy a file/dir from container to host. \\
\texttt{giveup} & Give up this case to terminate it immediately. Use this to stop solving the ARVO container. \\
\texttt{diff\_index} & Parse unified diff and build a simple line-level index without extra anchoring. Output JSON has per-file \{anchors\_old, insert\_points=[], per\_line\{line-\textgreater\{roles, matched\}\}\}; all line numbers are OLD-file coordinates from the diff. \\
\texttt{static\_analysis} & Run comprehensive analysis. For binaries: run Ghidra RE then static tools; for sources: run static tools directly. Uses cppcheck/clang-tidy/infer; aggregates findings to JSON and env state. \\
\texttt{chunk\_case} & Parse C/C++ sources under root\_dir with tree-sitter and save chunks JSON (index, file\_path, chunk\_kind, symbol, start/end\_line, source, ast\_type, imports). \\
\texttt{publish\_verified\_locations} & Verify/overwrite LLM locations by matching symbol+snippet in numbered snapshots; fallback to original lines; save verified JSON and snapshot index; updates \texttt{env.\_state['last\_llm\_json\_path']}. \\
\texttt{run\_san} & Run ARVO workflow that triggers ASAN/fuzzer and capture output. \\
\texttt{run\_gdb} & Run gdb in the ARVO container and return a backtrace. \\
\texttt{llm\_analyze} & Send crash/ASAN log to the LLM to predict likely bug locations (JSON expected). Supports refine mode with source slices. \\
\texttt{mark\_diff} & Load anchored diff (anchors\_old + insert\_points) and mark chunks touched by these lines; updates chunks JSON (adds \texttt{diff}, \texttt{diff\_hit\_lines}). \\
\texttt{compare\_llm\_metrics} & Compute detection rate (chunk-level) and localization rate (diff-line-level), plus strict localization by exact interval equality; updates JSON flags accordingly. \\
\texttt{gdb\_script} & Run GDB non-interactively with \texttt{-batch} and provided commands; returns GDB output. \\
\texttt{extract\_json} & Extract JSON array from a raw LLM response; optionally merge with last predictions; writes temp JSON and updates state. \\
\texttt{extract\_modified\_lines} & Parse a unified diff and return modified (file, line) list. \\
\texttt{compare\_patch} & Compare LLM predicted spans with the patch; set solved by match rate. \\
\texttt{pipeline} & End-to-end: ASAN, GDB, LLM analyze, (optional) compare with patch. \\
\texttt{delegate} & Delegate a task to an executor LLM agent (autonomous, equipped for CTF-style tasks). \\
\texttt{static\_analysis\_config} & Enable/disable static analysis inclusion in crash log (\texttt{enabled} flag). \\
\texttt{baseline\_llm\_analyze} & Single-shot baseline without ASAN/GDB or static context—ask LLM to guess once (no refine/verify/postprocess). \\
\end{longtable}

\subsection{Other case studies}

In this section, we present several additional demonstration cases that were not included in the main body of the paper. These examples aim to further illustrate the internal workflow, reasoning strategies, and decision-making processes of the \texttt{T2L-Agent} across diverse scenarios. By showcasing these supplementary cases, we hope to enhance the reader's understanding of how \texttt{T2L-Agent} performs trace-to-line localization in practice and highlight its robustness across varying bug types and complexity levels.


\begin{figure}[H]
  \centering
  \includegraphics[width=.48\textwidth,height=\textheight,keepaspectratio]{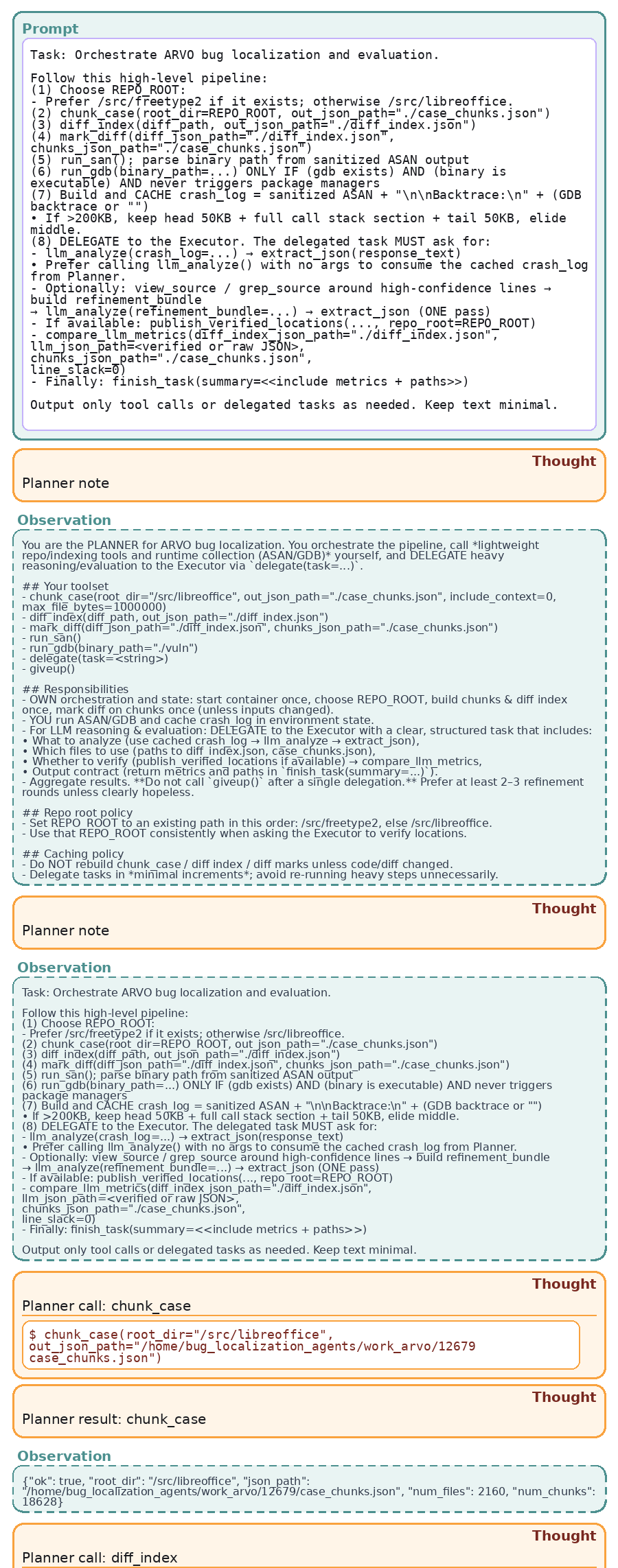}\hfill
  \includegraphics[width=.48\textwidth,height=\textheight,keepaspectratio]{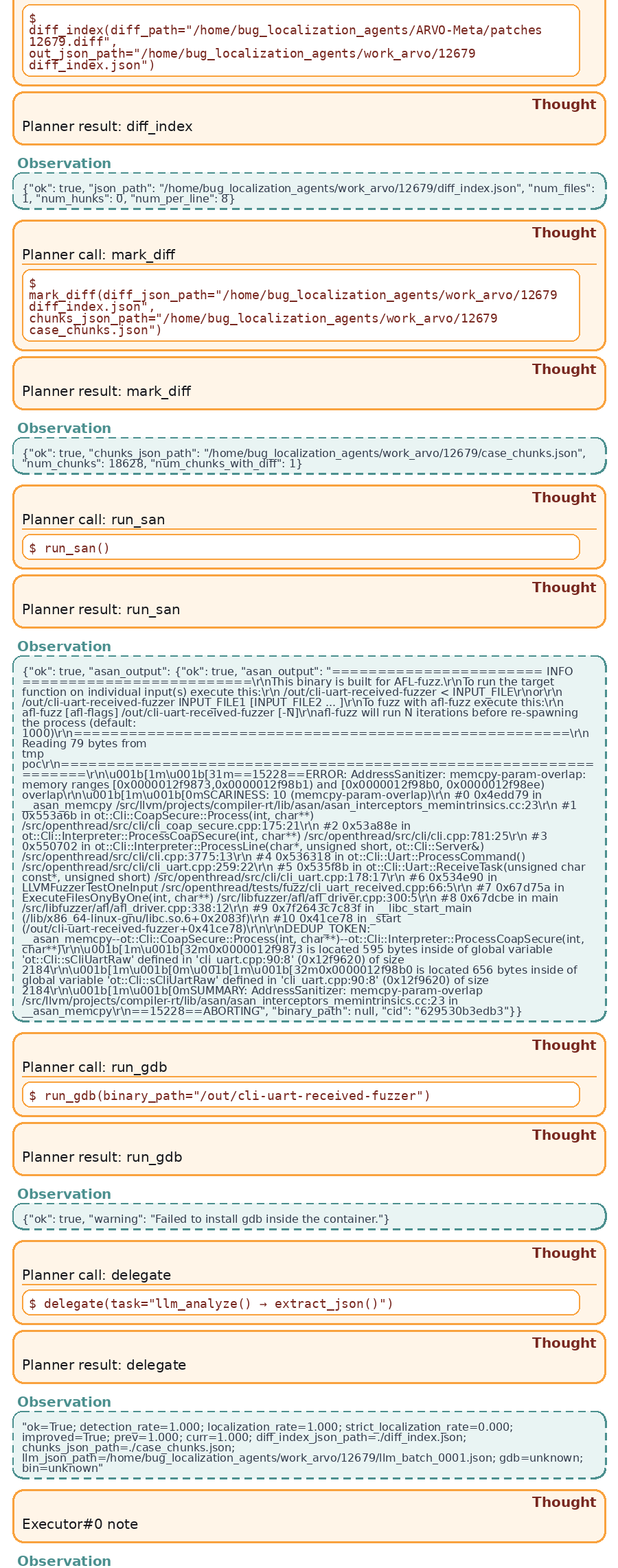}

\end{figure}

\begin{figure}[H]
  \centering
  \includegraphics[width=.48\textwidth,height=\textheight,keepaspectratio]{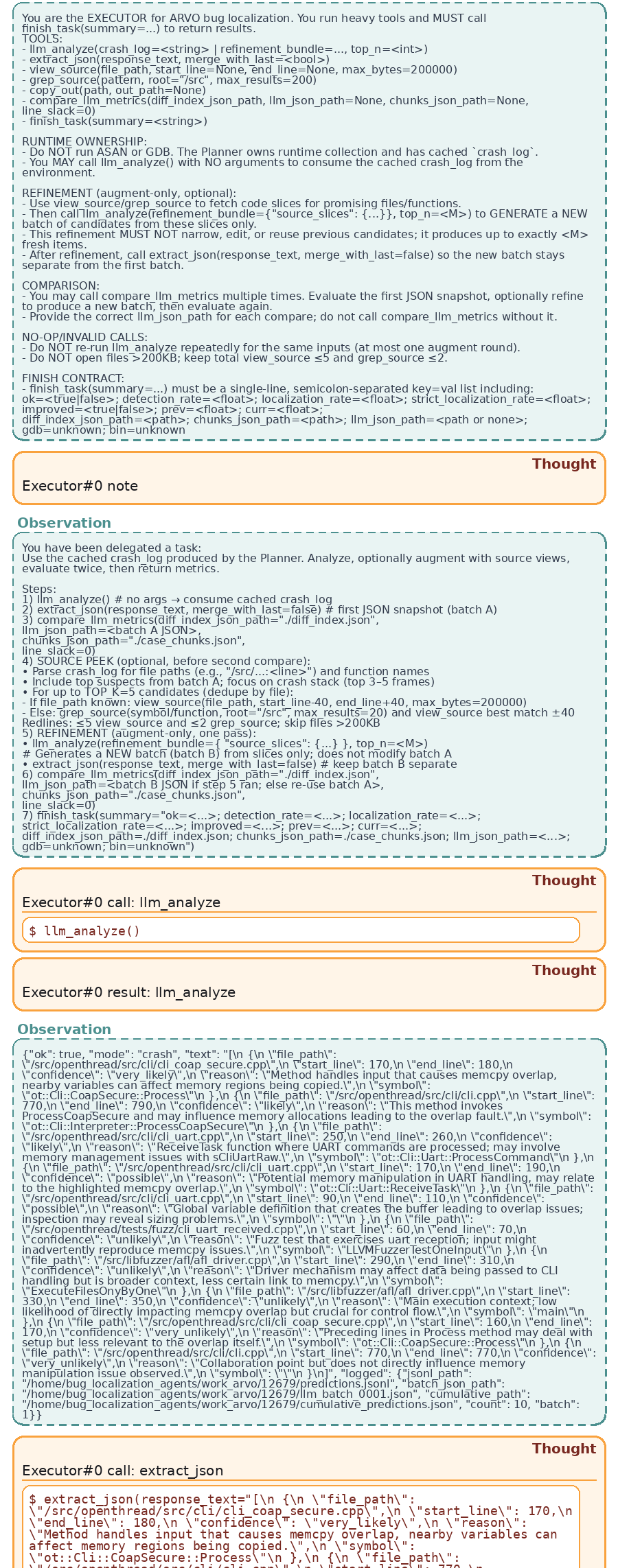}\hfill
  \includegraphics[width=.48\textwidth,height=\textheight,keepaspectratio]{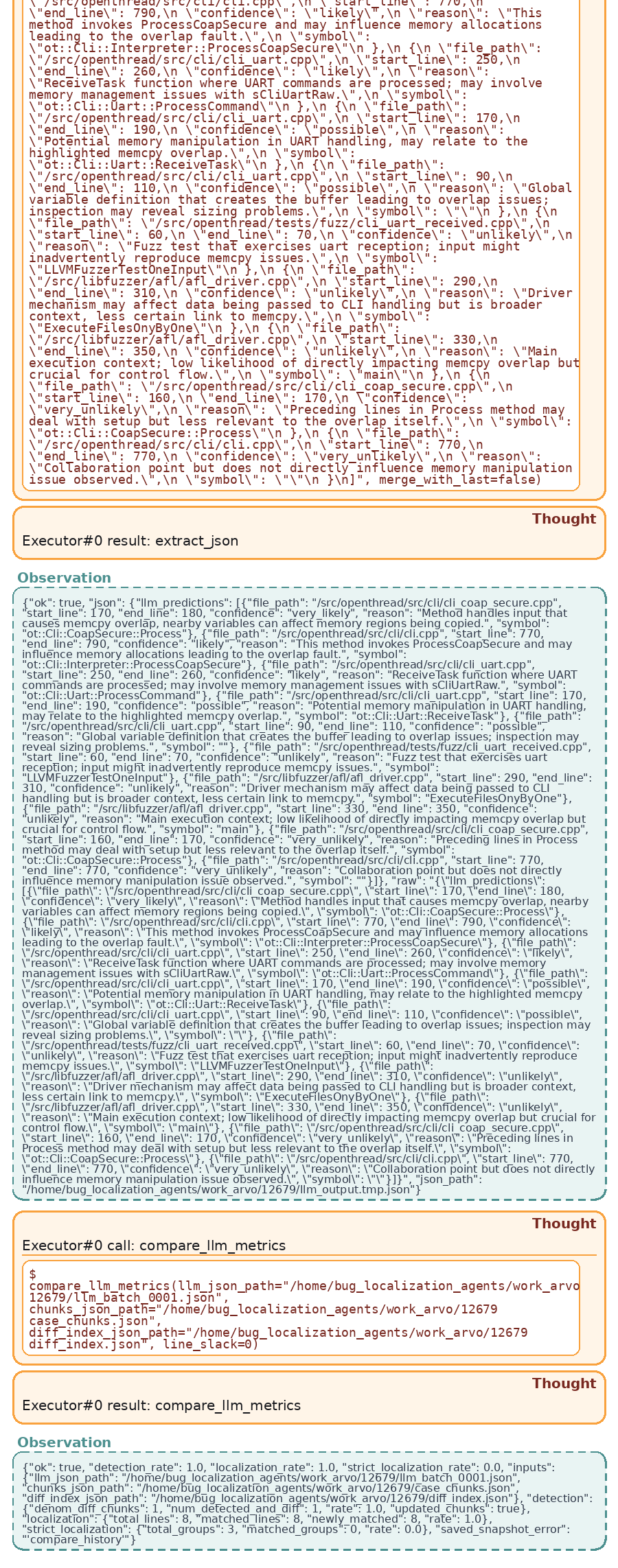}
    \caption{GPT-4o mini Divergence Tracing for case 12679.}
    \label{fig:12679_topn}
\end{figure}

\begin{figure}[H]
  \centering
  \includegraphics[width=.48\textwidth,height=\textheight,keepaspectratio]{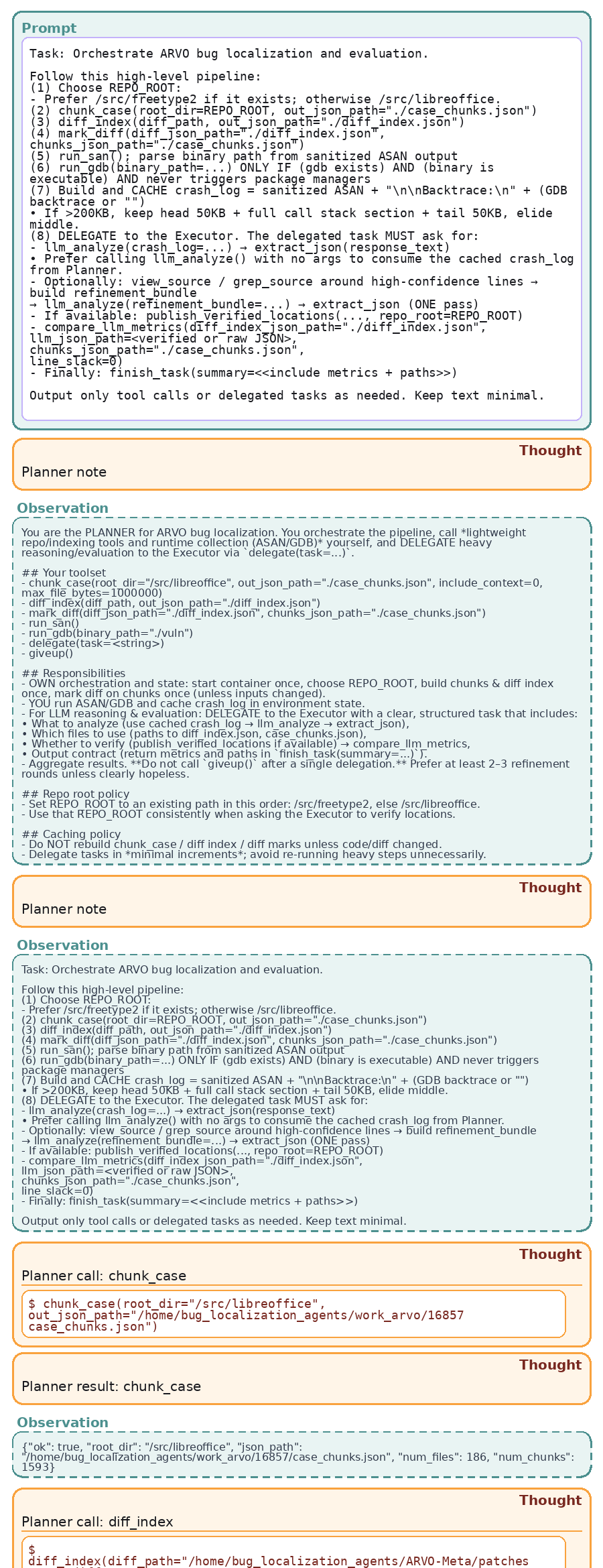}\hfill
  \includegraphics[width=.48\textwidth,height=\textheight,keepaspectratio]{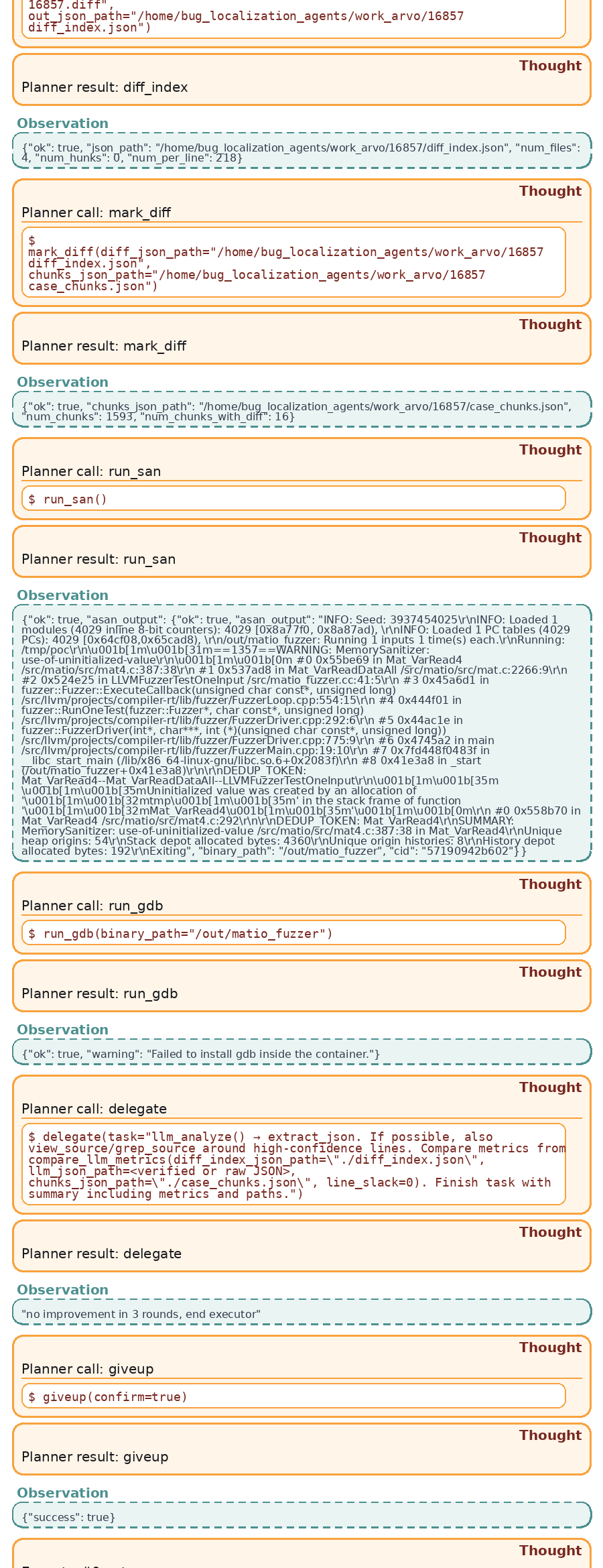}

\end{figure}

\begin{figure}[H]
  \centering
  \includegraphics[width=.48\textwidth,height=\textheight,keepaspectratio]{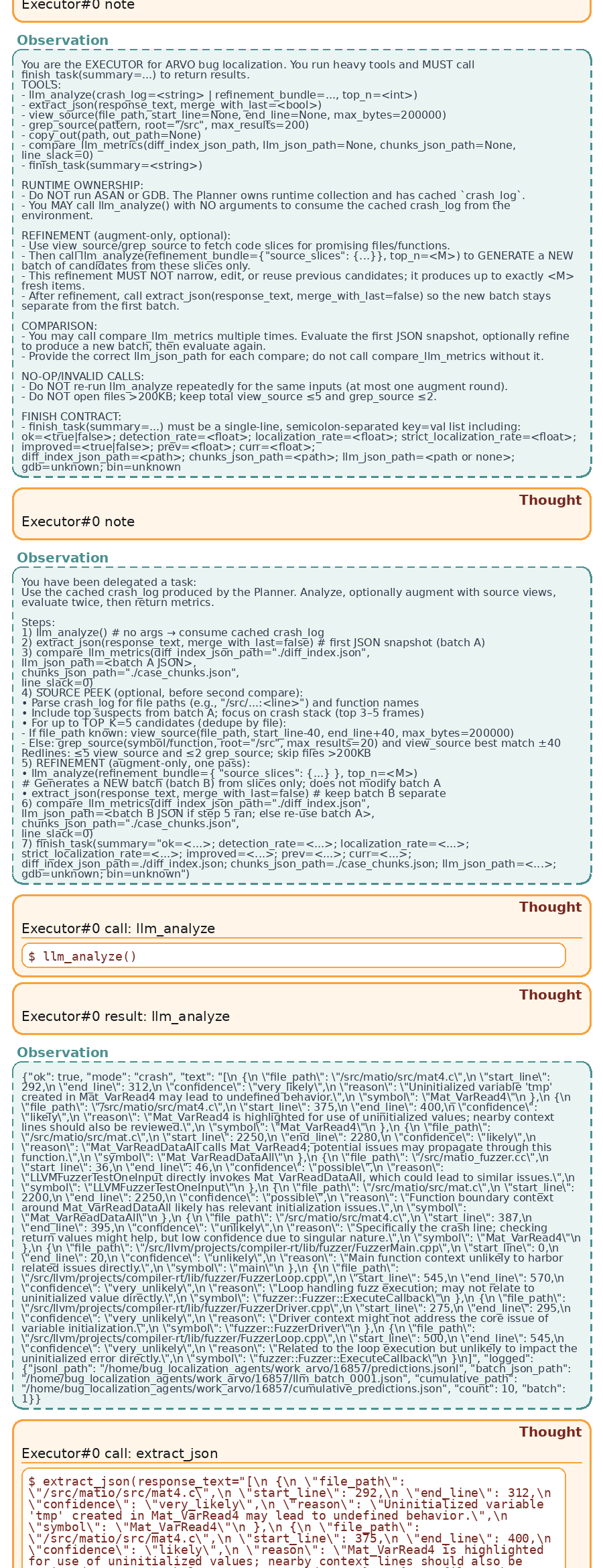}\hfill
  \includegraphics[width=.48\textwidth,height=\textheight,keepaspectratio]{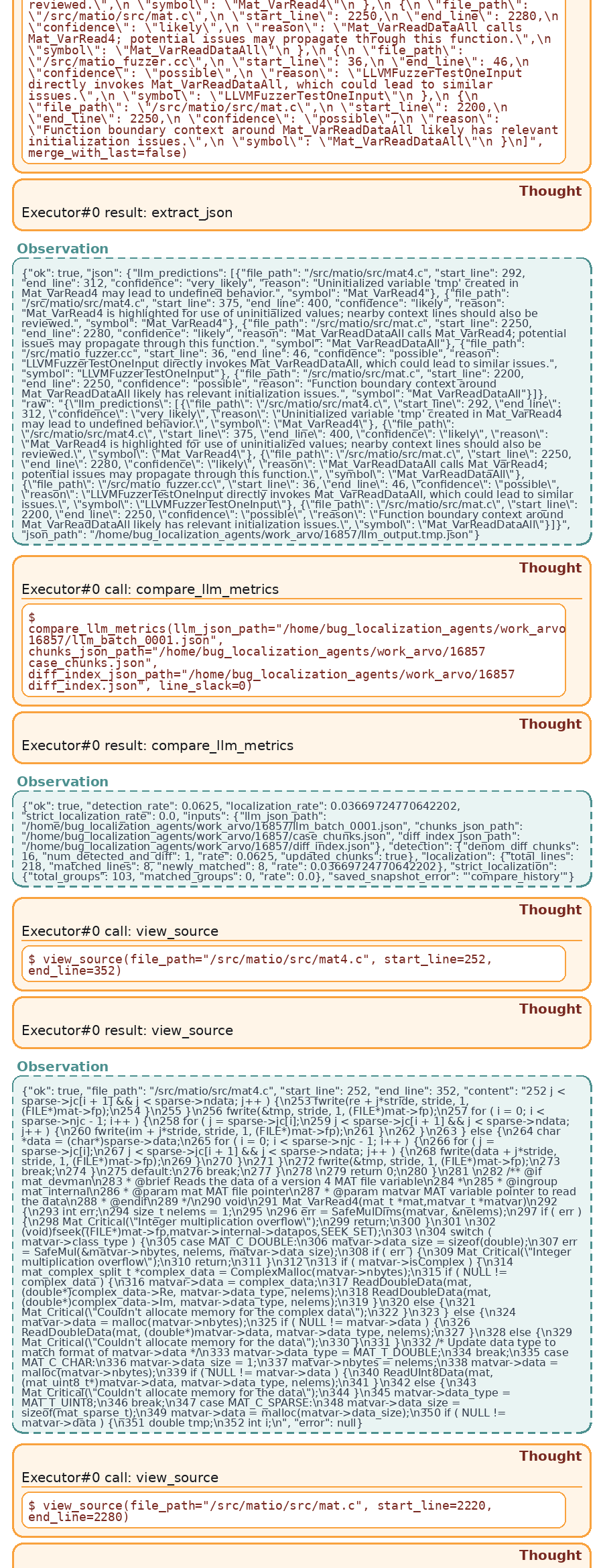}

\end{figure}

\begin{figure}[H]
  \centering
  \includegraphics[width=.48\textwidth,height=\textheight,keepaspectratio]{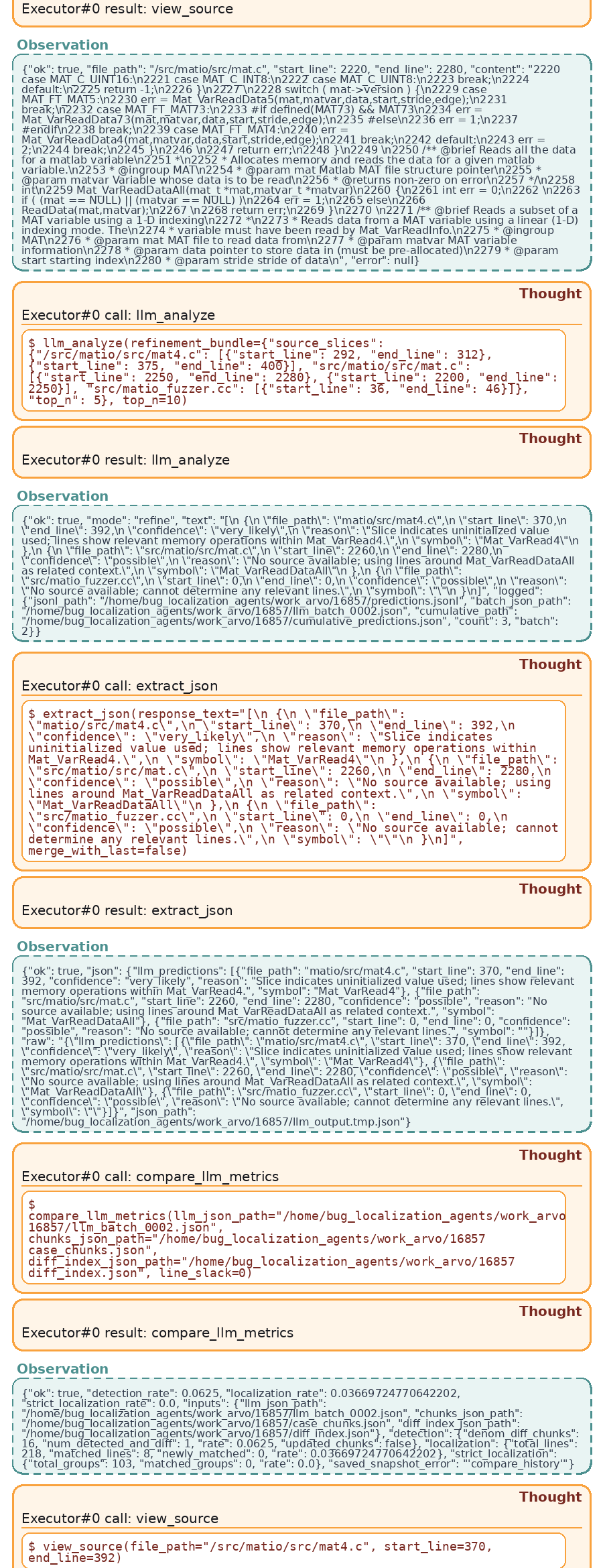}\hfill
  \includegraphics[width=.48\textwidth,height=\textheight,keepaspectratio]{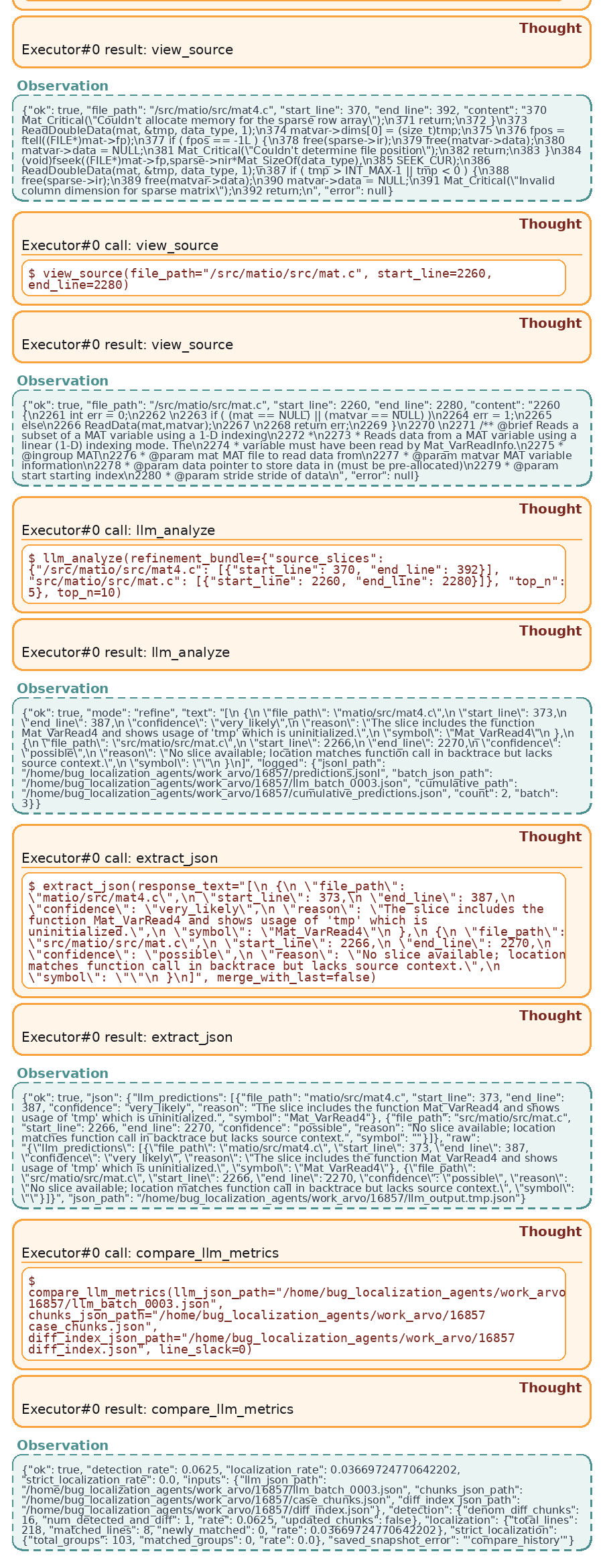}
    \caption{GPT-4o mini Divergence Tracing for case 16857.}
    \label{fig:16857_topn}
\end{figure}

\begin{figure}[H]
  \centering
  \includegraphics[width=.48\textwidth,height=\textheight,keepaspectratio]{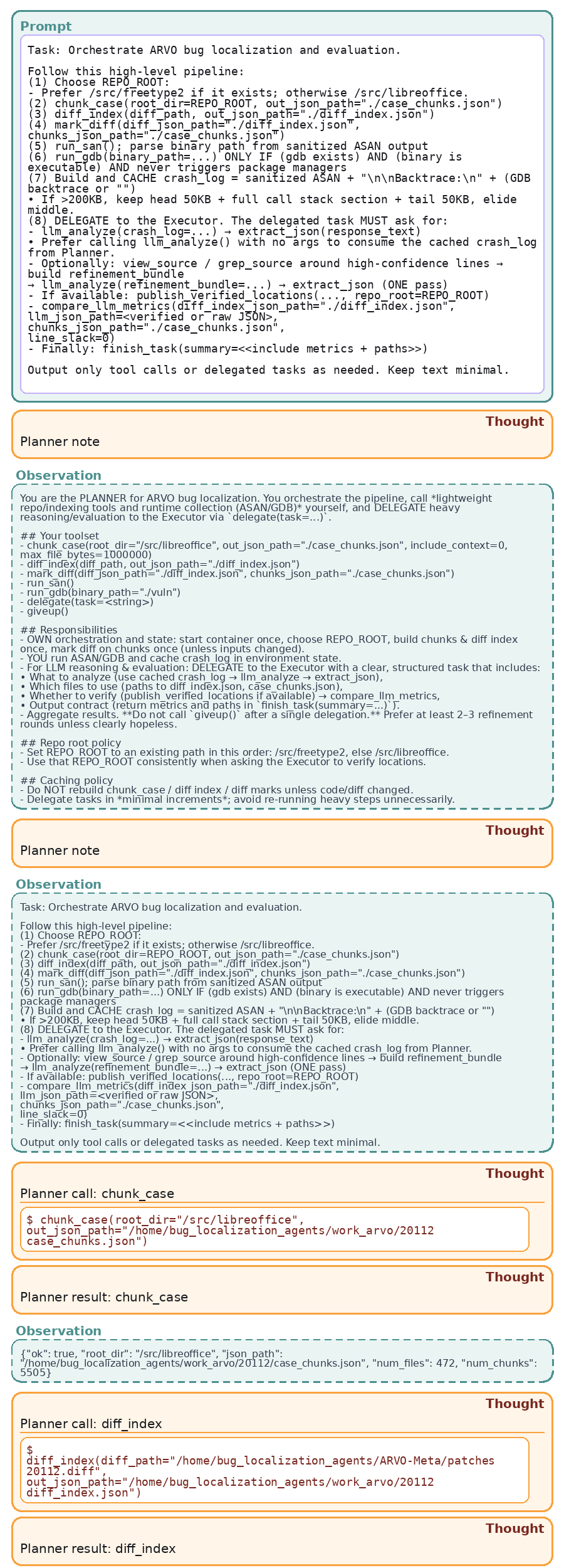}\hfill
  \includegraphics[width=.48\textwidth,height=\textheight,keepaspectratio]{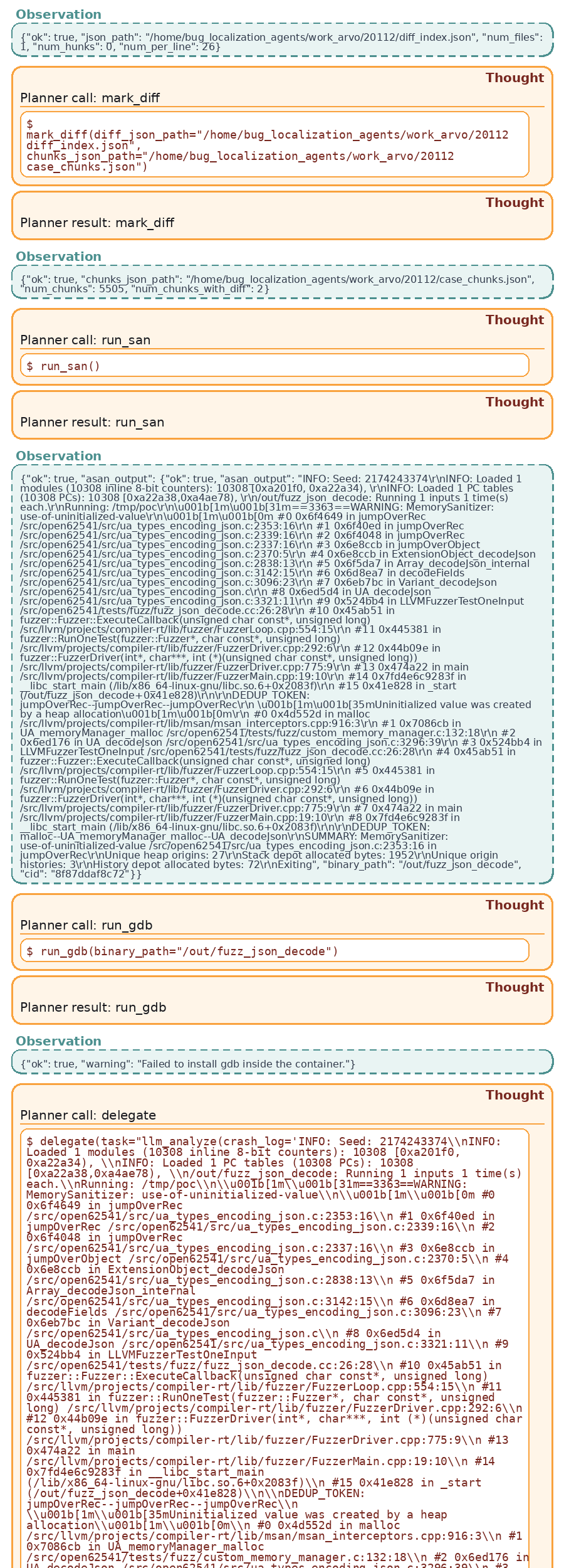}

\end{figure}

\begin{figure}[H]
  \centering
  \includegraphics[width=.48\textwidth,height=\textheight,keepaspectratio]{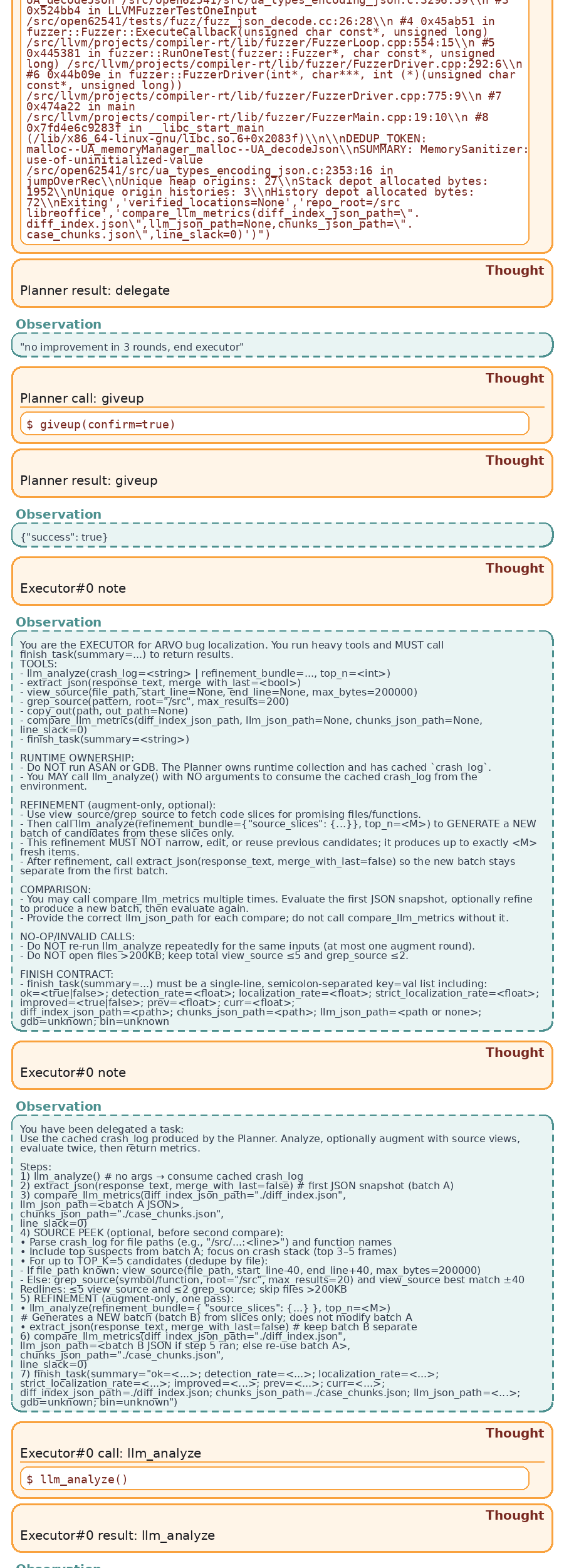}\hfill
  \includegraphics[width=.48\textwidth,height=\textheight,keepaspectratio]{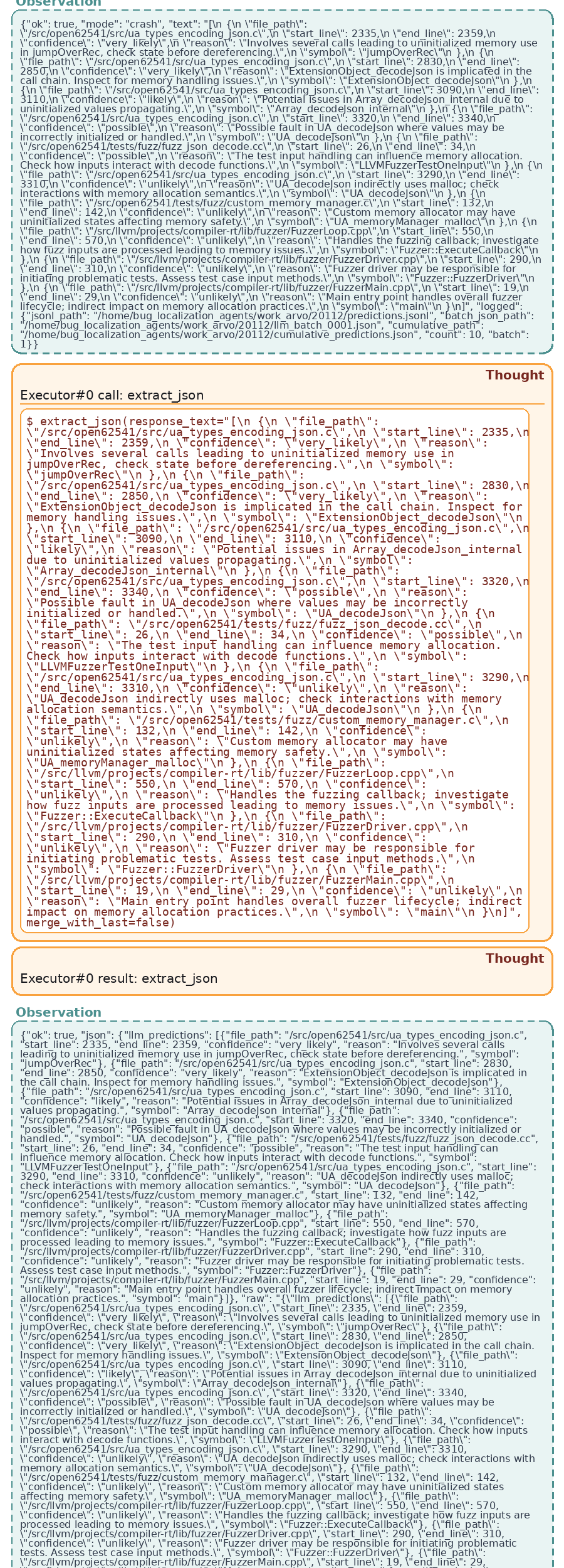}

\end{figure}

\begin{figure}[H]
  \centering
  \includegraphics[width=.48\textwidth,height=\textheight,keepaspectratio]{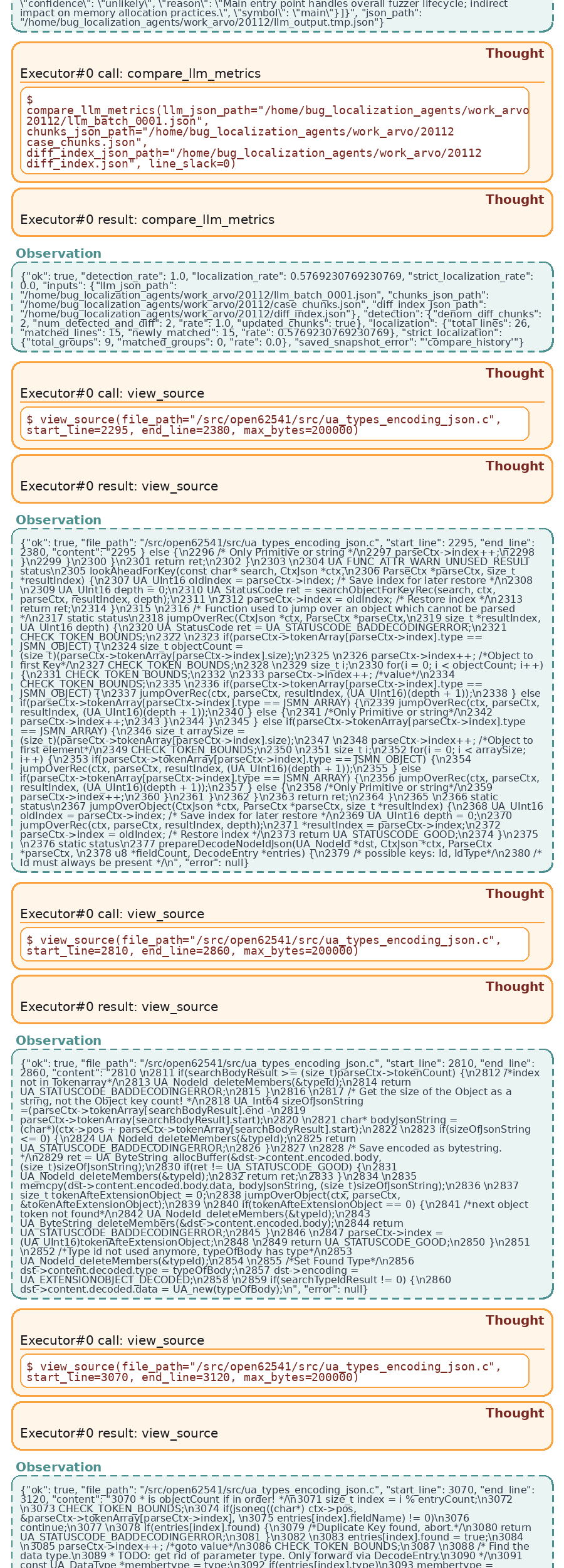}\hfill
  \includegraphics[width=.48\textwidth,height=\textheight,keepaspectratio]{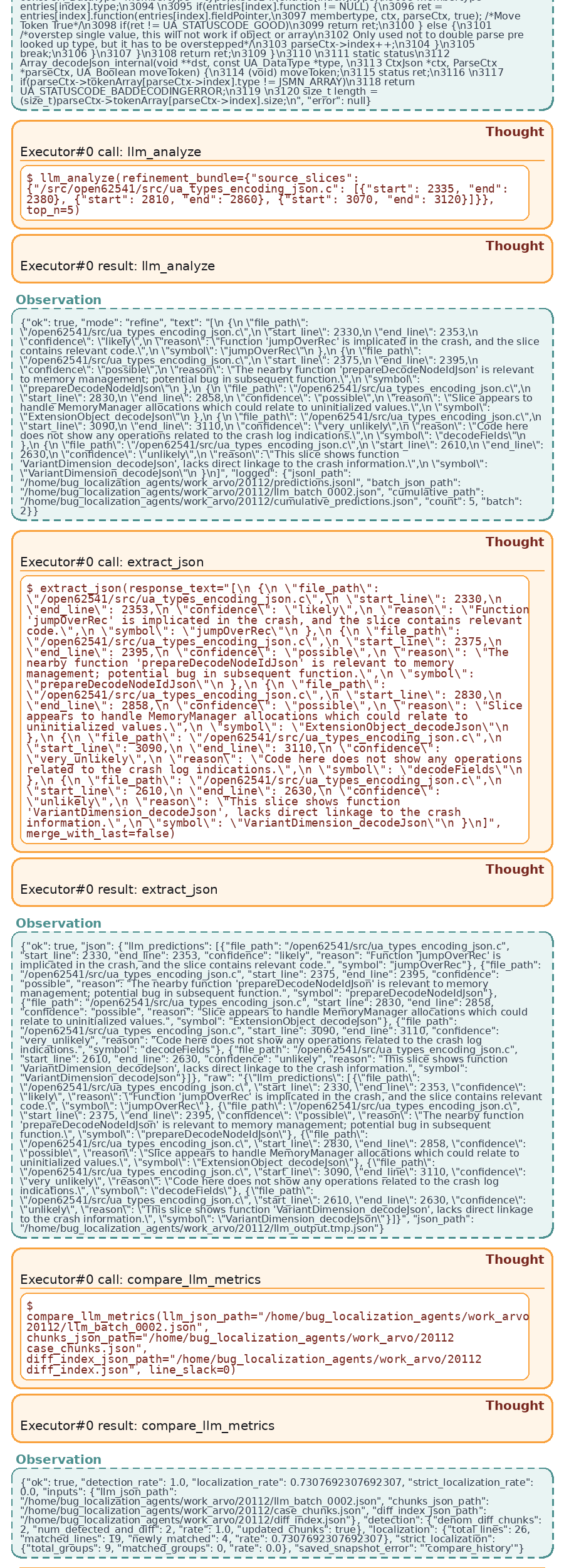}
\end{figure}

\begin{figure}[H]
  \centering
  \includegraphics[width=.48\textwidth,height=\textheight,keepaspectratio]{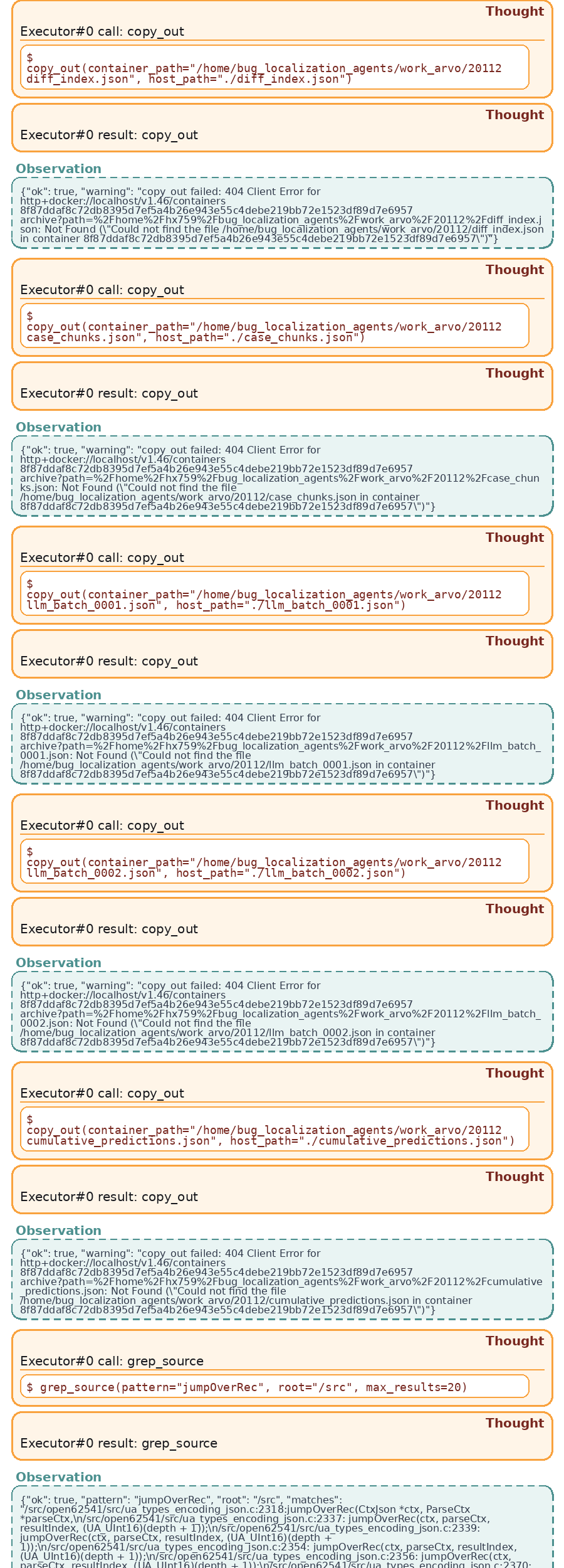}\hfill
  \includegraphics[width=.48\textwidth,height=\textheight,keepaspectratio]{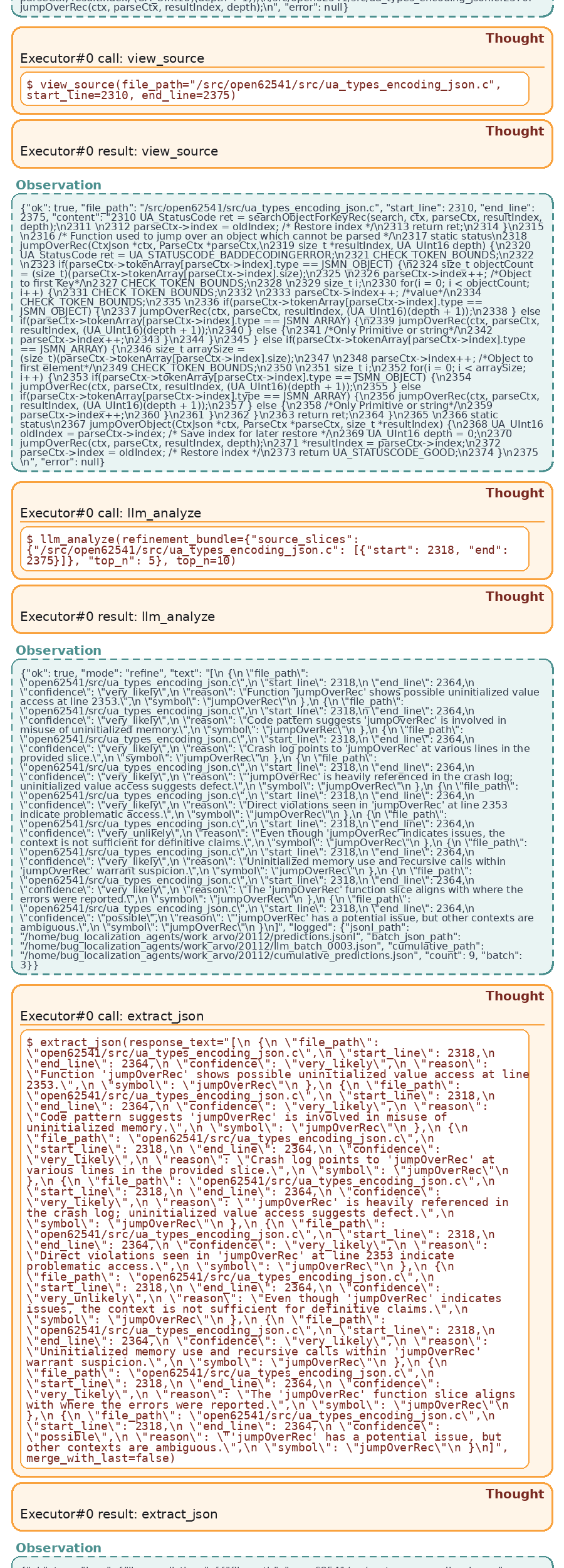}
    \caption{GPT-4o mini Divergence Tracing for case 20112.}
\end{figure}

\begin{figure}[H]
  \centering
  \includegraphics[width=.48\textwidth,height=\textheight,keepaspectratio]{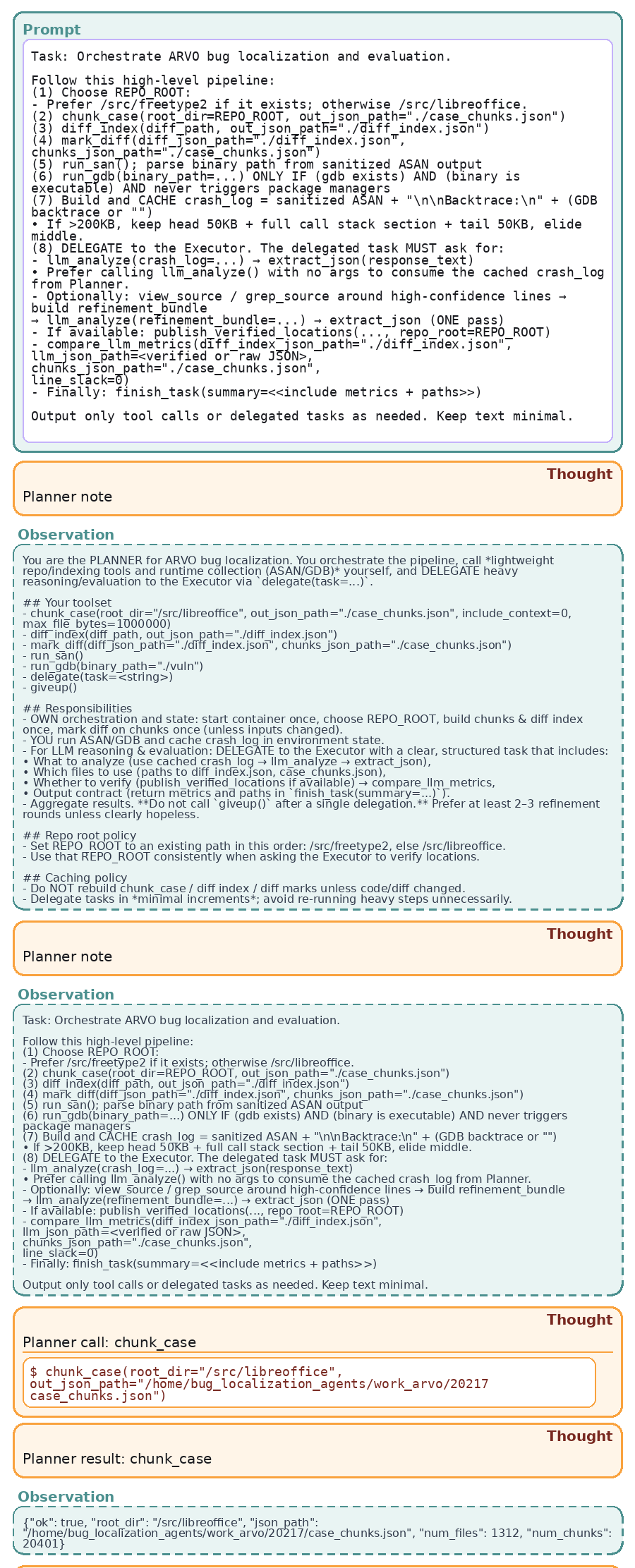}\hfill
  \includegraphics[width=.48\textwidth,height=\textheight,keepaspectratio]{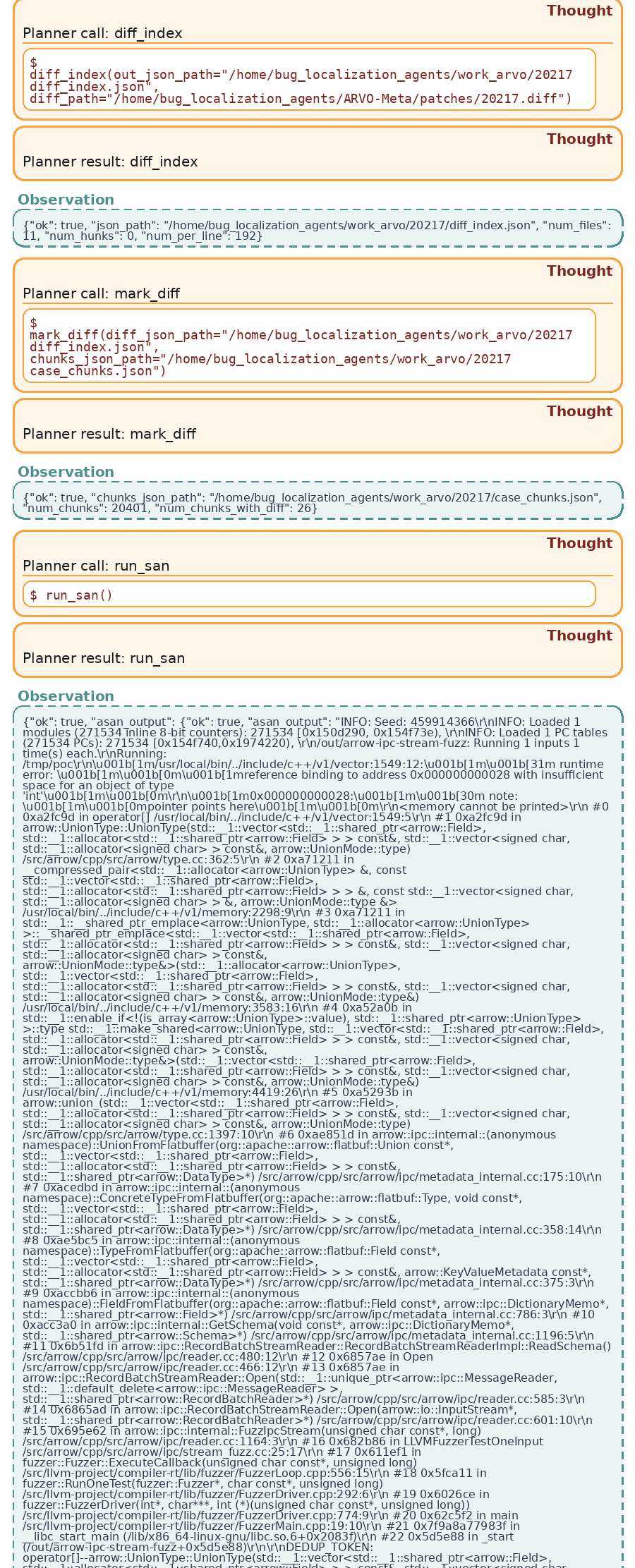}

\end{figure}

\begin{figure}[H]
  \centering
  \includegraphics[width=.48\textwidth,height=\textheight,keepaspectratio]{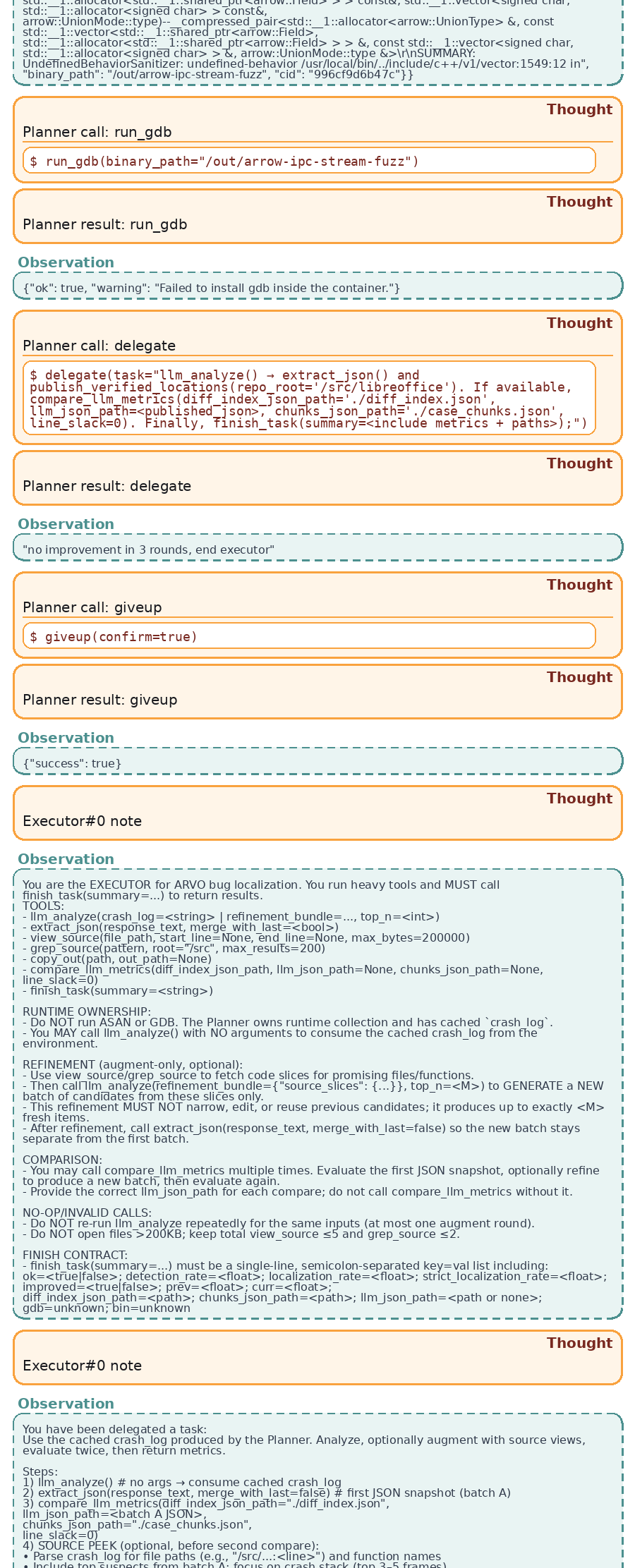}\hfill
  \includegraphics[width=.48\textwidth,height=\textheight,keepaspectratio]{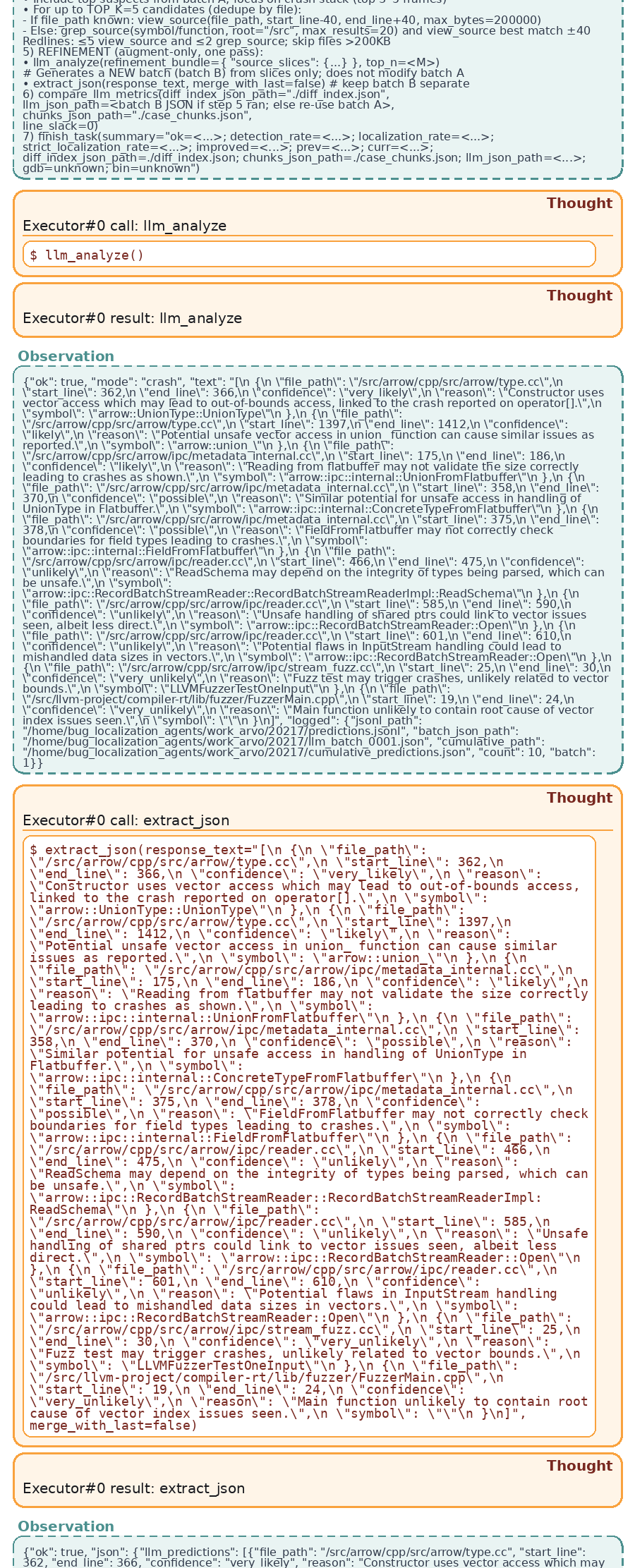}

\end{figure}

\begin{figure}[H]
  \centering
  \includegraphics[width=.48\textwidth,height=\textheight,keepaspectratio]{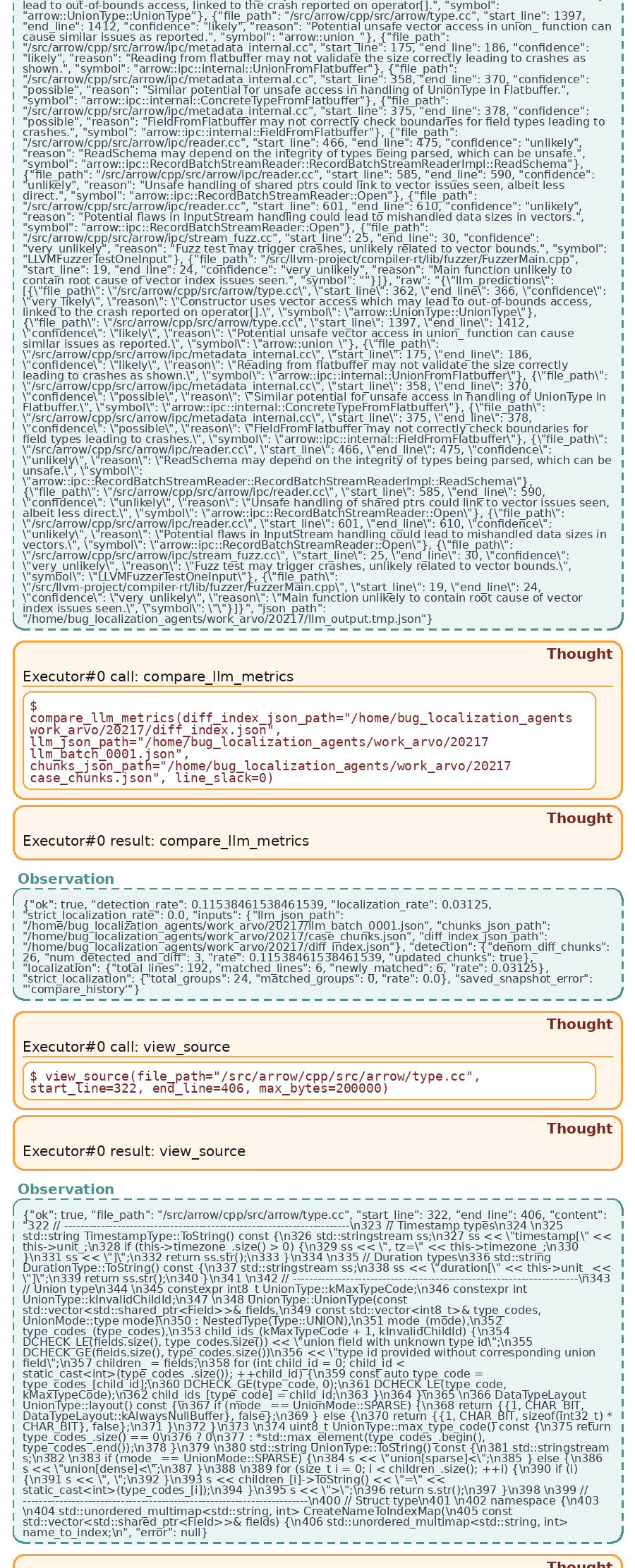}\hfill
  \includegraphics[width=.48\textwidth,height=\textheight,keepaspectratio]{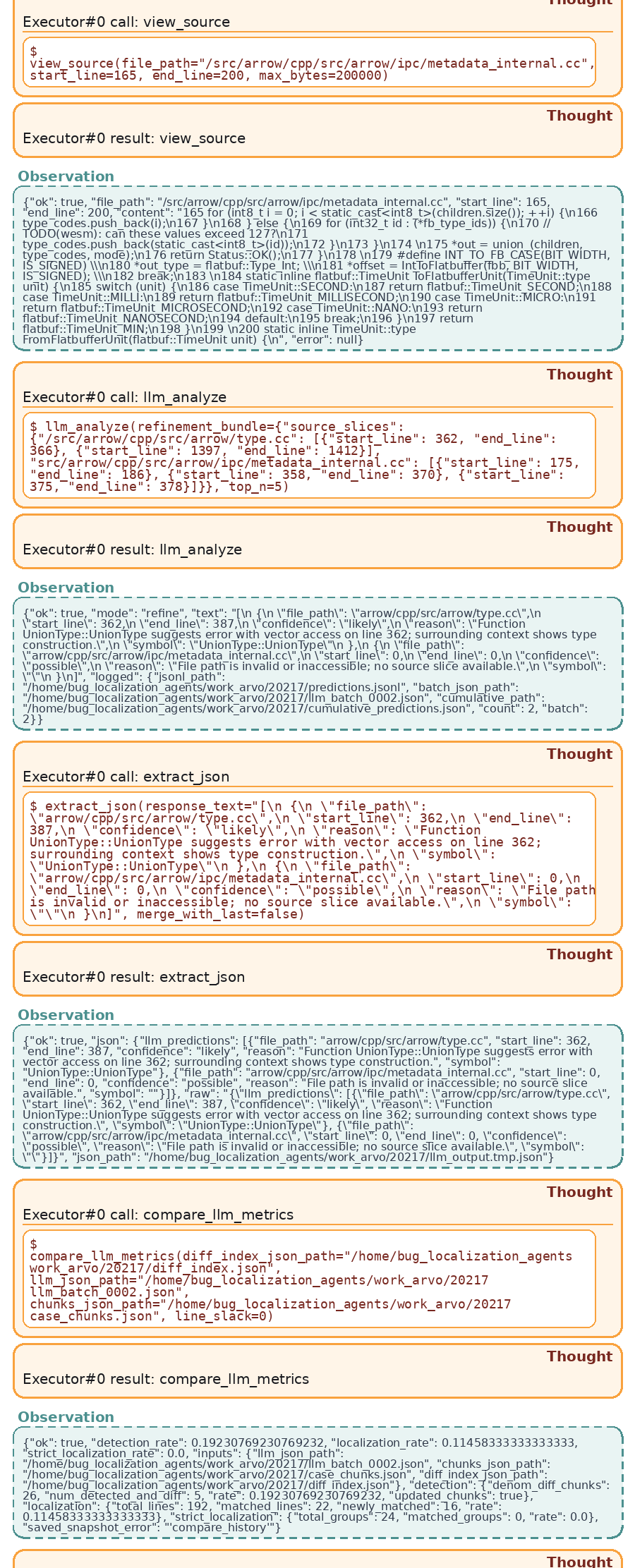}
\end{figure}

\begin{figure}[H]
  \centering
  \includegraphics[width=.48\textwidth,height=\textheight,keepaspectratio]{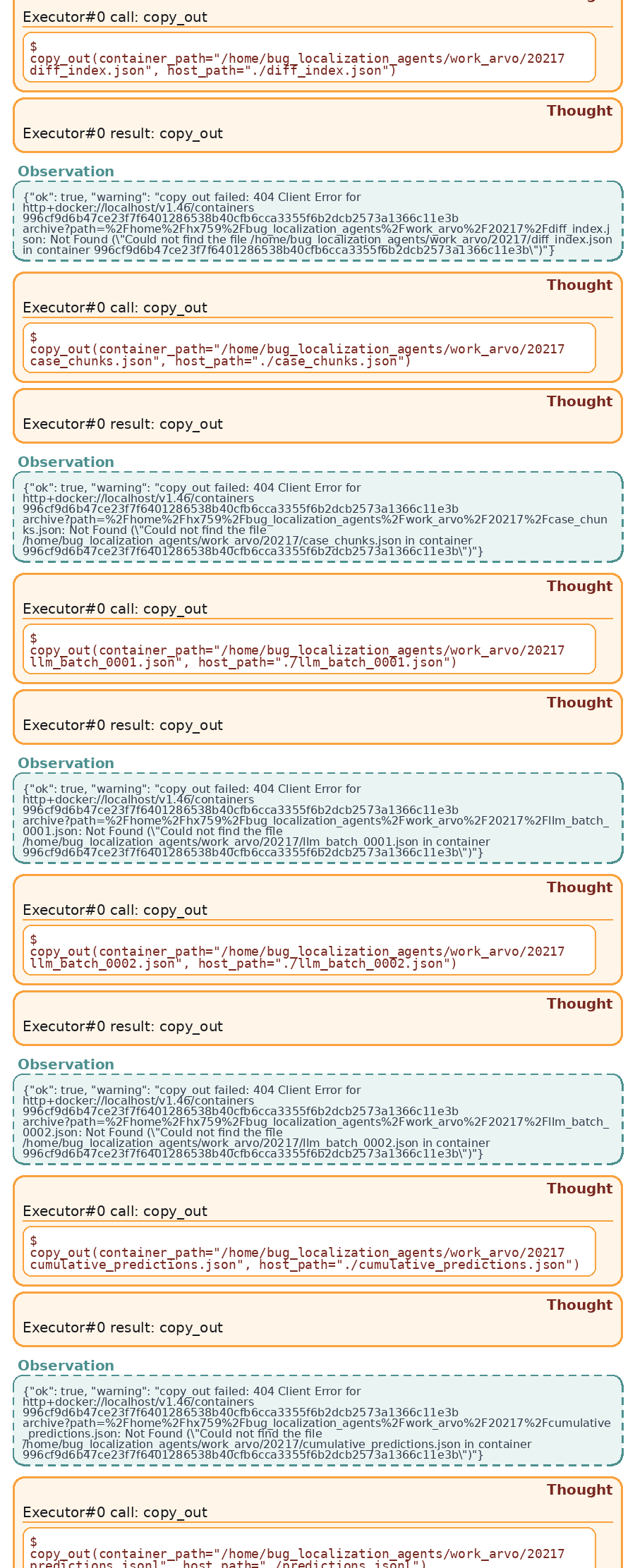}\hfill
  \includegraphics[width=.48\textwidth,height=\textheight,keepaspectratio]{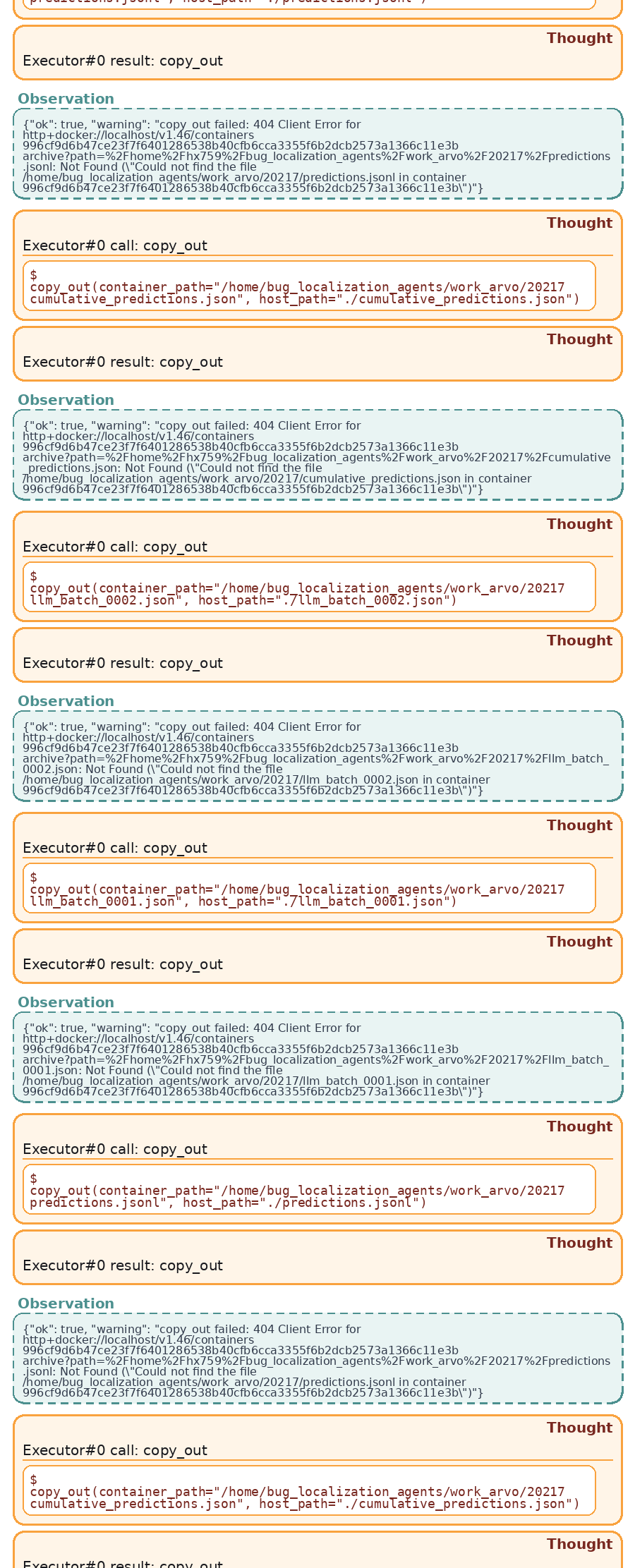}
\end{figure}

\begin{figure}[H]
  \centering
  \includegraphics[width=.48\textwidth,height=\textheight,keepaspectratio]{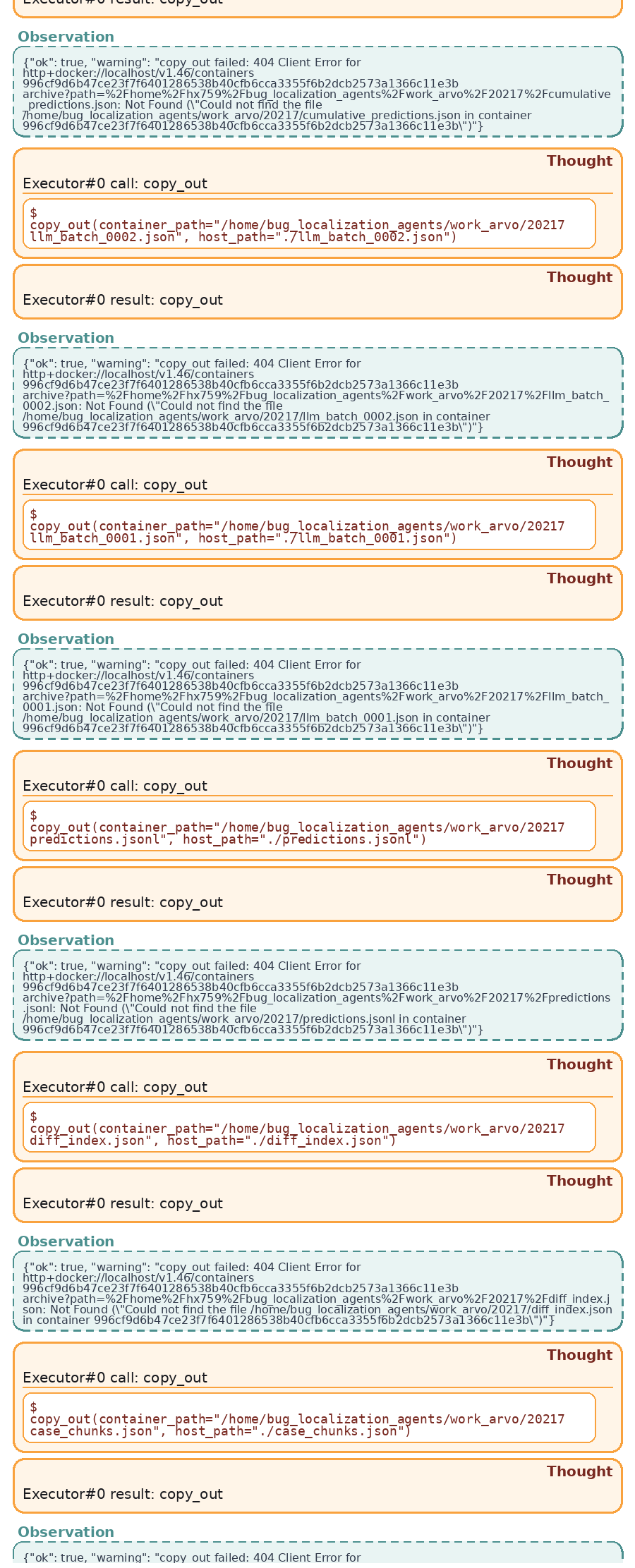}\hfill

    \caption{GPT-4o mini Divergence Tracing for case 20217.}
\end{figure}

\begin{figure}[H]
  \centering
  \includegraphics[width=.48\textwidth,height=\textheight,keepaspectratio]{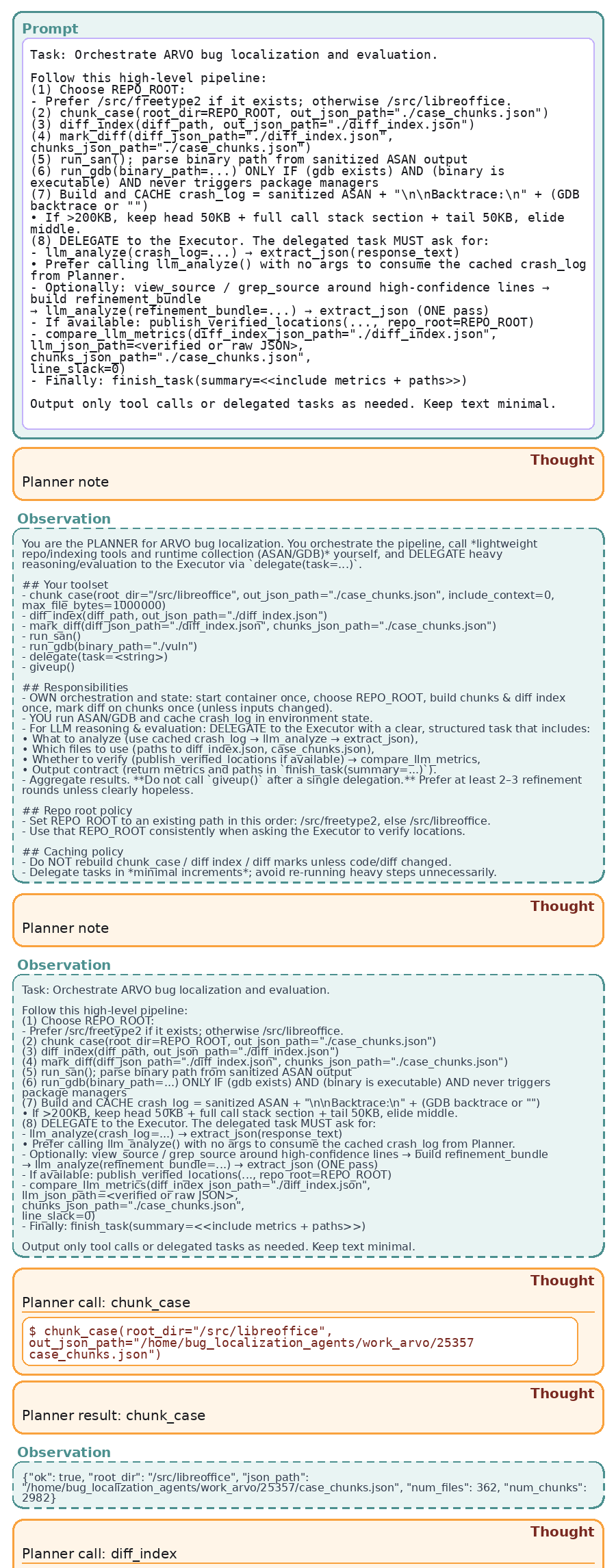}\hfill
  \includegraphics[width=.48\textwidth,height=\textheight,keepaspectratio]{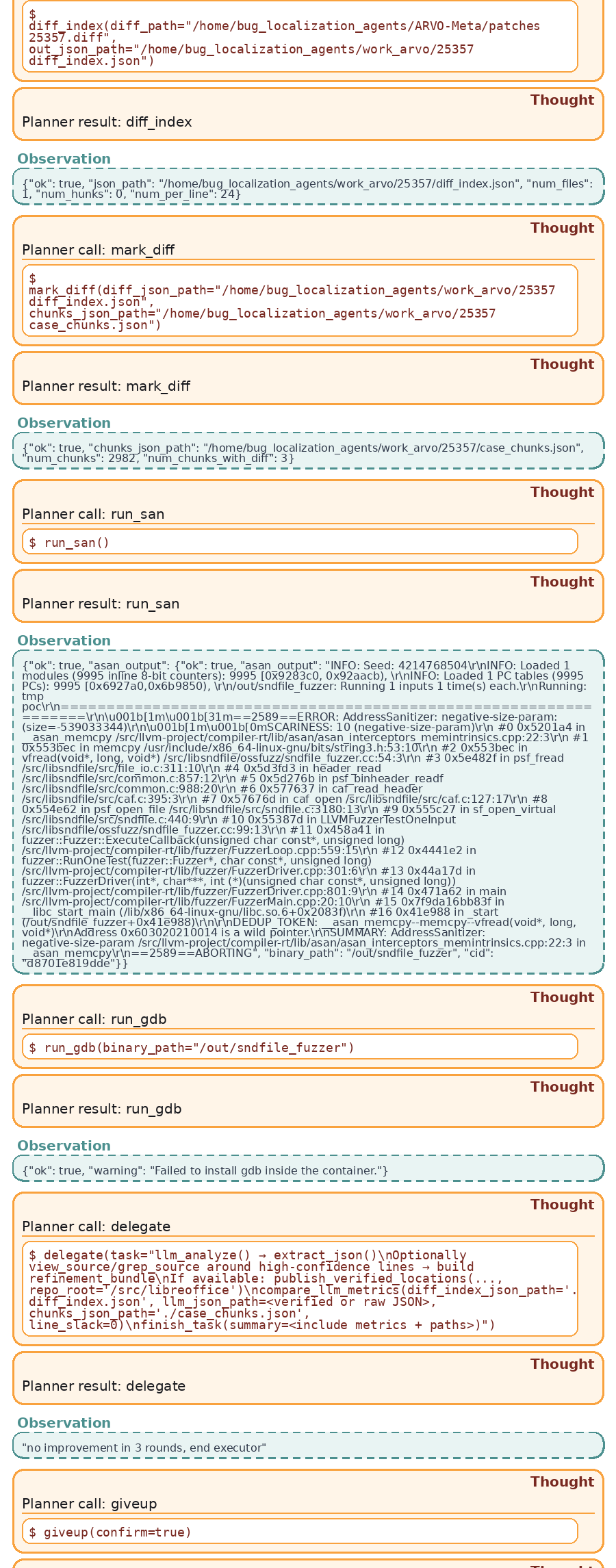}

\end{figure}

\begin{figure}[H]
  \centering
  \includegraphics[width=.48\textwidth,height=\textheight,keepaspectratio]{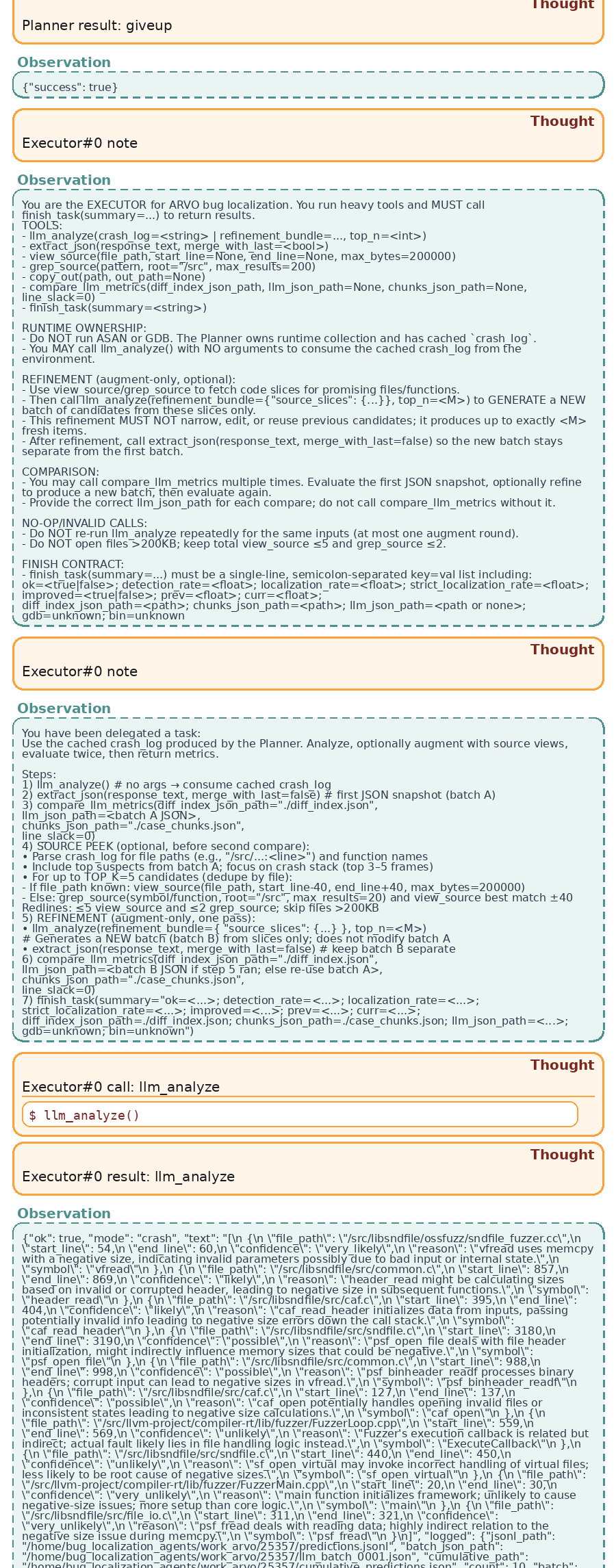}\hfill
  \includegraphics[width=.48\textwidth,height=\textheight,keepaspectratio]{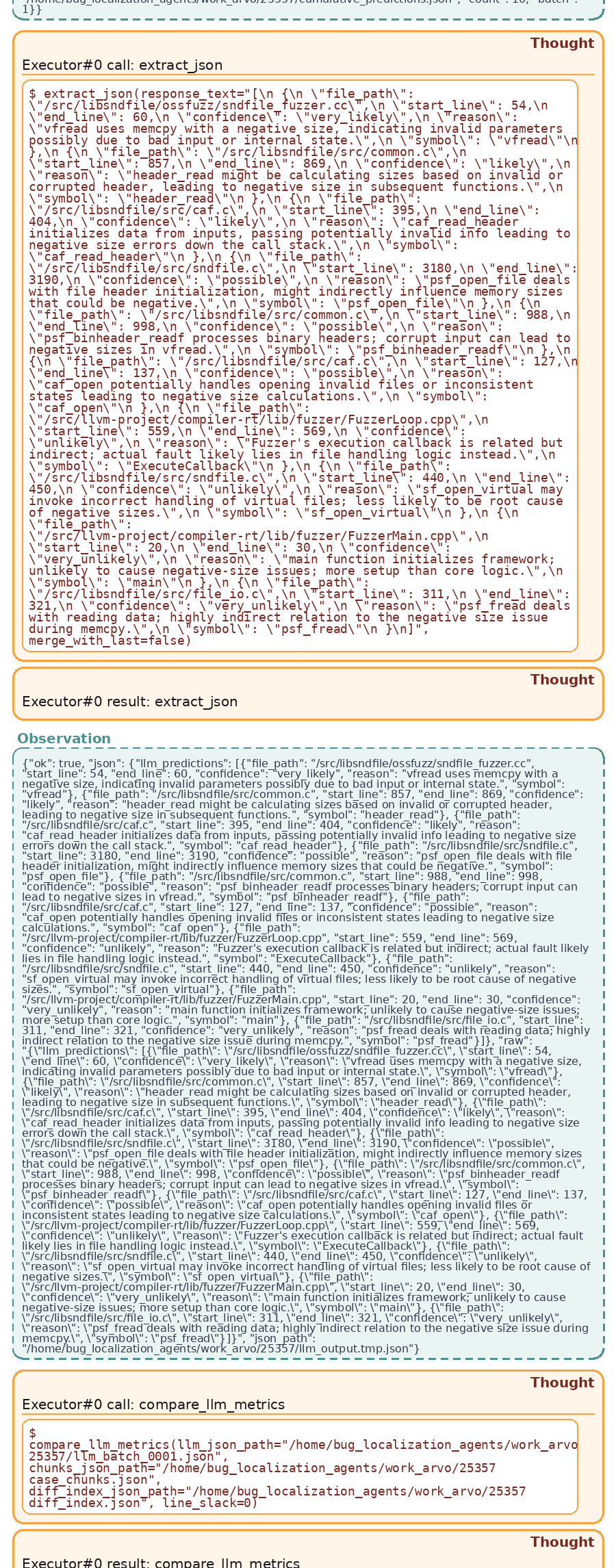}

\end{figure}

\begin{figure}[H]
  \centering
  \includegraphics[width=.48\textwidth,height=\textheight,keepaspectratio]{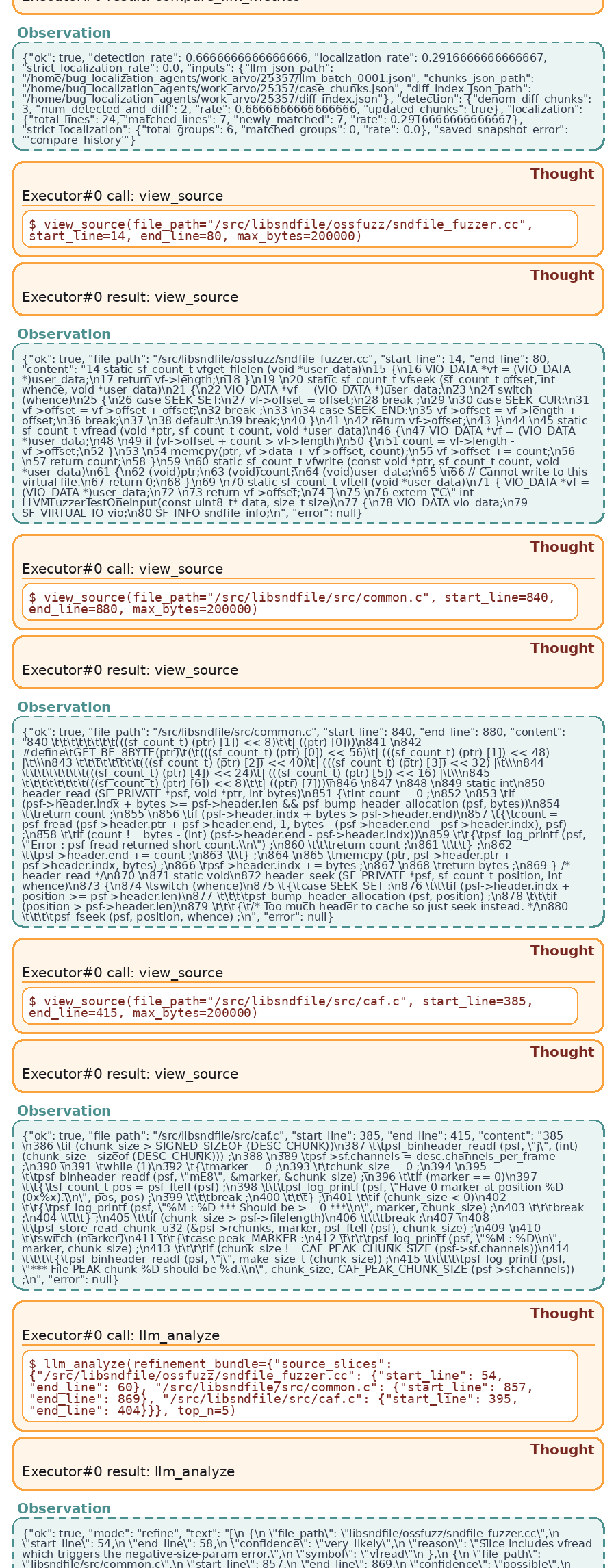}\hfill
  \includegraphics[width=.48\textwidth,height=\textheight,keepaspectratio]{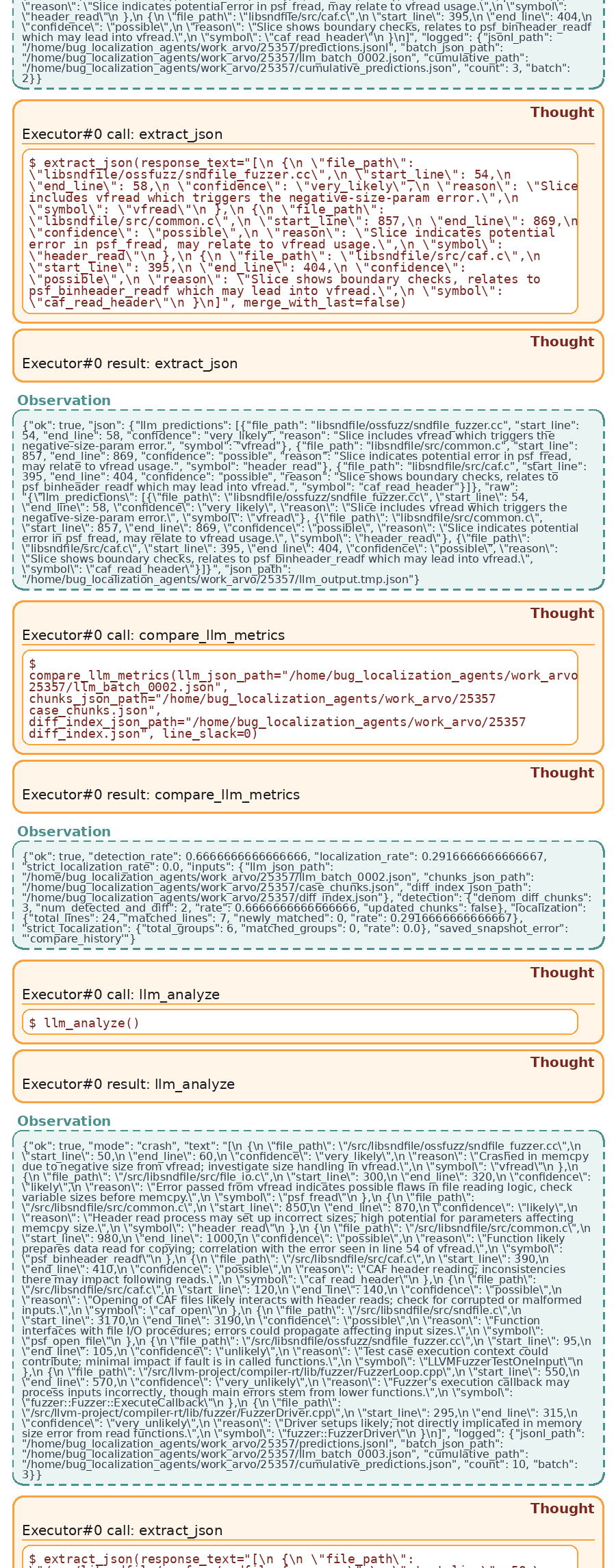}
\end{figure}

\begin{figure}[H]
  \centering
  \includegraphics[width=.48\textwidth,height=\textheight,keepaspectratio]{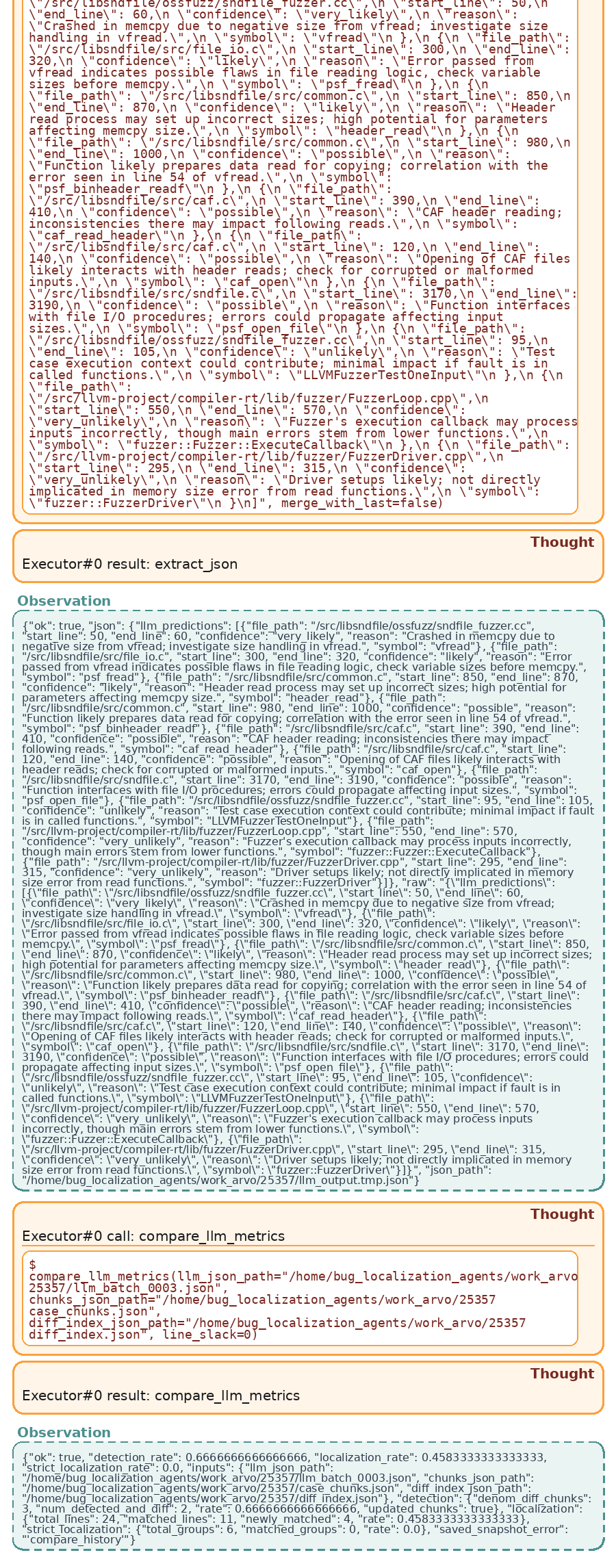}\hfill
    \caption{GPT-4o mini Divergence Tracing for case 25357.}
\end{figure}

\begin{figure}[H]
  \centering
  \includegraphics[width=.48\textwidth,height=\textheight,keepaspectratio]{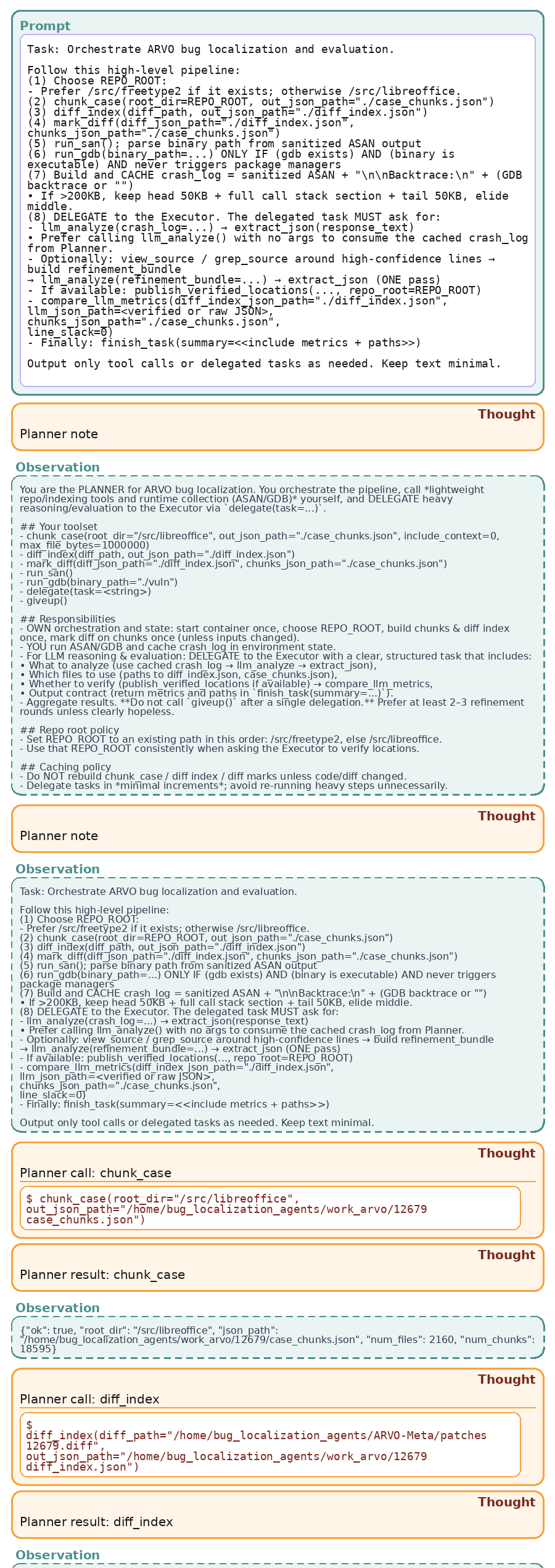}\hfill
  \includegraphics[width=.48\textwidth,height=\textheight,keepaspectratio]{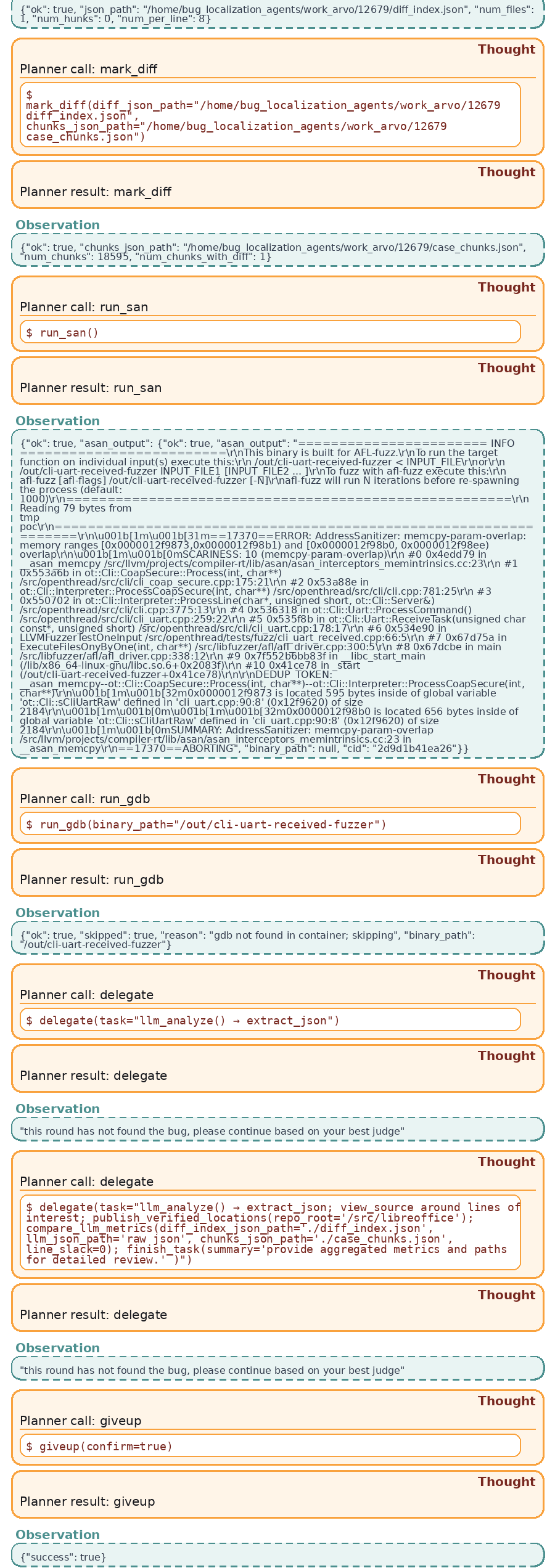}

\end{figure}

\begin{figure}[H]
  \centering
  \includegraphics[width=.48\textwidth,height=\textheight,keepaspectratio]{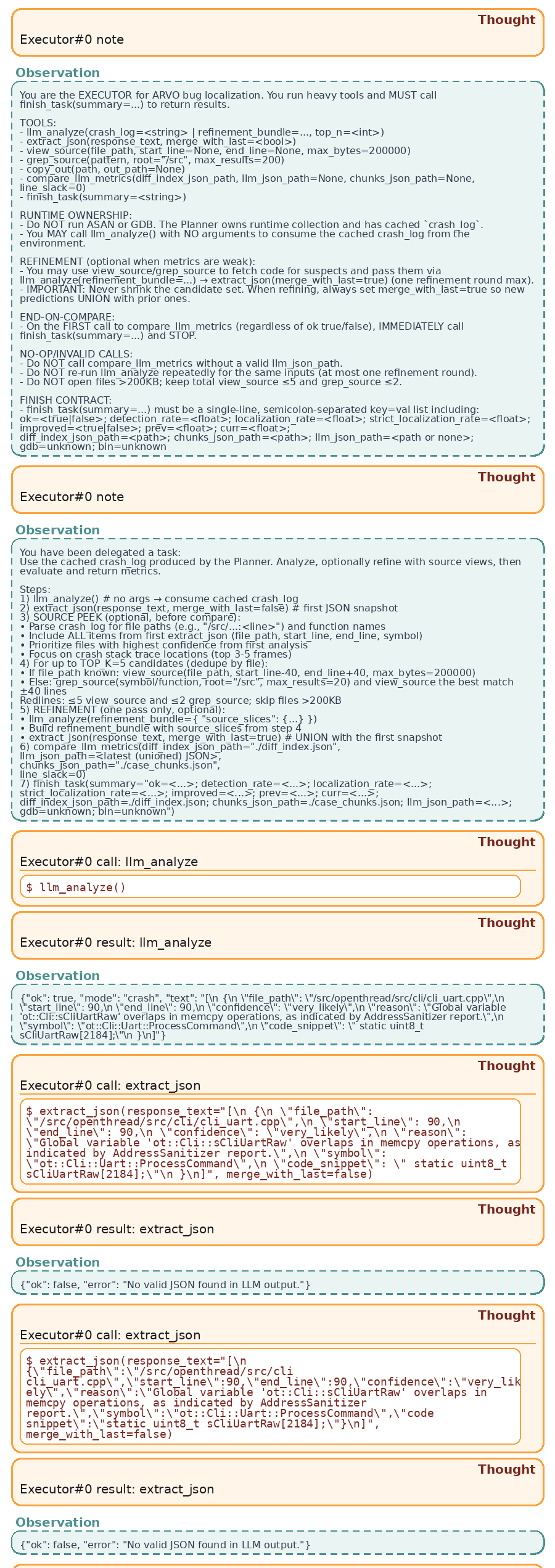}\hfill
  \includegraphics[width=.48\textwidth,height=\textheight,keepaspectratio]{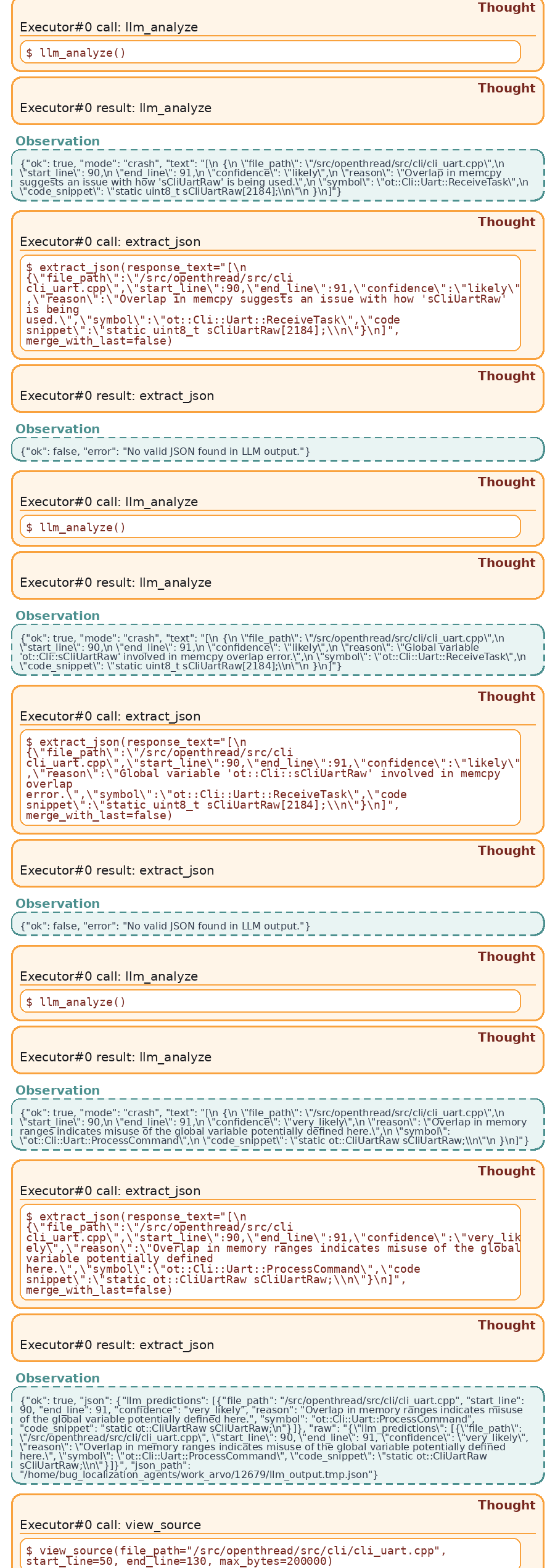}

\end{figure}

\begin{figure}[H]
  \centering
  \includegraphics[width=.48\textwidth,height=\textheight,keepaspectratio]{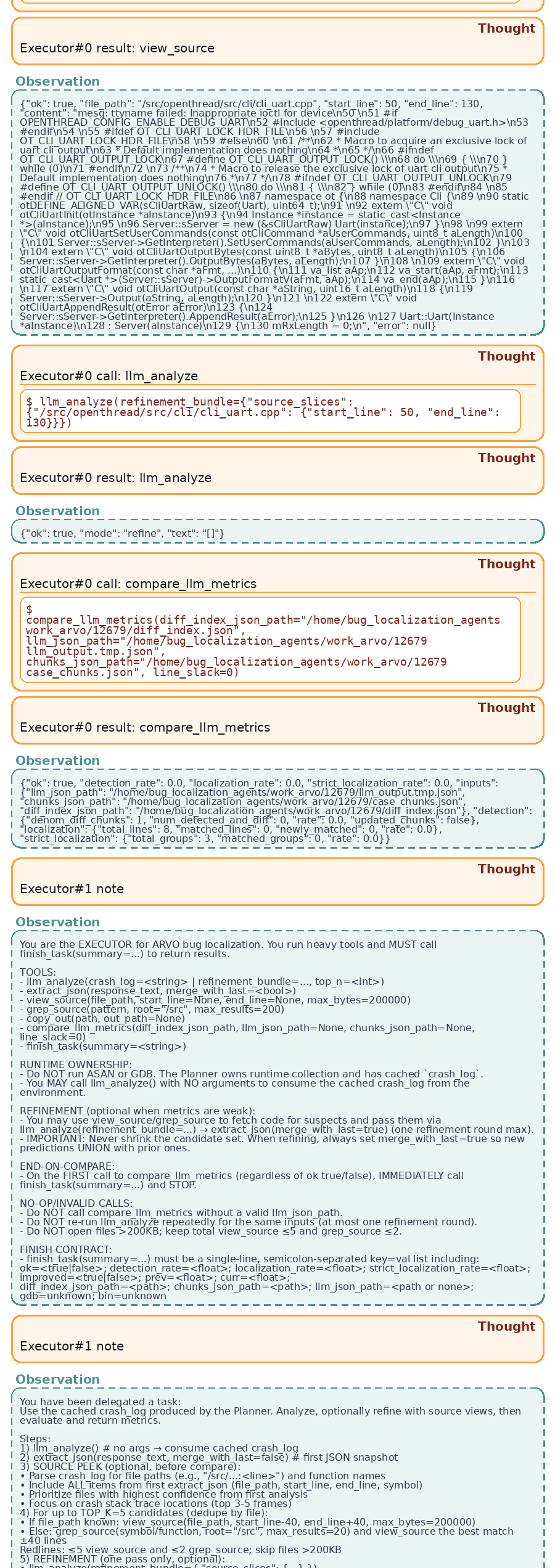}\hfill
  \includegraphics[width=.48\textwidth,height=\textheight,keepaspectratio]{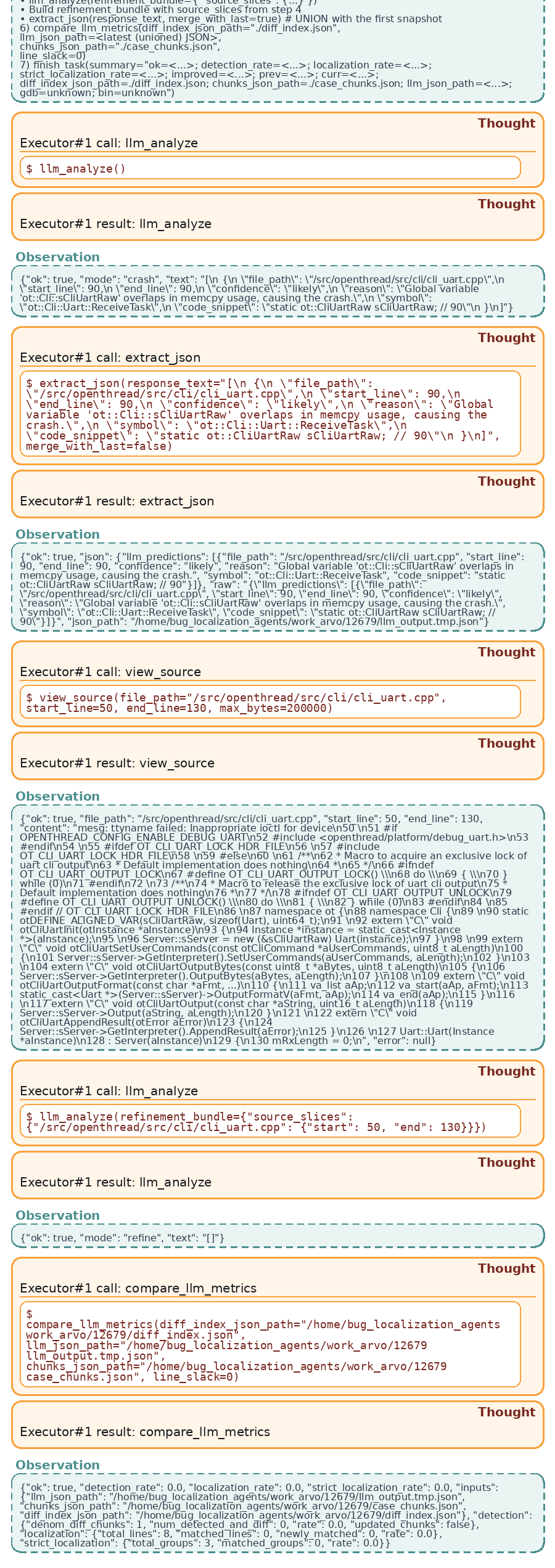}
  \caption{GPT-4o mini Baseline for case 12679.}
\end{figure}

\end{document}